\documentclass[twocolumn, trackchanges]{aastex7}

\usepackage{amsmath}
\usepackage{soul}
\usepackage{float}

\defcitealias{Sheehan2018}{SE18}

\begin{document}

\title{Physical and Chemical Characterization of GY 91's Multi-ringed Protostellar Disk with ALMA}

\author[orcid=0000-0003-1369-602X]{Sally D. Jiang}
\affiliation{Department of Astronomy, Columbia University, 538 W. 120th Street, Pupin Hall, New York, NY 10027, USA}
\email[show]{sj3323@columbia.edu}

\author[orcid=0000-0001-6947-6072]{Jane Huang}
\affiliation{Department of Astronomy, Columbia University, 538 W. 120th Street, Pupin Hall, New York, NY 10027, USA}
\email[show]{jane.huang@columbia.edu}

\author[0000-0002-1483-8811]{Ian Czekala} 
\affiliation{School of Physics \& Astronomy, University of St. Andrews, North Haugh, St. Andrews KY16 9SS, UK}
\email{ic95@st-andrews.ac.uk}
\author[0000-0002-8623-9703]{Leon Trapman}
\affiliation{Department of Astronomy, University of Wisconsin-Madison, 
475 N Charter St, Madison, WI 53706}
\email{ltrapman@wisc.edu}
\author[0000-0003-3283-6884]{Yuri Aikawa}
\affiliation{Department of Astronomy, Graduate School of Science, The University of Tokyo, 7-3-1 Hongo, Bunkyo-ku, Tokyo 113-0033, Japan}
\email{aikawa@astron.s.u-tokyo.ac.jp}

\author[0000-0003-2253-2270]{Sean M. Andrews}
\affiliation{Center for Astrophysics \textbar\ Harvard \& Smithsonian, 60 Garden Street, Cambridge, MA 02138, USA}
\email{sandrews@cfa.harvard.edu}

\author[0000-0001-7258-770X]{Jaehan Bae}
\affiliation{Department of Astronomy, University of Florida, Gainesville, FL 32611, USA}
\email{jbae@ufl.edu}
\author[0000-0003-4179-6394]{Edwin A. Bergin}
\affiliation{Department of Astronomy, University of Michigan, 323 West Hall, 1085 S. University Avenue, Ann Arbor, MI 48109, USA}
\email{ebergin@umich.edu}
\author[0000-0003-1413-1776]{Charles J.\ Law}
\altaffiliation{NASA Hubble Fellowship Program Sagan Fellow}
\affiliation{Department of Astronomy, University of Virginia, Charlottesville, VA 22904, USA}
\email{cjl8rd@virginia.edu}

\author[0000-0003-1837-3772]{Romane Le Gal}
\affiliation{Université Grenoble Alpes, CNRS, IPAG, F-38000 Grenoble, France}
\affiliation{Institut de Radioastronomie Millimetrique (IRAM), 300 rue de la piscine, F-38406 Saint-Martin d’Hères, France}
\email{romane.le-gal@univ-grenoble-alpes.fr}

\author[0000-0002-7607-719X]{Feng Long}
\altaffiliation{NASA Hubble Fellowship Program Sagan Fellow}
\affiliation{Kavli Institute for Astronomy and Astrophysics, Peking University, Beijing 100871, China}
\affiliation{Lunar and Planetary Laboratory, University of Arizona, Tucson, AZ 85721, USA}
\email{long.feng@pku.edu.cn}
\author[0000-0002-1637-7393]{Fran\c{c}ois M\'enard}
\affiliation{Univ. Grenoble Alpes, CNRS, IPAG, 38000 Grenoble, France}
\email{francois.menard@univ-grenoble-alpes.fr}
\author[0000-0001-8798-1347]{Karin I. \"Oberg} \affiliation{Center for Astrophysics \textbar\ Harvard \& Smithsonian, 60 Garden Street, Cambridge, MA 02138, USA}
\email{koberg@cfa.harvard.edu}
\author[0000-0001-8642-1786]{Chunhua Qi}
\affiliation{Institute for Astrophysical Research, Boston University, 725 Commonwealth Avenue, Boston, MA 02215, USA}
\email{cqi1@bu.edu}

\author[0000-0003-1534-5186]{Richard Teague}
\affiliation{Department of Earth, Atmospheric, and Planetary Sciences, Massachusetts Institute of Technology, Cambridge, MA 02139, USA}
\email{rteague@mit.edu}
\author[0000-0003-1526-7587]{David J. Wilner}\affiliation{Center for Astrophysics \textbar\ Harvard \& Smithsonian, 60 Garden Street, Cambridge, MA 02138, USA}
\email{dwilner@cfa.harvard.edu}
\author[0000-0002-0661-7517]{Ke Zhang}
\affiliation{Department of Astronomy, University of Wisconsin-Madison, 
475 N Charter St, Madison, WI 53706}
\email{ke.zhang@wisc.edu}

\begin{abstract}
 GY 91, commonly categorized as a Class I young stellar object, is notable for disk dust substructures that have been hypothesized to trace early planet formation. Using the ALMA 12-m and ACA arrays, we present new Band 7 dust continuum and molecular line observations of GY 91 at an angular resolution of $\sim0\farcs3$ (40 au). We report detections of CS $J=6-5$, N$_2$H$^+$ $J=3-2$, C$^{18}$O $J=3-2$, H$_2$CS $J_{K_a, K_c} = 8_{1,7}-7_{1,6}$, H$_2$CO $J_{K_a, K_c} = 4_{0,4}-3_{0,3}$, and H$_2$CO $J_{K_a, K_c} = 4_{2,3}-3_{2,2}$, as well as a tentative detection of $^{13}$C$^{18}$O $J=3-2$. We observe azimuthal asymmetry in CS and H$_2$CS emission, as well as radially structured H$_2$CO $4_{0,4}-3_{0,3}$ emission outside the dust continuum. C$^{18}$O and H$_2$CO 4$_{0,4}-3_{0,3}$ show significant cloud contamination, while CS and N$_2$H$^+$ are good tracers of Keplerian rotation originating from the disk. Envelope emission does not appear to contribute significantly either to the continuum or molecular line observations. GY 91's chemical properties appear in large part to resemble those of Class II disks, although observations of additional molecular probes should be obtained for a fuller comparison. With CS, we estimated a dynamical stellar mass of 0.58 $M_\odot$, which is higher than previous estimates from stellar evolutionary models (0.25 $M_\odot$). Using both radiative transfer modeling of the dust continuum and comparison of the C$^{18}$O and N$_2$H$^+$ fluxes to literature thermochemical models, we estimate a disk mass of $\sim0.01$ $M_\odot$.

\end{abstract}

\keywords{}

\section{Introduction}
High-resolution millimeter/sub-millimeter continuum observations have resulted in an updated view of protoplanetary disk evolution. Class II disks show diverse and nearly ubiquitous dust substructures such as rings, gaps, and spirals \citep[e.g.,][]{Andrews2018, Long2018, Cieza2019}. While the exact origins of these substructures are contested, theories have pointed to processes such as planet-disk interactions \citep[e.g.,][]{Goldreich1980, Zhang2018, Baruteau2014, Paardekooper2023}, magnetic disk winds \citep[e.g.,][]{Suriano2018}, dust evolution related to snowline locations \citep[e.g.,][]{Zhang2015}, or secular gravitational instability \citep{Takahashi2014}. Meanwhile, the prevalence of dust substructures in younger Class I disks is less clear. While the eDisk survey \citep{Ohashi2023} found that few disks in their sample showed clear substructure, \citet{Hsieh2025} and \citet{Vioque2025} report high detection rates in Class I disks. This difference may be due to the star-forming regions targeted or to the eDisk sample skewing younger than the other samples. If these substructures are due to planet-disk interactions, detections in Class I disks may place constraints on the timescales of planet formation and would suggest that giant planets are assembled faster than previously thought, in under 1 Myr \citep[e.g.,][]{Sheehan2018, Segura-Cox2020}. Alternatively, these substructures could represent sites that are favorable for future planet formation due to high concentrations of solids \citep[e.g.,][]{Morbidelli2020}. In either case, studies of Class I disks with substructures can provide a window into an important transitional point in disk evolution. 

The young stellar object (YSO) GY 91 (also known as ISO-Oph 54) is well-known for its disk substructures and has commonly been categorized as Class I based on its bolometric temperature (372 K) and mid-IR SED slope ($\alpha_{IR}$ = 0.45) \citep{Enoch2009, Sheehan2018}. It is located in the Rho Ophiuchi cloud complex, which is at a distance of 140 parsecs \citep{Canovas2019}. GY 91 itself does not have a parallax measurement from \textit{Gaia} \citep{Gaia2016, Gaia2021} due to high extinction ($A_V$ = 53, \citet{Cieza2021}). The millimeter continuum disk consists of at least 4 pairs of axisymmetric rings and gaps \citep{Sheehan2018, Cieza2021}. From radiative transfer modeling of the SED and millimeter continuum, \citet{Sheehan2018} estimated an envelope-to-disk mass ratio of $\sim0.12$ and an age of 0.5 Myr. However, \citet{McClure2010} argued that GY 91 should more appropriately be categorized as an envelope-free Class II YSO with a high amount of foreground cloud material being responsible for the infrared excess mimicking an envelope contribution. \citet{vanKempen2007} also classified GY 91 as an envelope-free source based on a single-dish HCO$^+$ survey of YSOs. 

Most of the published ALMA observations of GY 91 so far have been high-resolution continuum observations, which resolve out larger-scale emission \citep{Hsieh2024,Sheehan2018,Cieza2021}. The only spatially resolved molecular line observations published to date have been of $^{12}$CO $J=2-1$ with a synthesized beam size of $0\farcs13$ (18 au) and maximum recoverable scale of $0\farcs67$ (95 au), while the gas disk extends past a radius of 100 au \citep{Antilen2023}. \citet{Antilen2023} noted the presence of significant cloud emission in $^{12}$CO, with disk emission only being detected via the Keplerian line wings that trace smaller radii.

To investigate the physical and chemical properties of GY 91 further, we obtained new ALMA molecular line and dust continuum observations of GY 91 using multiple antenna configurations in order to be sensitive to emission on both disk and envelope scales. We targeted a suite of molecular lines, including common bright disk gas tracers such as C$^{18}$O, N$_2$H$^+$, CS, and H$_2$CO \citep[e.g.,][]{Anderson2019, LeGal2019, Pegues2020, Oberg2021}, in order to study the structure, kinematics, and chemistry of the gas around GY 91. Because the high foreground extinction poses a challenge to observing GY 91, we targeted transitions with upper state energy levels $>25$ K with the aim of preferentially tracing the warmer disk gas over colder foreground cloud material. 

We describe the observational setup and data reduction process in \autoref{sec:observations}, provide an overview of line detections and spatial distributions in \autoref{sec:detections}, estimate the dynamical stellar mass in \autoref{sec:dyn_mass}, gauge envelope contributions to the observed emission in \autoref{sec:envelope}, and estimate the disk mass in \autoref{sec:radmodel}. Finally, we discuss the results in \autoref{sec:discussion} and present conclusions in \autoref{sec:conclusion}.

\begin{table*}
    \centering
        \caption{Observational Setup for GY 91}
    \label{tab:arrayconfig}
    \begin{tabular}{cccccccc} \hline \hline
        Spectral Setting&  Obs Date&  Antenna &  Baselines &  Time on target &  Flux/Bandpass Calibrator & Phase Calibrator \\
        &  &  Count &  (m) &  (min)&   &  \\\hline 
         7a&  2022 March 10 &  43 (12-m) &  15-284&  18& J1517-2422& J1625-2527\\  
         &  2022 April 24  &  10 (ACA 7-m)&  8.9-45&  42& J1924-2914& J1700-2610\\  
         &  2022 April 25 &  10 (ACA 7-m)&  8.9-45 &  42&  J1924-2914& J1700-2610\\  
         &  2022 June 14 &  42 (12-m)&  15.1-1300&  26&  J1517-2422& J1700-2610\\  
         &  2022 June 30 &  46 (12-m)&  15.1-1300&  26&   J1517-2422&J1700-2610\\  \hline
         7b&  2022 March 08 &  42 (12-m)&  15-284&  33&  J1517-2422& J1625-2527\\  
         &  2022 March 27 &  8 (ACA 7-m)&  8.9-45&  47&  J1924-2914& J1700-2610\\  
         &  2022 April 19 &  10 (ACA 7-m)&  8.9-45&  47& J1924-2914& J1700-2610\\  
         &  2022 April 22 &  10 (ACA 7-m)&  8.9-45&  47&  J1924-2914& J1700-2610\\ 
         &  2022 April 23 &  10 (ACA 7-m)&  8.9-45&  47&  J1924-2914& J1700-2610\\ 
         &  2022 June 10 &  42 (12-m)& 15.1-1200&  49&  J1517-2422& J1700-2610\\ 
         &  2022 June 11 &  43 (12-m)&  15.1-1200& 49&  J1517-2422& J1700-2610\\ \
         
    \end{tabular}
\end{table*}

\begin{table*}
    \centering
\caption{Spectral Window Setup}
\label{tab:spectralconfig}
    \begin{tabular}{cccccccc}\hline \hline
        Spectral Setting&  Targeted Line& $E_u$ & Rest Frequency*& Resolution & Bandwidth & Channel & \\
        &  &  (K)&(GHz) & (MHz) & (MHz)& Number & \\\hline 
         7a&  H$_2$CS $J_{K_a, K_c}=8_{1,7}-7_{1,6}$ (ortho)
         & 73.4 &278.8876613& 0.141&117.19-125& 960 - 1024& \\
         &  N$_2$H$^+$ $J=3-2$&  26.8&279.5117491&0.141&117.19-125& 960 - 1024& \\
         &  H$_2$CO $J_{K_a, K_c} = 4_{0,4}-3_{0,3}$ (para)&34.9&  290.623405& 0.141 &117.19-125& 960 - 1024& \\
         &  H$_2$CO $J_{K_a, K_c} = 4_{2,3}-3_{2,2}$ (para) &82.1&  291.2377664& 0.141&117.19-125& 960 - 1024& \\
         &  CS $J=6-5$&  49.4&293.9120865& 0.141&234.38-250& 1920 - 2048&\\
         &  Continuum&  & 281 & 31.25&2000& 128& \\ \hline 
        7b &  $^{13}$C$^{18}$O $J=3-2$&  30.2&314.11964530& 0.141&234.38-250& 1920 - 2048& \\
        &  C$^{18}$O $J=3-2$ &  31.6&329.33055250& 0.141&234.38-250& 1920 - 2048& \\
        &  Continuum &  & 316& 31.25&2000& 128& \\
        &   Continuum &  & 328.2& 31.25&2000& 128& \\
    \end{tabular}
    \tablecomments{*Rest frequencies and $E_u$ obtained from the Cologne Database for Molecular Spectroscopy (CDMS) \citep{Muller2001, Muller2005, Endres2016}}
\end{table*}

\section{ALMA Observations and Data Reduction}\label{sec:observations}

Observations of GY 91 were taken in ALMA Project 2021.1.01588.S (P.I.: J. Huang) in Cycle 8. In order to maintain sensitivity to larger-scale envelope emission while still spatially resolving the disk, GY 91 was targeted using two 12-m antenna configurations (which we refer to as ``compact'' for the one with shorter maximum baselines and ``extended'' for the one with longer maximum baselines) and the 7-meter Atacama Compact Array (ACA). Further information about the observational setup is shown in \autoref{tab:arrayconfig}. Two Band 7 spectral settings were used, covering both the continuum and the following seven lines: CS $J=6-5$, N$_2$H$^+$ $J=3-2$, H$_2$CS $J_{K_a, K_c}=8_{1,7}-7_{1,6}$, H$_2$CO $J_{K_a, K_c}=4_{0,4}-3_{0,3}$, H$_2$CO $J_{K_a, K_c}=4_{2,3}-3_{2,2}$, C$^{18}$O $J=3-2$, and $^{13}$C$^{18}$O $J=3-2$. We refer to the lower frequency setting as ``7a'' and the higher frequency setting as ``7b.'' The maximum recoverable scale (MRS)\footnote{Defined by the ALMA Technical Handbook to be $\frac{0.983\lambda}{L_5}$, where $\lambda$ is the observing wavelength and $L_5$ is the 5th percentile baseline length}
is $7\farcs3$ (1020 au) for setting 7a and $7\farcs2$ (1010 au) for 7b. The spectral setup is explained in more detail in \autoref{tab:spectralconfig}. 

Initial calibration of all data was performed by NAASC (North American ALMA Science Center) staff using the CASA 6.2 pipeline \citep{CASA2022}. Further self-calibration and imaging were completed using CASA 6.5.
We first flagged all channels where line emission was expected to be present and created spectrally averaged continuum measurement sets for each spectral setting. The continuum from each execution was then imaged using the \texttt{tclean} algorithm with Briggs weighting and a \texttt{robust} value of 0.5, using the Hogbom deconvolver for the ACA and compact 12-m observations and the multiscale deconvolver for the extended 12-m observations. To correct for slight (sub-beam) positional offsets between observations from different configurations, we followed the procedure described in \citet{Andrews2018}. We fit a Gaussian to the continuum to measure the position of the disk centers, applied phase shifts to move the disk center to the phase center, and then relabeled the phase centers to match the coordinates of one of the 12-m extended configuration execution block's phase center. The measurement sets were then reimaged and manually checked to ensure the centers were properly aligned. Finally, the measurement sets for the different configurations were concatenated to create a combined set of continuum visibilities for each spectral setting. 

Self-calibration and imaging processes were then performed separately for the two spectral settings. Two iterations of phase-only self-calibration were performed, first with a scan-length solution interval and then with a 120 s interval. Amplitude calibration was conducted once with a scan-length solution interval. The final continuum image for each spectral setting was produced using multiscale CLEAN with scales of [$0''$, $0\farcs2$, $0\farcs4$, $1''$] and a \texttt{robust} value of 0.5. 

\floattable
\begin{table*}[htp!]
    \centering
\caption{Imaging Summary}
\label{tab:imaging}
    \begin{tabular}{cccccc}\hline \hline
    Transition & Synthesized beam & Tapered &Per-channel rms\tablenotemark{a}  & Moment 0 rms  & Flux\tablenotemark{c}\\
    &(arcsec $\times$ arcsec, (deg)) & ($0\farcs2$) & (mJy beam$^{-1}$) &(mJy beam$^{-1}$ km s$^{-1}$)&(mJy km s$^{-1}$)\\
    \hline
         H$_2$CS $J_{K_a, K_c}=8_{1,7}-7_{1,6}$ & $0\farcs43\times 0\farcs34$, $(-88\fdg2)$ & $\checkmark$ & 2.0& 2.5 & 51 $\pm$ 8 \\
         N$_2$H$^+$ $J=3-2$ & $0\farcs43 \times 0\farcs34$ $(-88.4)$ & $\checkmark$ & 2.2& 3.2 & 282 $\pm$ 9\\
         H$_2$CO $J_{K_a, K_c} = 4_{0,4}-3_{0,3}$ & $0\farcs42 \times 0\farcs34$ $(85\fdg7)$ & $\checkmark$ & 2.3& 3.5 & 322$\pm$ 12\tablenotemark{d} \\
         H$_2$CO $J_{K_a, K_c} = 4_{2,3}-3_{2,2}$ & $0\farcs41 \times  0\farcs34$ $(-89\fdg 9)$ & $\checkmark$ & 2.0& 3.1 & 58 $\pm$ 8 \\
         CS $J=6-5$& $0\farcs30 \times 0\farcs23$ $(-85\fdg 1)$ & $\times$ & 1.9& 3.0 & 768 $\pm$ 9 \\
         $^{13}$C$^{18}$O $J=3-2$& $0\farcs46\times 0\farcs36$ ($75\fdg0$) & $\checkmark$ & 1.7& 2.7 & 17 $\pm$ 7 \\ 
         C$^{18}$O $J=3-2$& $0\farcs37\times0\farcs24$ ($68\fdg6$) & $\times$ & 2.2& 4.2  & 696 $\pm$ 12\tablenotemark{d} \\         
         1.05 mm continuum (7a) & $0\farcs32\times0\farcs25$ $(82\fdg6)$& $\times$ & - & 0.02\tablenotemark{b} & 155.1 $\pm$ 0.2\\
         0.93 mm continuum (7b)& $0\farcs37\times0\farcs24$ ($70\fdg1$) & $\times$ & - & 0.03\tablenotemark{b} & 223.6 $\pm$ 0.2\\
    \end{tabular}
    \tablenotetext{a}{Channels have a velocity width of 0.25 km s$^{-1}$}
    \tablenotetext{b}{Continuum rms units are in mJy beam$^{-1}$}
        \tablenotetext{c}{Error estimate do not include 10\% systematic flux calibration uncertainty}
    \tablenotetext{d}{Significant cloud contamination}
\end{table*}

We then applied the gain tables from the phase and amplitude self-calibration of the continuum to the line spectral windows. The continuum was fit with a first-order polynomial and subtracted from the spectral windows using \texttt{uvcontsub}. Line images were produced with \texttt{tclean} using the multiscale deconvolver and the \texttt{AUTO-MULTITHRESH} auto-masking algorithm \citep{Kepley2020} to create masks for the irregularly-shaped line emission. The automasking parameters were motivated by NRAO's CASA Guide-recommended values.\footnote{\url{https://casaguides.nrao.edu/index.php/Automasking_Guide_CASA_6.5.4}} To improve the signal-to-noise (S/N) of the weaker lines (H$_2$CS, N$_2$H$^+$, both H$_2$CO transitions, and $^{13}$C$^{18}$O), a $0\farcs2$ $uv$-taper was applied in \texttt{tclean}. This procedure multiplies the weights by a 2D Gaussian with a user-specified FWHM, thereby downweighting data at longer baselines, increasing surface brightness sensitivity, and degrading the angular resolution \citep{Briggs1995}. Finally, all images were primary beam-corrected. The image properties are summarized in \autoref{tab:imaging}. \footnote{The visibilities and images can be downloaded from Zenodo: \url{https://zenodo.org/records/17965437}}

\begin{figure*}[htp!]
    \centering
    \includegraphics[width=\linewidth]{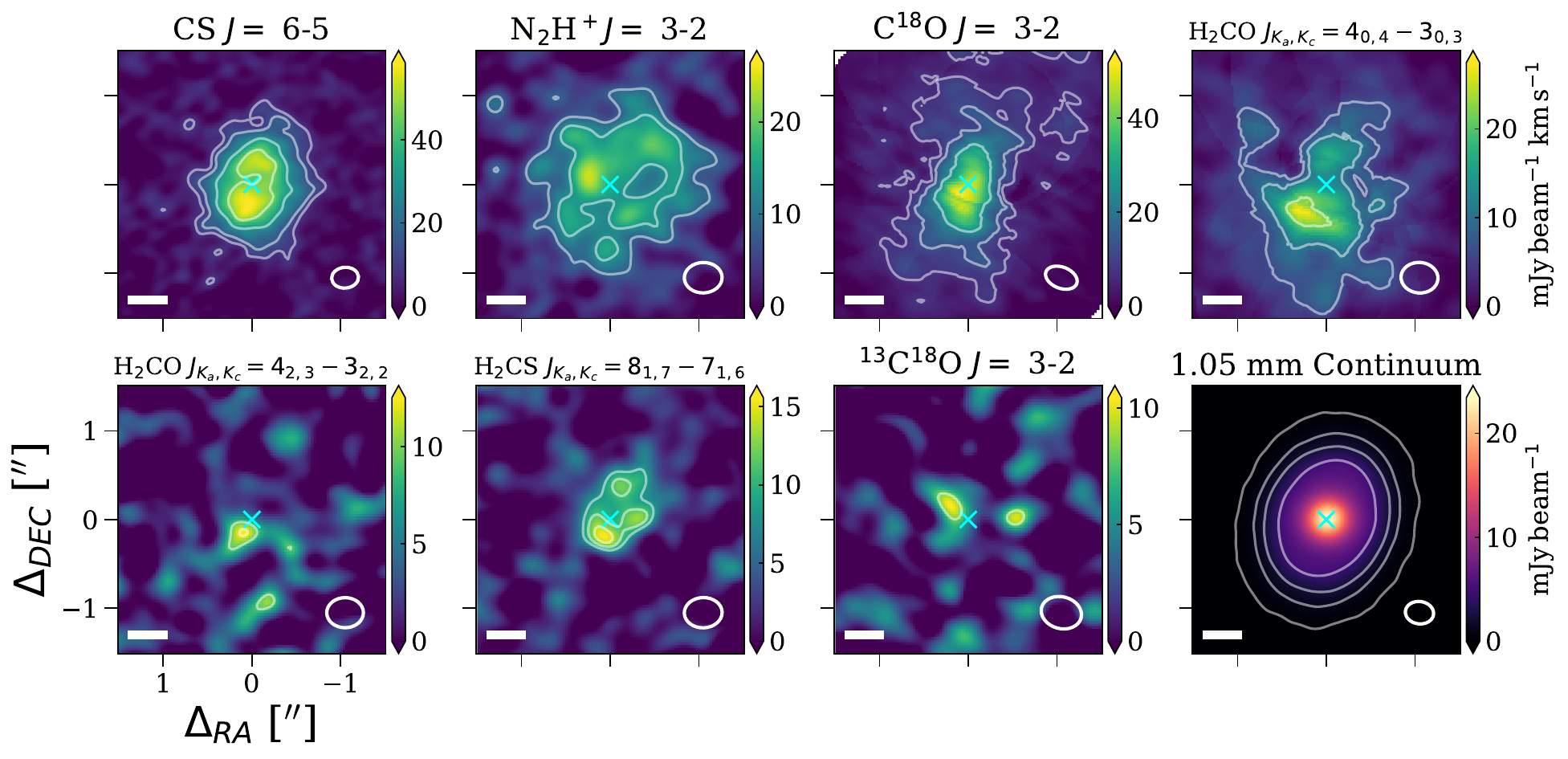}
    \caption{Zeroth moment maps of molecular line emission and the 1.05 mm dust continuum image. The synthesized beam size is indicated in the lower right of each image. CS contours are at the 3, 5, 10, and 15$\sigma$ levels, where $\sigma$ is the rms listed in Table \ref{tab:imaging}. The N$_2$H$^+$, H$_2$CO $4_{0,4}-3_{0,3}$, and C$^{18}$O contour levels are at 3, 5, and 7$\sigma$. H$_2$CS, H$_2$CO $4_{2,3}-3_{2,2}$, and $^{13}$C$^{18}$O have contours at 3, 4, and 5$\sigma$. The dust continuum contours are at 10, 50, 100, and 200$\sigma$. A Keplerian mask was used to compute the zeroth-moment maps of H$_2$CO $4_{0,4}-3_{0,3}$ and C$^{18}$O. A 50 au scale bar is shown in the bottom left of each panel. The location of the continuum peak is marked with a cyan cross.}
    \label{fig:moment0}    
\end{figure*}

\begin{figure*}[htp!]
    \centering
    \includegraphics[width=\linewidth]{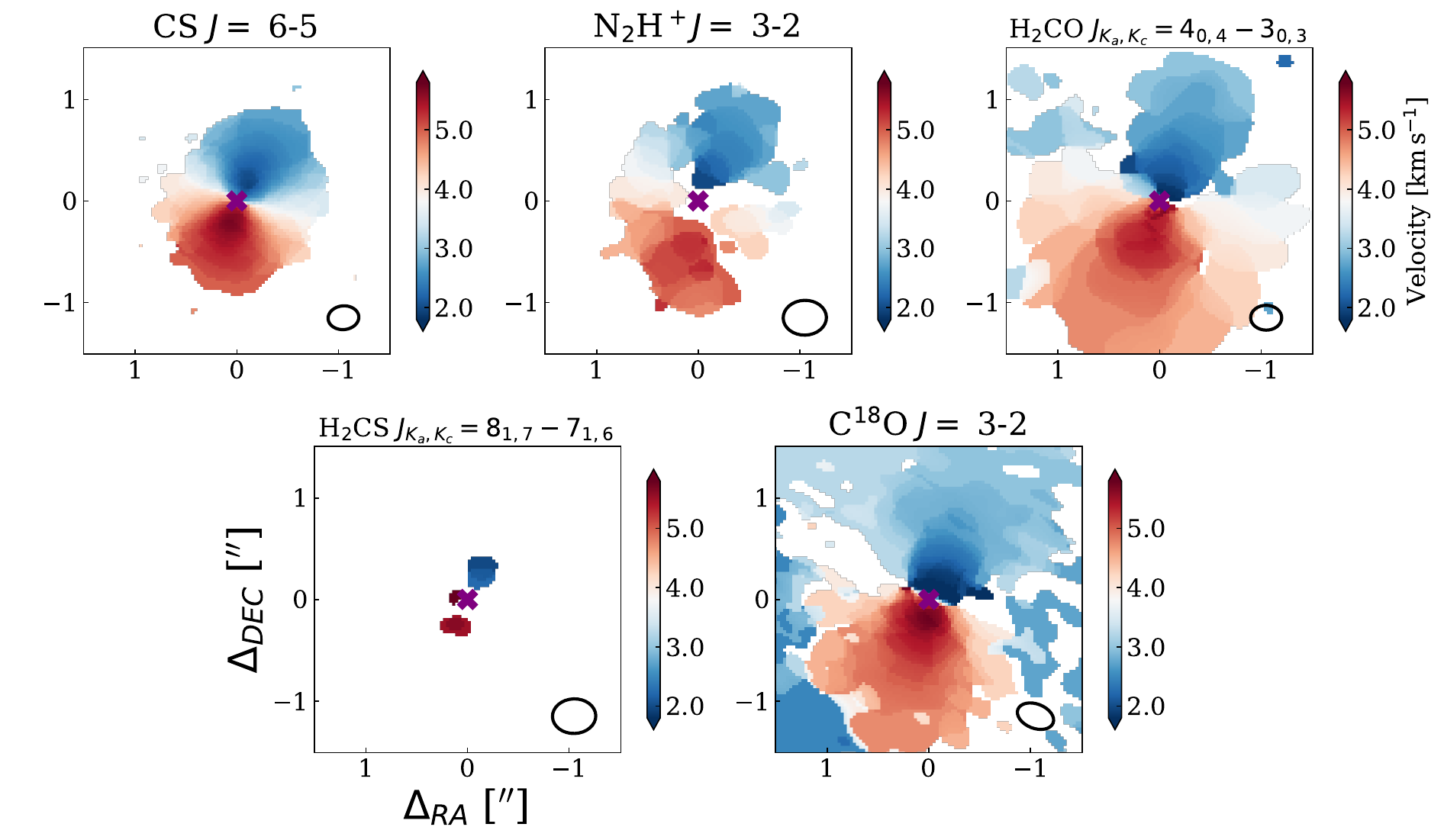}
    \caption{Moment 1 maps of CS, N$_2$H$^+$, H$_2$CO 4$_{0,4}$-3$_{0,3}$, H$_2$CS, and C$^{18}$O. The synthesized beam is shown in the lower right corner of each panel. }
    \label{fig:moment1}    
\end{figure*}

\begin{figure*}[htp!]
    \centering
      \includegraphics[width=\linewidth]{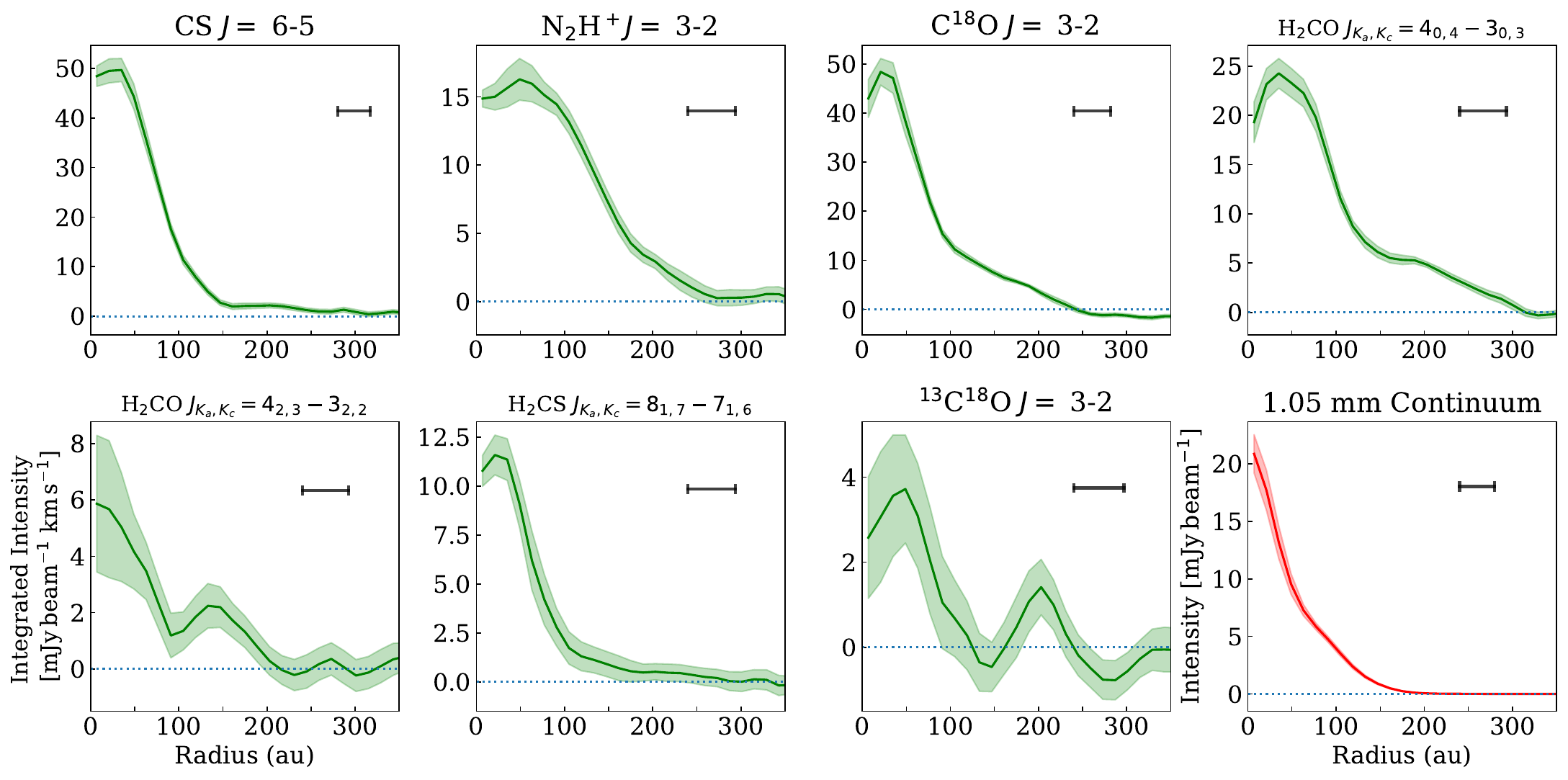}
    \caption{Deprojected, azimuthally-averaged radial profiles for each line and the 1.05 mm dust continuum. The shaded areas show the statistical $1\sigma$ uncertainty. The horizontal black lines show the scale of the synthesized beam.} 
    \label{fig:radprof}    
\end{figure*}

\begin{figure*}[htp!]
    \centering
    \includegraphics[width=\linewidth]{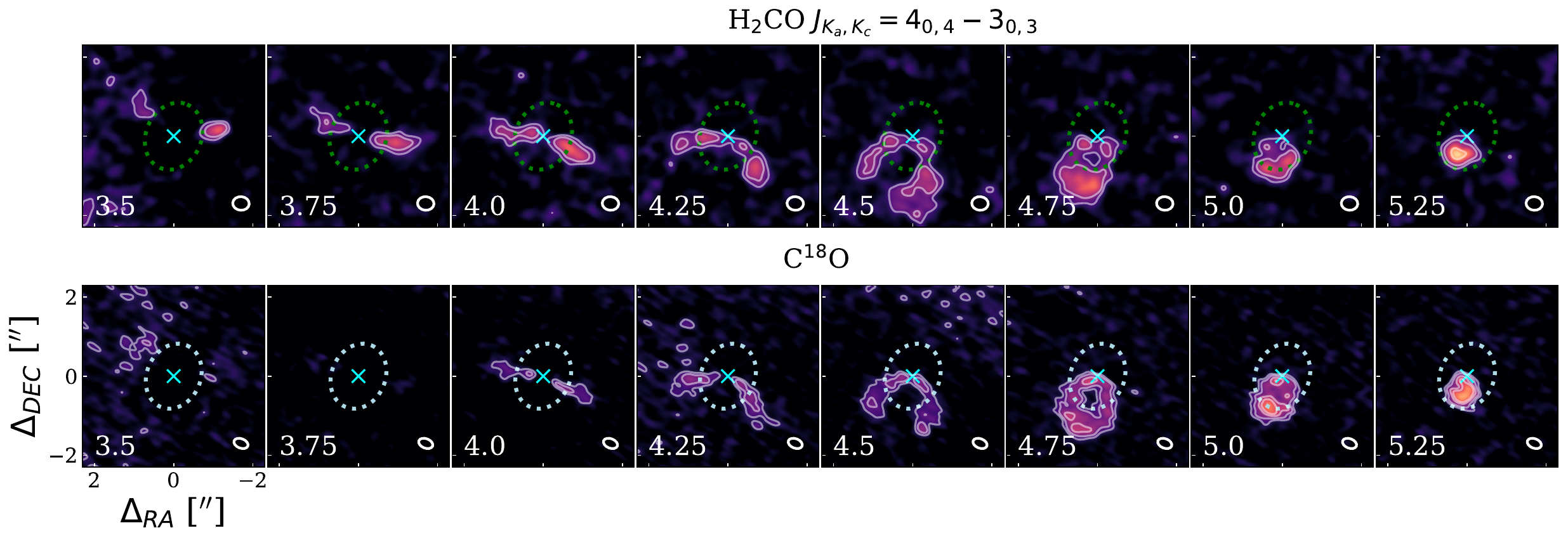}
    \caption{Subset of channels of H$_2$CO $4_{0,4}-3_{0,3}$ and C$^{18}$O where an annular gap in emission is visible. Ellipses are drawn with the same P.A. and inclination as the GY 91 disk to denote the radius at which the break in emission is observed. The semi-major axis of the H$_2$CO ellipse is 122 au, while that for C$^{18}$O is 118 au. Contours correspond to 3, 5, 10, and 15$\sigma$, where $\sigma$ is the rms listed in \autoref{tab:imaging}.}
    \label{fig:channelmap_rings}    
\end{figure*}

\section{Molecular Line Characterization \label{sec:detections}}
\subsection{Image overview}\label{subsec:emissiondistr}
We present the individual channel maps for each line in \autoref{appsec:channelmaps}. Matched filtering was used to quantify the significance of line detection (see \autoref{appsec:matchedfilter}). The 1.05 mm continuum image and integrated intensity maps (zeroth-moment maps) for the molecular lines are shown in \autoref{fig:moment0}. Although high-resolution ALMA images ($<0\farcs13$) have shown that the GY 91 disk has multiple dust gaps \citep{Sheehan2018, Cieza2021}, our new continuum image appears mostly smooth due to our lower resolution ($\sim0\farcs3$). The integrated intensity maps were created for most lines by collapsing the image cube from 1 to 7 km s$^{-1}$, which represents the velocity range over which CS (the strongest line without significant foreground contamination) is detected above the $3\sigma$ level.  C$^{18}$O was integrated over a slightly wider velocity range (0 to 7.5 km s$^{-1}$). Cloud contamination is visible in the channel maps (\autoref{appsec:channelmaps}) for both C$^{18}$O (2.5 km s$^{-1}$ $<$ $v_{lsrk}$ $<$ 3.75 km s$^{-1}$) and H$_2$CO $4_{0,4}-3_{0,3}$ (2.75 km s$^{-1}$ $<$ $v_{lsrk}$ $<$ 3.5 km s$^{-1}$). To address this, a Keplerian mask was used to limit the zeroth-moment computation to the spatial regions where disk emission is expected. The mask was generated with the \texttt{keplerian\textunderscore mask} software \citep{Teague2020keplerianmask} using the inclination ($36\fdg4$) and position angle ($157\fdg6$) constrained from parametric modeling of the high-resolution millimeter continuum in \autoref{appsec:paramodel}, and the stellar mass (0.58 $M_\odot$) and systemic velocity (3.87 km s$^{-1}$) obtained from modeling the CS kinematics in \autoref{sec:dyn_mass}, respectively. We used a radius of $2''$ for the Keplerian mask and convolved it with a Gaussian with the same size as the synthesized beam. 

In the C$^{18}$O and H$_2$CO $4_{0,4}-3_{0,3}$ channel maps (\autoref{appsec:channelmaps}) and integrated intensity maps (\autoref{fig:moment0}), the northwest side of the disk shows cloud contamination, while the southeast side exhibits minimal contamination. The integrated intensity maps for CS and H$_2$CS (\autoref{fig:moment0}) show slightly asymmetric emission, with the southeast of the disk being brighter than the northwest. H$_2$CS does not show any evidence of large-scale non-disk emission, and CS shows only very faint large-scale emission in a couple channels (3.25-3.5 km s$^{-1}$) of the channel maps (\autoref{appsec:channelmaps}), so their asymmetric emission may reflect genuine underlying azimuthal asymmetries in their column densities (see \autoref{sec:discussion} for further commentary).

For the lines with sufficient S/N (CS, N$_2$H$^+$, H$_2$CO $4_{0,4}-3_{0,3}$, C$^{18}$O and H$_2$CS), we created intensity-weighted velocity maps (moment 1 maps) with the same velocity range as the integrated intensity maps. These moment 1 maps are shown in \autoref{fig:moment1}. For most of these, a $5\sigma$ pixel clip was applied so that values were only calculated for strongly-detected emission. A $4\sigma$ clip was applied for H$_2$CS due to its lower S/N. Keplerian rotation tracing the disk is visible in the maps for CS, N$_2$H$^+$, H$_2$CO $4_{0,4}-3_{0,3}$, and C$^{18}$O, with the northwest side of the disk being blueshifted with respect to the systemic velocity (3.87 km s$^{-1}$), and the southeast side being redshifted. The H$_2$CO $4_{0,4}-3_{0,3}$ and C$^{18}$O appear asymmetric due to foreground absorption and the presence of large-scale emission (see also \autoref{sec:envelope}), whereas CS and N$_2$H$^+$ trace the disk more cleanly. The kinematics of H$_2$CS are less clear due to the lower S/N, but it shows blueshifted emission to the northwest of the star and redshifted emission in the southeast, similar to the other lines. Furthermore, its compact radial extent in the integrated intensity map (\autoref{fig:moment0}) suggests that H$_2$CS also likely originates predominantly in the disk.

Radial profiles for all lines and the 1.05 mm dust continuum are shown in \autoref{fig:radprof}. The deprojected, azimuthally averaged radial profile for each line was calculated from the integrated intensity maps using \texttt{gofish} \citep{Teague2019_gofish}. Because the blueshifted (northwest) sides of the C$^{18}$O and H$_2$CO $4_{0,4}-3_{0,3}$ disk emission exhibit heavy contamination, we calculated the radial profiles by creating integrated intensity maps of the redshifted (southeast) side (4 km s$^{-1}$ $<$ v$_{lsrk}$ $<$ 7 km s$^{-1}$) and calculating the radial profiles using an azimuthal mask over a particular position angle range. 
For H$_2$CO, we selected the region from 157$\pm$75$^\circ$. For C$^{18}$O, we selected a slightly smaller region ($\pm$60$^\circ$) due to the stronger cloud contamination.


The radial profiles of CS and H$_2$CS are similar, with both lines peaking at a radius of $\sim$25 au and most of the emission originating from within $\sim150$ au. The CS emission additionally shows a faint emission shelf out to $\sim300$ au, which is not seen in H$_2$CS (\autoref{fig:radprof}). This difference may be due to the higher S/N of the CS emission. N$_2$H$^+$ peaks further out at $\sim50$ au and extends to $\sim250$ au. While the radial profiles of C$^{18}$O and H$_2$CO $4_{0,4}-3_{0,3}$ appear to have a central dip, this may be an artifact from having to exclude part of the disk from the radial profile calculation due to foreground absorption. H$_2$CO $4_{0,4}-3_{0,3}$ shows a small decrement at $\sim150$ au and bump at $\sim200$ au, outside the millimeter continuum (R$_{90}$ = 117 au; see \autoref{appsec:paramodel}). The C$^{18}$O profile shows a tail over the same radial range where H$_2$CO has an emission bump. This substructure is also seen in some of the individual channels, where there appears to be an emission gap along the wings of the Keplerian pattern 
(\autoref{fig:channelmap_rings}). We also see fluctuations in the radial profile of H$_2$CO $4_{2,3}-3_{2,2}$ (\autoref{fig:radprof}), but this is likely due to the low S/N .

\subsection{Column Density Estimates}
Fluxes for each line were calculated after applying a Keplerian mask to the image cubes to minimize contributions from cloud contamination. To calculate flux errors, we used the Keplerian mask to obtain 200 flux samples of random regions off-source and then took the standard deviation. The fluxes are listed in Table \ref{tab:imaging}.

Assuming that the line emission is optically thin and in local thermal equilibrium and that any background emission is small compared to the line emission, 
the disk-averaged column density can be estimated with the following formula \citep[e.g.,][]{Goldsmith1999, Bisschop2008}:
\begin{equation}
N = \frac{4\pi S_\nu \Delta \nu Q(T)}{A_{ul} \Omega h c g_u}e^{E_u/T},
\end{equation}
where $S_\nu \Delta \nu$ is the integrated flux over the disk, $Q(T)$ is the partition function, $A_{ul}$ is the Einstein coefficient, $g_u$ is the upper state degeneracy, and $\Omega$ is the solid angle subtended by the region in which the emission is measured. For our flux measurements, $\Omega$ corresponds to the ellipse spanning the Keplerian mask used to extract the disk fluxes. To assess whether the optically thin approximation is reasonable, we checked the peak brightness temperatures (calculated using the full Planck equation) of each line from the image cubes. Most have peak brightness temperatures $<10$ K, below the typical gas temperatures of $\sim20-50$ K probed by ALMA molecular emission \citep[e.g.][]{Law2021_profile}. The peak brightness temperatures of C$^{18}$O (12 K) and CS (17 K) are colder than the expected gas temperatures, although not to as great a degree as the other targeted molecules. Therefore the optically thin approximation generally seems to be reasonable, although the CS column density may be somewhat underestimated depending on the actual temperature.    

Based on the range of excitation temperatures estimated for the targeted molecules in other systems \citep[e.g.][]{LeGal2019, Qi2019, Pegues2020, Law2021_profile}, we calculate the column densities assuming gas temperatures of 20 and 50 K. The H$_2$CO and H$_2$CS column densities are calculated assuming a standard ortho-to-para ratio of 3 (although \citet{Guzman2018} and \citet{Terwisscha2021} found that the ortho-to-para ratio of H$_2$CO might be lower in the cold regions of disks). For H$_2$CO, we only use the para $4_{0,4}-3_{0,3}$ transition to estimate the column density because the para $4_{2,3}-3_{2,2}$ emission is only marginally detected in the image cube. While the weaker H$_2$CO line is well-detected with the matched filter method, and ratios of impulse responses for different transitions of the same species have been used as proxies for the flux ratios \citep[e.g.,][]{Carney2017, Loomis2018}, this kind of estimate could be problematic for our H$_2$CO observations given the strong cloud contamination. For CS, H$_2$CS, and N$_2$H$^+$, the fluxes are taken from \autoref{tab:imaging}. Because the blueshifted side of the disk is heavily cloud-contaminated in C$^{18}$O and H$_2$CO $4_{0,4}-3_{0,3}$, we estimated the total disk emission for these two lines by measuring the integrated flux redward of the systemic velocity and multiplying by 2, obtaining a value of 690 mJy km s$^{-1}$ for C$^{18}$O and 430 mJy km s$^{-1}$ for H$_2$CO. A $10\%$ systematic flux calibration uncertainty is assumed. The spectroscopic parameters and disk-averaged column densities for lines that are firmly detected in the images are listed in \autoref{tab:columndensities}.

\begin{deluxetable*}{ccccccccc}

\tablecaption{Disk-averaged column densities\label{tab:columndensities}}
 \tablehead{\colhead{Targeted Line}& \colhead{$E_u$} &\colhead{ Rest Frequency}& \colhead{$g_u$} &\colhead{ $A_{ul}$}& \colhead{$Q$ (20 K)} & \colhead{$N$ (20 K) }& \colhead{$Q$ (50 K)}&\colhead{ $N$ (50 K)}  \\
        & \colhead{ (K)}&\colhead{(GHz)} &  & \colhead{(s$^{-1}$)}&  & \colhead{(cm$^{-2}$)} & & \colhead{(cm$^{-2}$)}}
    \startdata
      H$_2$CS $J_{K_a, K_c}=8_{1,7}-7_{1,6}$ (ortho)
         & 73.4 &278.8876613&51 & $3.18\times10^{-4}$&155 &$(5.1\pm1.0)\times10^{12}$&425&$(1.5\pm0.3)\times10^{12}$\\
         N$_2$H$^+$ $J=3-2$&  26.8&279.5117491& 63&$1.26\times10^{-3}$ & 83.6 &$(2.3\pm0.2)\times10^{11}$&204.3 & $(2.5\pm0.3)\times10^{11}$\\
         H$_2$CO $J_{K_a,K_c} = 4_{0,4}-3_{0,3}$ (para)&34.9&  290.623405& 9& $6.90\times10^{-4}$ &13.6 & $(4.3\pm0.4)\times10^{12}$& 51.7& $(5.7\pm0.6)\times10^{12}$\\
          CS $J=6-5$&  49.4&293.9120865&13&$5.23\times10^{-4}$&17.4 &$(4.7\pm0.5)\times10^{12}$&42.9& $(2.6\pm0.3)\times10^{12}$\\
        C$^{18}$O $J=3-2$ &  31.6&329.33055250&7 & $2.17\times10^{-6}$ &7.9 & $(3.4\pm0.3)\times10^{14}$ &19.3 & $(3.2\pm0.3)\times10^{14}$\\
    \enddata
    \tablecomments{Values for $A_{ul}$, $E_u$, $g_u$, and $Q$ (interpolated) taken from Cologne Database for Molecular Spectroscopy (CDMS) \citep{Endres2016}. The H$_2$CO and H$_2$CS column density estimates correspond to an ortho-to-para ratio of 3:1.}
\end{deluxetable*}

\section{Stellar Host Dynamical Mass Estimate}\label{sec:dyn_mass}
To better understand the physical properties of the YSO and disk, it is essential to obtain an estimate of the stellar mass. We modeled the Keplerian rotation traced by CS to estimate the dynamical stellar mass \citep[e.g.,][]{Simon2000, Rosenfeld2012}. CS was chosen for this analysis because of its high S/N and minimal cloud contamination.

\subsection{Image plane modeling}
We used the \texttt{bettermoments} package \citep{Teague2018bettermoments, Teague2019bettermoments} to create a CS moment 1 map (masking pixels with S/N $<5$) and estimate the associated uncertainties. The moment map was downsampled by the beam size so that adjacent pixels are independent. The moment map was then modelled with the \texttt{eddy} \citep{Teague2019_eddy} package as a geometrically thin, Keplerian disk. The line-of-sight velocity is expressed as

\begin{equation}\label{eq:velocity}
    v_0 = \sqrt{\frac{GM_{\ast}}{r}} \times \cos \phi \times \sin i + v_{\rm sys}
\end{equation} 
where $r$ is the radial separation from the star in the disk frame in units of arcseconds, $\phi$ is the polar angle in the disk frame, and $v_{\rm sys}$ is the systemic velocity.

\begin{figure*}[htp!]
    \centering
    \includegraphics[width=\linewidth]{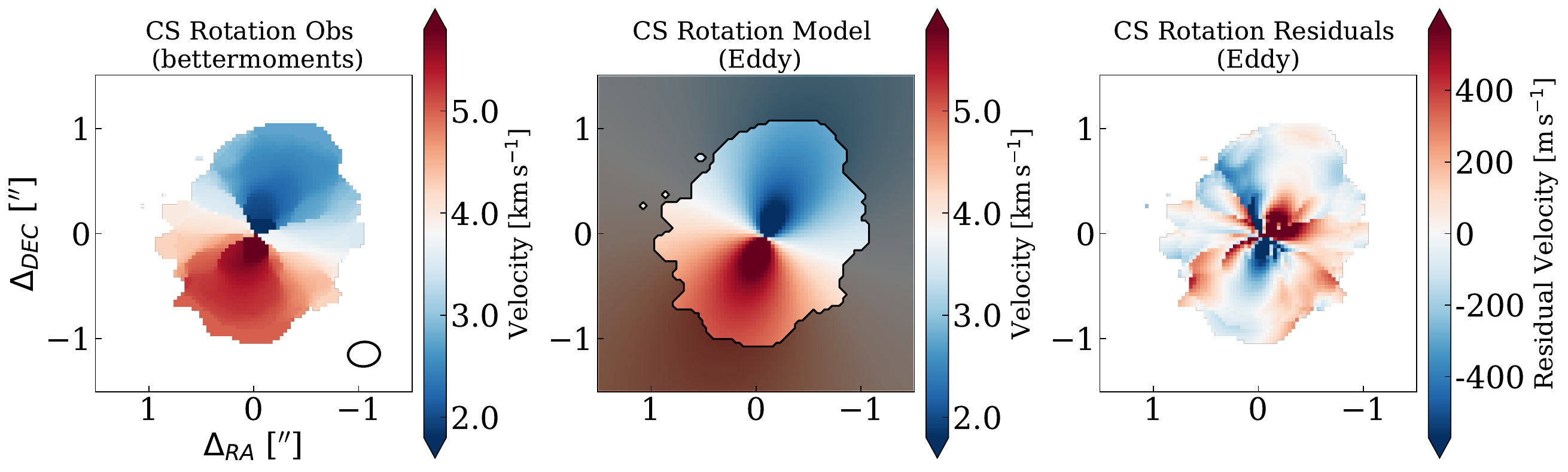}
    \caption{Left: CS moment 1 map generated from \texttt{bettermoments} without down-sampling. The synthesized beam is shown in the lower right corner. 
    Middle: Model line-of-sight velocity map based on best-fit parameters from \texttt{eddy}. 
    Right: Residuals from subtracting the model map in the center panel from the observed map in the left panel.  }
    \label{fig:eddy}    
\end{figure*}

Given the degeneracy between the mass of the central object ($M_{\ast}$) and the disk inclination ($i$), we fixed $i$ to $36\fdg4$ and the position angle (P.A.) to $157\fdg6$ based on the parametric continuum modeling presented in \autoref{appsec:paramodel}. The distance was also fixed to 140 pc. Thus, the model has four free parameters: the disk center offsets from the phase center $\Delta x$ and $\Delta y$, $M_\ast$, and $v_\mathrm{sys}$. We set the priors to the \texttt{eddy} default uniform priors. The package then uses \autoref{eq:velocity} to generate model moment 1 maps and samples the posteriors using the affine invariant Markov Chain Monte Carlo package \texttt{emcee} \citep{emcee}. We used 128 walkers over 2000 steps, with the first 1000 steps set aside as burn-in. The posterior medians and 68\% confidence intervals are listed in \autoref{tab:eddyvalues}. We present a full-resolution moment-1 map (i.e., without any downsampling), a model moment 1 map generated using the best-fit model values, and the residuals in \autoref{fig:eddy}. 
 
\floattable
\begin{table}[htp!]
    \caption{\texttt{eddy} stellar dynamical mass model parameters \label{tab:eddyvalues}}
    \begin{tabular}{lcc} \hline \hline
        Parameter& Value\\
          \\ 
          \hline 
        $\Delta x$ (arcsec) & $-0.0733 \pm 0.003$ \\
         $\Delta y$ (arcsec) & $-0.0322\pm 0.005$ \\
         $M_\ast$ ($M_{\sun}$) & $0.5809\pm0.006$ \\
         $v_{\text{sys}}$ (m/s) & $3870.2\pm6.9$ \\
         \hline
    \end{tabular}
\end{table}
The best-fit model yields a dynamical stellar mass of $0.58\pm0.006$ $M_\sun$. As noted in \citet{Keppler2019}, the formal uncertainties from \texttt{eddy} are very small and may be underestimated due to systematic error. To assess the robustness of our dynamical mass estimate, we compare our \texttt{eddy} results to those of the \texttt{DiskJockey} dynamical mass modeling code in the following section. 

\subsection{Visibility modeling}
To obtain an independent estimate of the dynamical stellar mass, we also used the \texttt{DiskJockey} code \citep{Czekala2015} to model the CS kinematics. Whereas \texttt{eddy} uses analytic expressions for the line-of-sight velocity to generate synthetic moment 1 maps to compare to the observed moment 1 maps, \texttt{DiskJockey} uses the \texttt{RADMC3D} radiative transfer code \citep{Dullemond2012} to generate line image cubes and then the corresponding visibilities to compare to observed visibilities. The advantages of \texttt{eddy} are that it is orders of magnitude less computationally intensive than \texttt{DiskJockey} and does not require parametrization of surface density or temperature profiles, which may be very complex for certain disks. The advantages of \texttt{DiskJockey} are that uncertainties are better quantified by fitting in the $uv$ plane because the visibilities are independent and that it directly accounts for the effects of $uv$ sampling (note, for instance, that \texttt{eddy} will generally have strong residuals at the disk center due to beam dilution, as seen in Figure \ref{fig:eddy}). 

The \texttt{DiskJockey} model setup is described in detail in \citet{Czekala2015}. The disk is assumed to be geometrically thin, axisymmetric, and in hydrostatic equilibrium. The temperature is modeled as a power-law, 

\begin{equation}
T(r) = T_\mathrm{10} \left(\frac{r}{\text{10 au}}\right)^q,
\end{equation}
where $r$ is the radial distance from the star in cylindrical coordinates, and $T_\mathrm{10}$ and $q$ are free parameters. Because the CS radial profile exhibits a slight dip in the center, we selected the Nuker profile \citep{Lauer1995} option to model the gas surface density:

\begin{equation}
\Sigma_g(r) \propto \left(\frac{r}{r_T}\right)^{-\gamma}\left(1+\left(\frac{r}{r_T}\right)^{\alpha}\right)^{\frac{\gamma-\beta}{\alpha}}
\end{equation}
The Nuker profile breaks the disk into two regions, with $r_T$ denoting the radius at which the transition between the two regions occurs and $\alpha$ controlling the sharpness of the transition (higher values have sharper transitions). The $\gamma$ parameter determines the steepness of the profile in the inner region, with a negative value corresponding to a cavity. The $\beta$ parameter determines the steepness of the profile in the outer region. \citet{Tripathi2017} illustrates how the Nuker profile varies with different choices of values. \texttt{DiskJockey} determines the normalization constant for the gas surface density profile based on the input for the total disk gas mass, $M_\mathrm{gas}$.
Additional free parameters include $M_\ast$, $v_\mathrm{sys}$, the microturbulence level $\xi$ (which contributes to line broadening), and the disk offsets from the phase center ($\Delta x$, $\Delta y$). We fixed the source distance to 140 pc. The P.A. and inclination were fixed to the values derived from modeling the high resolution continuum (\autoref{appsec:paramodel}). Thus, our model has a total of 12 free parameters. 

We assume that the CS:H$_2$ abundance is $10^{-8}$, motivated by chemical models \citep{Gross2025}. We note that in general, CS is not well-represented by a constant abundance. However, \texttt{DiskJockey} uses the CS abundance and total gas density to calculate the CS number densities, which are then fed into RADMC-3D for a local thermal equilibrium calculation. Thus, the gas surface density profile and the CS abundance are completely degenerate. For this reason, we emphasize that the $M_\mathrm{gas}$ estimate from \texttt{DiskJockey} should not be treated as a meaningful estimate of the disk mass. 

\begin{figure*}[h]
    \centering
    \includegraphics[width=\linewidth]{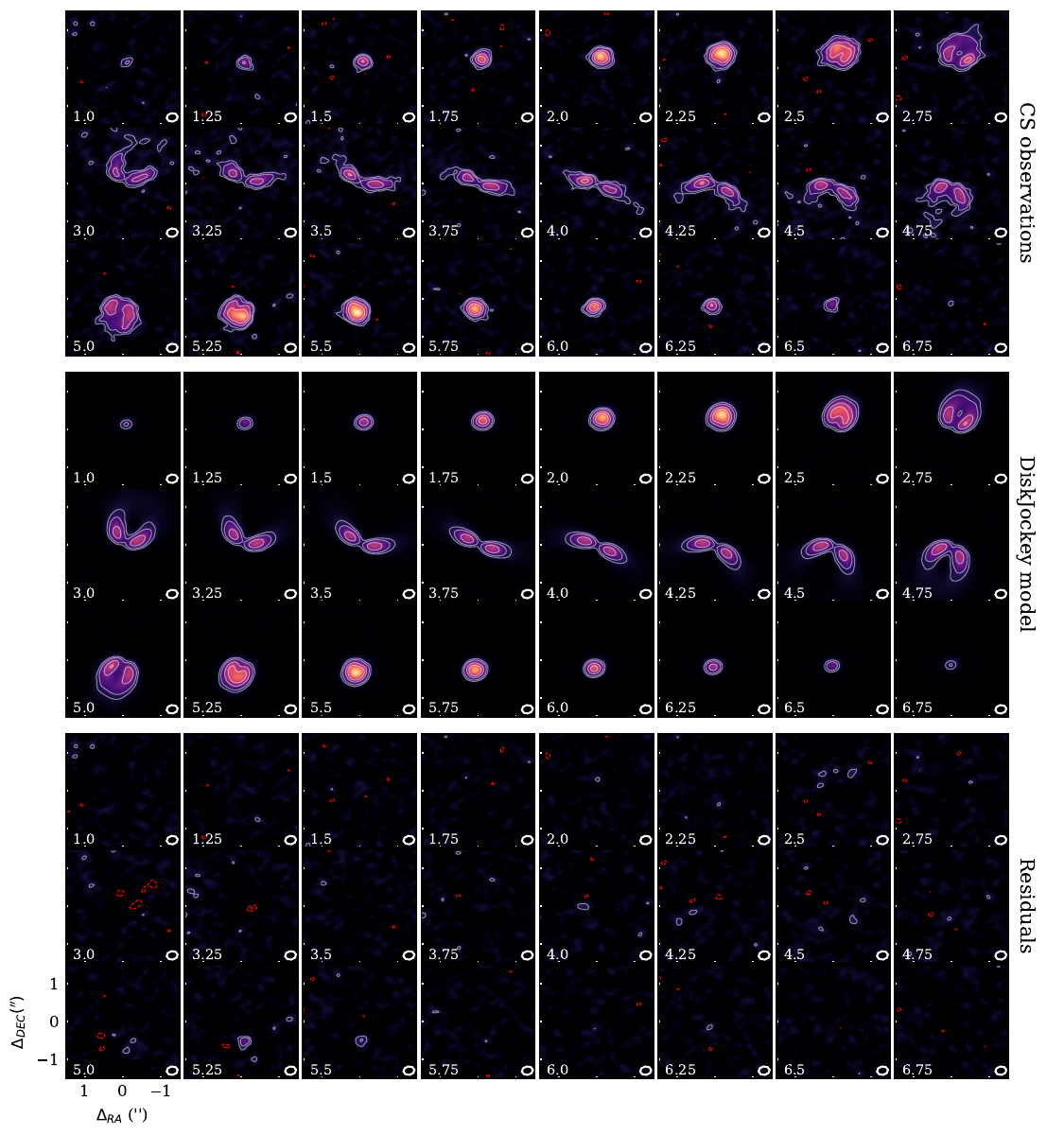}
    \caption{A comparison of channel maps of the CS observations, best-fit \texttt{DiskJockey} model, and residuals, all on the same color scale. The white solid contours correspond to the [3, 5, 10, 15]$\sigma$ levels and the red dashed contours correspond to the $-3\sigma$ level, where $\sigma$ is the rms listed in Table \ref{tab:imaging}. The synthesized beam is plotted in the lower right corner of each panel, while the LSRK velocity is in the lower left corner.}
    \label{fig:diskjockey}    
\end{figure*}

\texttt{DiskJockey} uses a Julia implementation of the \texttt{emcee} code to sample the posterior distributions. Uniform priors (listed in \autoref{tab:diskjockeyparameters}) were chosen based on visual inspection of the CS data as well as \texttt{DiskJockey} model results for other disks \citep{Czekala2015, Czekala2017}. We used 48 walkers over 9000 steps, including a burn-in period of 3000 steps. Following 
\citet{Czekala2017}, we assessed convergence by checking that the Gelman-Rubin statistic $\hat R$ \citep{Gelman2014} was $<1.1$ for each parameter. The exception was $\log \alpha$, which had a slightly higher value of $1.2$. The values for $\log \alpha$ are not well-constrained because of the moderate angular resolution of our observations, which cannot rule out the possibility of very sharp transitions in the Nuker profile. However, this is not important for replicating the kinematic pattern of the CS emission. \autoref{tab:diskjockeyparameters} reports the medians of the posterior distributions as well as the uncertainties calculated from the 16th and 84th percentiles. A comparison of the channel maps of the observations, best-fit model, and residuals are shown in \autoref{fig:diskjockey}. Some moderate residuals are visible in the southeast of the disk at velocities of $\sim5.25-5.5$ km s$^{-1}$ due to the slight asymmetry of the CS emission, but the model otherwise appears to reproduce the data well. The derived M$_\ast$ value of 0.573 $M_\ast$ agrees well with the \texttt{eddy} value, within two percent.

We also note that while the distances were fixed in the models, the stellar mass scales linearly with distance, so what we have estimated in practice is $M_\ast \left(\frac{140\,\mathrm{pc}}{d} \right)$. \citet{Canovas2019} find that the dispersion of the YSOs in Ophiuchus with \textit{Gaia} parallaxes is $\sim8$ pc. Since the distance we use for GY 91 is based on the median value for the region, this implies an uncertainty on the distance of GY 91 of $\sim6\%$. 
\floattable
\begin{table}[htp!]
    \caption{\texttt{DiskJockey} stellar dynamical mass model parameters}
    \label{tab:diskjockeyparameters}
    \begin{tabular}{lcc} \hline \hline
        Parameter& Priors & Value \\
          \\ 
          \hline 
         $\Delta x$ (arcsec) & Uniform($-0.3$, 0.3) &$-0.024\pm0.001$ \\
         $\Delta y$ (arcsec) &  Uniform($-0.3$, 0.3)&$0.006\pm0.001$\\
         $M_\ast$ ($M_\odot$)&Uniform(0.1, 2.0)&$0.573\pm0.002$\\
         $v_\mathrm{sys}$ (km s$^{-1}$)& Uniform(3.5, 4.0)&$3.858\pm0.002$\\
         $T_\mathrm{10}$ (K) & Uniform(10, 100)&$21.3\substack{+1.0\\-0.6}$\\
         $q$ &Uniform(0, 2) &$0.2\substack{+0.03\\-0.01}$\\
         Nuker $r_t$ (au) &Uniform(10, 200) &$32\pm3$\\
         Nuker $\gamma$&  Uniform($-5$, 0)& $-3.0\substack{+0.8\\-1.1}$\\
         Nuker $\log\alpha$ &  Uniform(0, 5)& $1.4\substack{+0.8\\-0.6}$\\
        Nuker $\beta$& Uniform(0, 10)& $3.91\substack{+0.05\\-0.06}$\\
         $\xi$ (km s$^{-1}$)& Uniform(0, 0.5) &$0.199\pm0.005$\\
         $\log M_\mathrm{gas}/M_\odot$&Uniform($-4$, $-0.1$)&$-3.67\substack{+0.06\\-0.07}$\\
         \hline
    \end{tabular}
\end{table}

\section{Gauging envelope contributions to observed emission}\label{sec:envelope}
\subsection{Dust continuum emission}
\cite{Federman2023} suggested using the ratio of the 870 $\mu$m 12-m and ACA continuum fluxes to probe the contributions of the envelope to the overall dust emission in YSOs. Due to its shorter baselines, the ACA is more sensitive to large-scale emission from envelopes, while 12-m observations resolve out more of the envelope emission and are often dominated by the smaller-scale disk emission. \cite{Federman2023} defined a ratio of $R<0.5$ as envelope-dominated and $R>0.5$ as disk-dominated emission, where $R$ is the ratio of the 12-m to ACA fluxes. 

Our closest observed continuum wavelength to the \citet{Federman2023} observations is at 930 $\mu$m. We split out our 930 $\mu$m ACA observations and the extended 12-m array configuration observations (corresponding to the observations dated 2022 June 10 and 11 with baselines of 15-1200 m in \autoref{tab:arrayconfig}) from our final self-calibrated continuum measurement set and imaged the ACA and 12-m observations separately. The MRS is $\sim3000$ au for the ACA-only observations and $\sim530$ au for the extended 12-m observations. As in \cite{Federman2023}, we estimated the total flux in each image by calculating the flux within ellipses of increasing sizes until the flux measurements leveled out. The flux in the image produced from the extended 12-m array observation is 221 mJy, while the flux measured from the ACA observations is 218 mJy. This yields $R\sim$ 1, suggesting that the emission is disk-dominated. While the ACA flux measurement is slightly lower, which may seem counterintuitive, the difference is only $\sim1\%$, which is in line with the small percentage variations between observations in different configurations that were noted in the exoALMA survey of Class II disks \citep{Loomis2025}.

\subsection{Molecular line emission}
We explored whether any of the large-scale emission observed in C$^{18}$O and H$_2$CO might originate from an envelope component, as opposed to a foreground cloud, by comparing observed moment 1 maps to synthetic moment 1 maps. We used the Flat Envelope Model with Rotation and Infall under Angular Momentum Conservation (FERIA) package \citep{Oya2022} to generate a synthetic image cube of a Keplerian disk with a ballistic infalling-rotating envelope. Based on \texttt{eddy} modeling of the Keplerian disk in CS emission (\autoref{sec:dyn_mass}), we set the protostellar mass to 0.58 $M_\odot$, the systemic velocity to 3.87 km s$^{-1}$, the disk inclination to $36\fdg4$, and the P.A. to $157\fdg6$. The radius of the centrifugal barrier was set to 300 au (approximately the extent of the emission detected from the Keplerian disk) and the envelope radius was set to 600 au (chosen to be large enough to visualize the envelope kinematics clearly). We adopted a flaring angle of $30^\circ$ and a density profile of $n(r)\propto r^{-1.5}$. The temperature profile is assumed to follow $T(r)\propto r^{-0.5}$ within the disk and is set to a value of 10 K at the envelope-disk boundary and throughout the envelope. We note, however, that the parameters that mainly control the appearance of the synthetic moment 1 map are the protostellar mass, orientation of the system, and the centrifugal and envelope radii. FERIA calculates relative intensities assuming optically thin emission, with the absolute intensity scale being arbitrary. The resulting image cube was convolved to the same angular resolution as the GY 91 observations. 

\begin{figure*}[htp!]
    \centering
    \includegraphics[width=\linewidth]{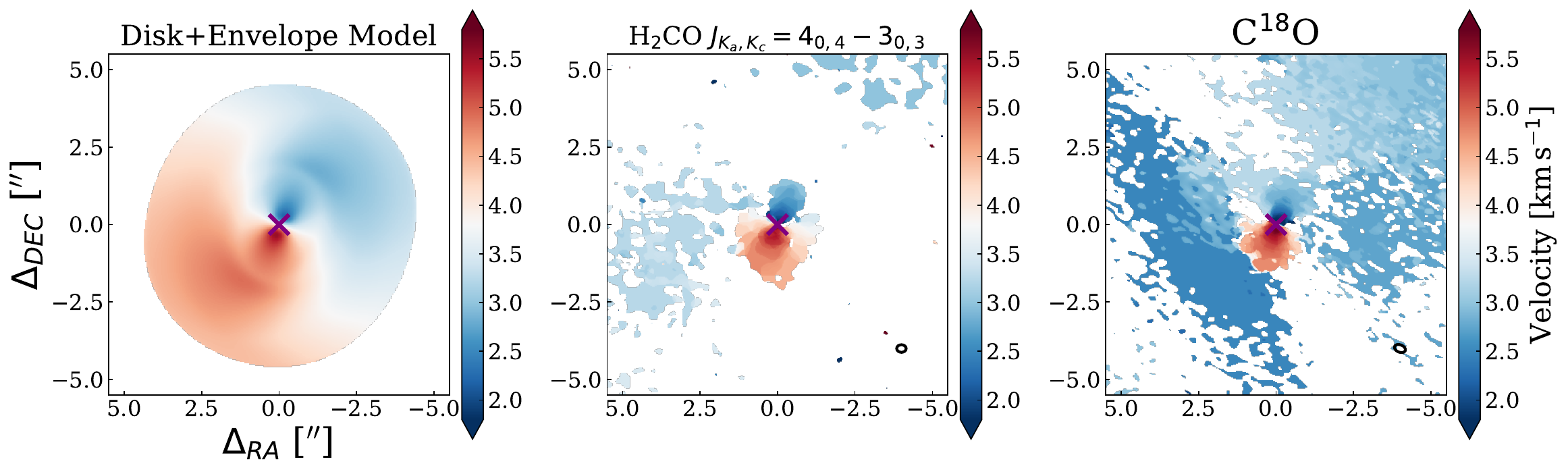}
    \caption{Left: FERIA model moment 1 map of disk and infalling envelope. Middle: Tapered (0.2$\arcsec$) moment 1 map of H$_2$CO 4$_{0,4}$-3$_{0,3}$. Right: Tapered (0.2$\arcsec$) moment 1 map of C$^{18}$O. The field of view is expanded compared to Figure \ref{fig:moment1} in order to show the large-scale non-disk emission better. The synthesized beam is shown in the lower right corner.}
    \label{fig:moment1_compare}    
\end{figure*}

The moment 1 map generated from the FERIA model is shown in \autoref{fig:moment1_compare}, along with moment 1 maps of H$_2$CO $4_{0,4}-3_{0,3}$ and C$^{18}$O generated from images with a $0\farcs2$ $uv$ taper applied in order to increase sensitivity to the larger-scale emission. The kinematics of the emission outside of the disk do not appear to match that of the infalling-rotating envelope in the FERIA model. We only observe large-scale emission at blueshifted velocities surrounding the disk, while the FERIA model predicts both redshifted and blueshifted envelope emission. As such, the observed large-scale emission does not exhibit clear envelope signatures. That said, infalling material in Class 0 and I systems often exhibit more complex morphologies than the axisymmetric model adopted by FERIA. Extended infalling streamers feeding material to disks from the cloud have been detected in YSOs ranging from Class 0 to Class II \citep[e.g.,][]{Ginski2021, Valdivia-Mena2022, Pineda2023, Taniguchi2024, Tanious2024}. However, the extended emission observed in H$_2$CO and C$^{18}$O does not exhibit the narrow and elongated morphologies commonly associated with streamers. Using single-dish CO maps, \citet{1990AA...231..137D} measured a systemic velocity of 3.03 km s$^{-1}$ for the $\rho$ Oph cloud complex containing GY 91, which is consistent with the dominant velocities of the large-scale H$_2$CO and C$^{18}$O emission that we observe. The large-scale emission around GY 91 therefore appears likely to be associated with foreground cloud material. 

\section{Disk mass estimates}\label{sec:radmodel}

\subsection{Radiative transfer modeling of dust continuum}
We used the Monte Carlo radiative transfer code MCFOST \citep{Pinte2006, Pinte2009} to model our 0.93 and 1.05 mm dust continuum observations. While our observations are at lower angular resolution than those of \cite{Sheehan2018} and \cite{Cieza2021}, our observations have better $uv$ coverage at short baselines, which yields better sensitivity at larger scales and more accurate fluxes. We elected not to attempt to fit the SED simultaneously, given that the heavy cloud contamination would likely have a significant influence on the infrared emission, as previously noted by \citet{McClure2010}. 

We adopt MCFOST's power-law disk model option. The disk surface density as a function of radius $r$ in cylindrical coordinates is parametrized as: 

\begin{equation}
    \Sigma(r)\ \propto \ r^{p}
\end{equation}
where $p$ is the surface density exponent. Given an input disk mass, MCFOST computes a normalization factor for the surface density. 
The inner radius of the disk ($R_\text{in}$) was fixed to 0.1 au, while the outer radius $R_\text{out}$ is left as a free parameter. 

The disk scale height is parametrized as 
\begin{equation}
h(r) = h_0\left(\frac{r}{R_\mathrm{ref}} \right)^\beta, 
\end{equation}
where $h_0$ is the disk scale height in au at the reference radius $R_\mathrm{ref}$, and $\beta$ is the flaring exponent. Since our millimeter continuum observations do not constrain the disk vertical structure, we fixed $h_0$ to 10 au, $R_\mathrm{ref}$ to 100 au, and $\beta$ to 1.2, motivated by values that have previously been inferred from radiative transfer modeling of protoplanetary disks \citep[e.g.,][]{Zhang2021}.

Dust grain sizes were assumed to have a power-law distribution $n(a) \propto a^{-3.5}$, following \cite{Mathis1977}. Grain sizes ranged from 0.01 to 1000 $\mu$m. We adopted the DSHARP dust composition \citep{Birnstiel2018} and used the \texttt{dsharp\_opac} package \citep{Birnstiel2018} to retrieve the wavelength-dependent optical constants, which originated from \cite{Henning1996, Draine2003, Warren2008}.

\citet{Doppmann2005} measured an effective temperature of 3300 K for GY 91 and a stellar luminosity of $L_\ast=1.7$ $L_\odot$. Approximating the star as a perfect blackbody (L$_\ast$ = 4$\pi$R$_\ast^2 \sigma_\mathrm{SB}$T$_\mathrm{eff}^{4}$), we derive a stellar radius of 3.99 $R_\odot$. These values of $T_\mathrm{eff}$ and $R_\ast$ are used as inputs for MCFOST. 

\begin{figure*}[htp!]
    \centering
    \includegraphics[width=\linewidth]{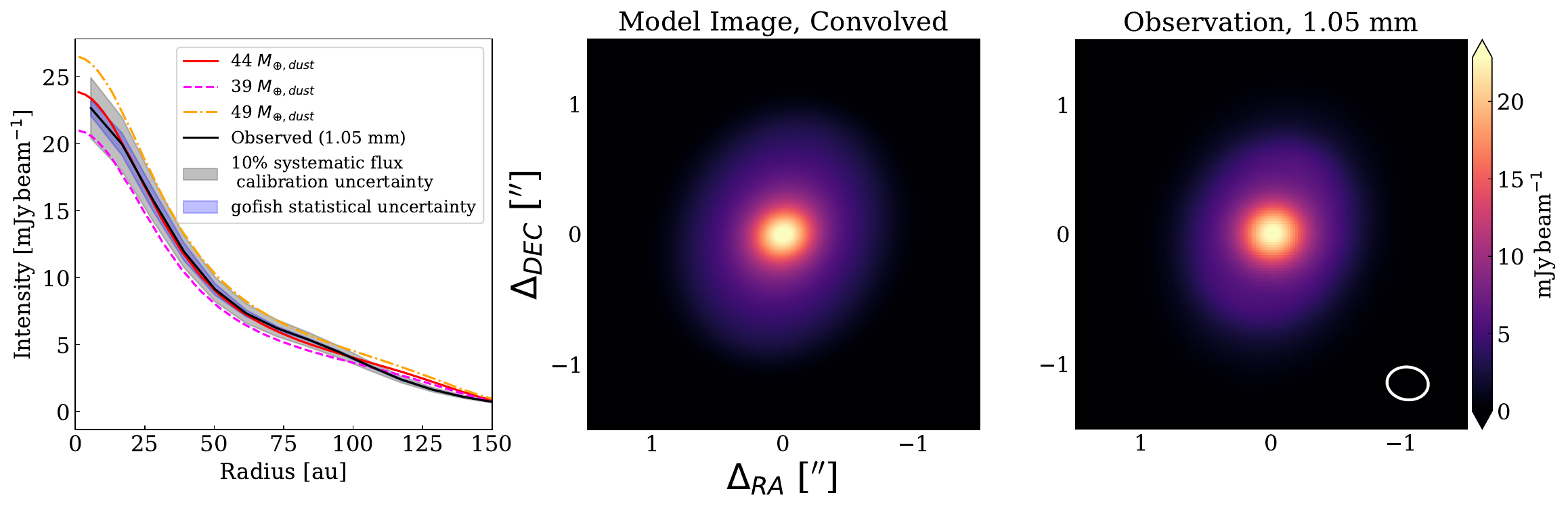}
    \includegraphics[width=\linewidth]{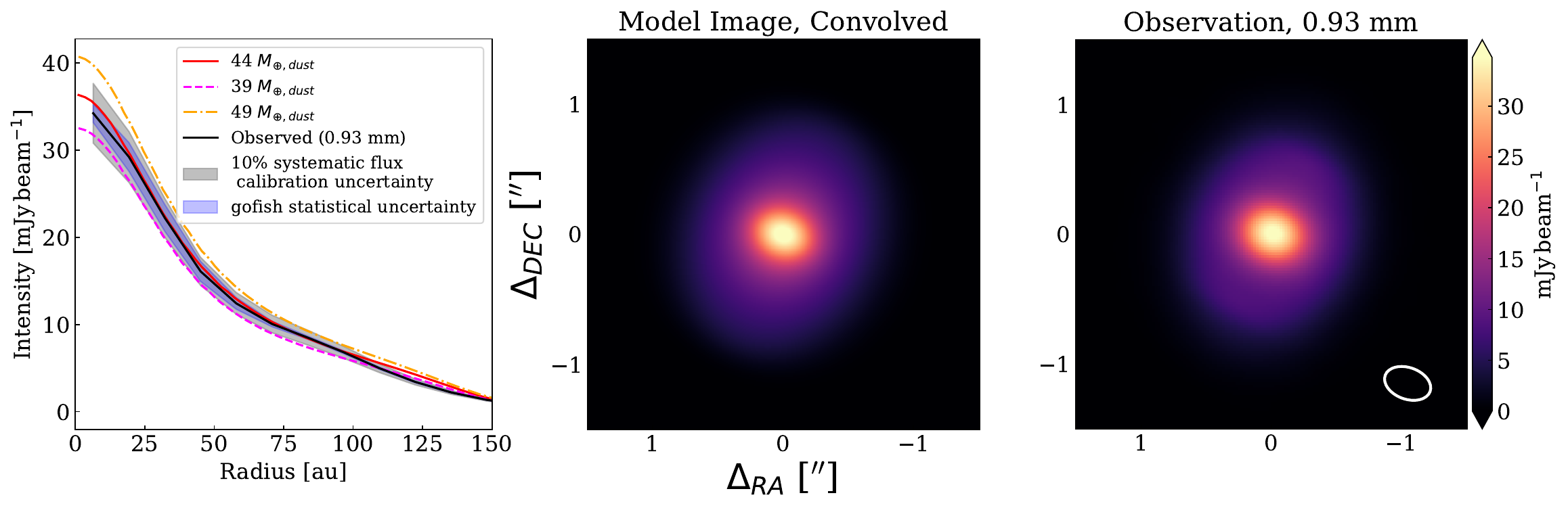}
    \caption{Left panels: Comparison between the radial profiles of the 1.05 (top)/0.93 (bottom) mm dust continuum observations. The blue shaded region indicates the uncertainty computed with \texttt{gofish} based on the scatter of the data and the gray shaded region shows the 10\% flux calibration uncertainty.  Different linestyles correspond to models with different masses. 
    Middle panels: Model image of 44 M$_\earth$ disk from MCFOST convolved with the observed beam. Right panels: Observed dust continuum image (1.05 mm on top, 0.95 mm on bottom) with the beam size indicated on the bottom right corner of each panel.}
    \label{fig:mcfost_model}    
\end{figure*}

Using the values obtained from parametric modeling of the dust continuum (\autoref{appsec:paramodel}), the P.A. was set to $157\fdg6$ and the inclination to $36\fdg4$. We generated different models by varying the dust disk mass ($M_{\rm dust}$), the surface density exponent $p$, and the outer radius $R_\mathrm{out}$. We then convolved the synthetic continuum image produced by MCFOST with a 2D Gaussian with the same dimensions as the synthesized beam to compare to the observations. We found that $M_\mathrm{dust}=44$ $M_\oplus$, $p=-0.4$, and $R_\mathrm{out}=140$ au provided a good visual match to the continuum radial profiles. The parameters of our best model are presented in \autoref{tab:disk_mcfost}. We also examined the range of dust masses that would be consistent with the radial profile, given the 10\% systematic flux calibration uncertainty, and found that masses of roughly 39-49 $M_\oplus$ were compatible. \autoref{fig:mcfost_model} shows a comparison between the model and the observed continuum radial profiles at 0.93 and 1.05 mm. Assuming an ISM-like gas-to-dust ratio of 100, a dust mass of 44 $M_\oplus$ corresponds to a total disk mass of 0.013 $M_\odot$ (4400 $M_\oplus$).

\floattable
\begin{table}[htp!]
    \caption{MCFOST best-fit dust continuum model parameters}
    \label{tab:disk_mcfost}
    \begin{tabular}{lcc} \hline \hline
        Parameter& Value\tablenotemark{a} \\
          \\ 
          \hline 
         $h_0$ (au) & 10\\
         $R_\mathrm{in}$ (au) & 0.1 \\
         $R_\mathrm{out}$ (au) & 140\\
         $R_\mathrm{ref}$ (au) & 100 \\
         Surface density exponent $p$ & $-0.4$ \\
         Flaring exponent $\beta$ & 1.2 \\
         $M_\mathrm{dust}$ ($M_{\earth}$) & 44  \\
         \hline
        \end{tabular}
        \tablenotetext{a}{Values for $p$, $M_\mathrm{dust}$, and $R_\mathrm{out}$ were varied, while other parameter values were held fixed.}
\end{table}

\subsection{Disk mass estimate from N$_2$H$^+$ and C$^{18}$O}
The combination of N$_2$H$^+$ and C$^{18}$O is useful for inferring disk masses because N$_2$H$^+$ helps to break the degeneracy between the CO:H$_2$ ratio and the total gas mass \cite[e.g.,][]{Anderson2019, Trapman2018}. We used the method presented in \citet{Trapman2025, Trapman2025_AGEPRO} to estimate the disk gas mass of GY 91 based on its N$_2$H$^+$ and C$^{18}$O integrated fluxes (as well as the 1.05 mm continuum to estimate the gas-to-dust ratio). \citet{Trapman2025,Trapman2025_AGEPRO} used the DALI thermochemical code \citep{Bruderer2012,Bruderer2013} to generate a grid of disk models with different values for the disk gas mass ($M_\mathrm{gas}$), characteristic radius ($R_c$), stellar luminosity, disk flaring parameter ($\psi$), gas-to-dust ratio, fraction of dust mass in large grains, midplane cosmic ray ionization rate ($\zeta_\mathrm{mid}$), and fractional CO abundance ($x_\mathrm{CO}$). They performed raytracing at different disk inclinations to generate line and continuum flux and emission size estimates for each model. They then generated a more finely sampled grid through interpolation. Posterior distributions for $M_\mathrm{gas}$, $R_c$, the gas-to-dust ratio, $x_\mathrm{CO}$, and $\zeta_\mathrm{mid}$ can be sampled using \texttt{emcee} \citep{emcee} by comparing the model fluxes and convolved emission sizes to the available observed fluxes and sizes. In this modeling framework, the posterior of $L_\ast$ is also sampled, but because the chemical models do not place strong constraints on $L_\ast$, its posterior distribution is essentially set by the choice of the prior. 

The N$_2$H$^+$ and 1.05 mm continuum fluxes were taken from \autoref{tab:imaging}. For C$^{18}$O, we used the same flux value as that used for calculating column densities in \autoref{sec:detections} (i.e., multiplying the flux from the redshifted side by two due to the cloud contamination on the blueshifted side). A 10\% systematic flux calibration uncertainty was assumed for the continuum and N$_2$H$^+$, while we adopted a larger uncertainty of $20\%$ for C$^{18}$O given the need to approximate the total flux by using the non-contaminated side of the disk. We used the C$^{18}$O radial profile (\autoref{fig:radprof}) to estimate $R_\mathrm{90}$, the disk radius that encloses $90\%$ of the C$^{18}$O flux, and obtained a value of 185 au. Again, given the impact of cloud contamination, we conservatively estimate an uncertainty of $\pm40$ au on the $R_\mathrm{90}$ value, corresponding to roughly $2\times$ the standard deviation of the synthesized beam. 
Based on the $L_\ast$ measurement from \citet{Doppmann2005}, we adopted a Gaussian prior for $L_\ast$ centered at 1.7 $L_\odot$ with a standard deviation of 0.5 $L_\odot$ (corresponding to $0.3L_\ast$). Following \citet{Trapman2025}, we set flat priors with bounds of $[-4, -0.3]$ for $\log M_\mathrm{gas}/M_\odot$, $[1,3]$ for the log of the gas-to-dust ratio, $[-6.5, -4]$ for $\log x_\mathrm{CO}$, $[15,200]$ au for $R_c$, and $[-19, -17]$ for $\log (\zeta_\mathrm{mid}/\mathrm{s}^{-1})$. 

We ran \texttt{emcee} with 128 walkers and 20,000 steps, including a burn-in of 5000 steps. The posterior medians and errors measured from the 16th and 84th percentiles are listed in \autoref{tab:gasmodel}. The median value for the gas mass (0.008 $M_\odot$) agrees within a factor of 2 of the disk mass estimate from the \texttt{MCFOST} model of the dust continuum. The uncertainties, though, from the N$_2$H$^+$ and C$^{18}$O modeling are very large. The uncertainties might be reduced either by observing the N$_2$H$^+$ $4-3$ transition or C$^{17}$O, which were also computed in the \citet{Trapman2025} model grids. ($^{13}$C$^{18}$O was not included in their grids).

\begin{figure}[hbt!]
    \centering
    \includegraphics[width=\linewidth]{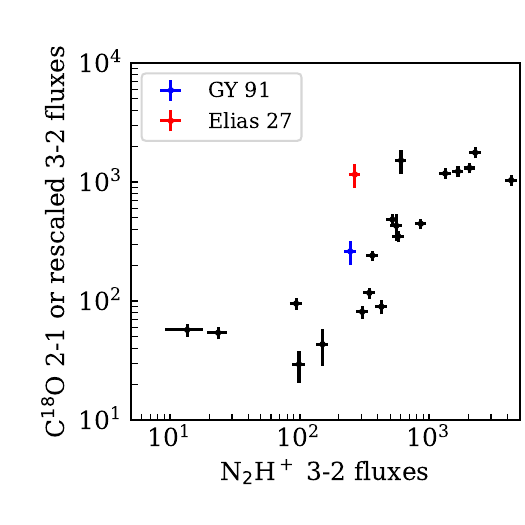}
    \caption{A comparison of GY 91's N$_2$H$^+$ and C$^{18}$O fluxes to a sample of Class II disks from \citet{Trapman2025}, scaled to a common distance of 150 pc. Following \citet{Trapman2025}, sources that only have C$^{18}$O 3-2 fluxes (including GY 91) are plotted with C$^{18}$O fluxes scaled down by a factor of 2.3 based on the $3-2$/$2-1$ flux ratios calculated from their model grid. The $1\sigma$ error bars include the statistical uncertainty and an assumed 10\% systematic flux calibration uncertainty. \label{fig:N2Hpcomparison}}
\end{figure}

\begin{table}
    \caption{Gas mass modeling results}
    \label{tab:gasmodel}
    \begin{tabular}{lcc} \hline \hline
        Parameter& Value\tablenotemark{a} \\
          \\ 
          \hline 
$M_\mathrm{gas}$ ($M_\odot$) & $0.008\substack{+0.04\\-0.005}$ \\
$\log(\text{gas-to-dust ratio})$ & $1.7\substack{+0.3\\-0.2}$\\
$\log x_\mathrm{CO}$&$-4.6\pm0.3$\\
$R_c$ (au) &$80\pm40$ \\
$\log(\zeta_\mathrm{mid}/\mathrm{s}^{-1})$&$-17.9\substack{+0.6\\-0.8}$\\
         \hline
        \end{tabular}
        \tablenotetext{a}{Values correspond to posterior medians, while uncertainties are computed from the 16th and 84th percentiles}
\end{table}
The \citet{Trapman2025, Trapman2025_AGEPRO} models were designed for Class II disks, so envelopes were not incorporated into the source structures. Not accounting for envelope backwarming could lead to underestimates in disk temperatures \citep[e.g.,][]{Butner1994, DAlessio1997}, which in turn would lead to overestimates in disk masses. To assess whether GY 91's behavior differs substantially from Class II disks, we compare its C$^{18}$O and N$_2$H$^+$ fluxes to those of Class II disks hosted by K and M stars in \autoref{fig:N2Hpcomparison}. The sources chosen for comparison are the disks with N$_2$H$^+$ and C$^{18}$O detections that were compiled and modelled by \citet{Trapman2025, Trapman2025_AGEPRO}. (We exclude the disks hosted by earlier type stars because \citet{Trapman2025} show that they have higher C$^{18}$O fluxes compared to K and M stars with similar N$_2$H$^+$ fluxes.) Generally speaking, the C$^{18}$O and N$_2$H$^+$ fluxes for disks hosted by K and M stars tend to increase together, with a possible flattening-out of the C$^{18}$O fluxes at the upper end of the N$_2$H$^+$ fluxes. GY 91 appears to fall in a similar part of the C$^{18}$O vs. N$_2$H$^+$ flux plot as the Class II disks hosted by K and M stars. It is also interesting to compare GY 91's fluxes to that of Elias 27. Elias 27 is commonly identified as a Class II disk, but its molecular emission shows that it is still partially embedded \citep[][]{Huang2018,Paneque2021}. \citet{Trapman2025} pointed out that Elias 27 exhibits an elevated C$^{18}$O flux compared to other Class II disks with similar N$_2$H$^+$ fluxes. They suggested that envelope backwarming could lead to less CO depletion compared to the other Class II disks. If this is the case, then GY 91 does not appear to show a similar influence from an envelope. 
\section{Discussion}\label{sec:discussion}
\subsection{Comparison with previous stellar and disk mass estimates for GY 91}\label{subsec:compare}

Based on a $T_\mathrm{eff}$ estimate of 3300 K from \citet{Doppmann2005} and an estimated age of 0.5 Myr, \citet{Sheehan2018} derived a stellar mass of 0.25 $M_\odot$ from the \citet{Baraffe2015} stellar evolutionary models. Our stellar dynamical mass measurement of 0.58 $M_\odot$ is more than twice as large as the estimate. In a study of pre-main sequence stars with spectral types similar to GY 91 (i.e., around M4), \citet{Pegues2021} found that the masses derived from stellar evolutionary models could be up to $80\%$ less than the dynamical mass estimate. Some of the explanations they proposed for the discrepancy were that the $T_\mathrm{eff}$ values were underestimated, that the sources could be unresolved binaries, or that a significant fraction of the stars are covered by starspots. One of these proposed explanations could also account for the discrepancy in the estimated stellar masses for GY 91. 

Assuming optically thin emission and a uniform disk temperature of 20 K, \citet{Cieza2021} used the \citet{Hildebrand1983} formula to convert millimeter continuum disk fluxes to dust masses and estimated a dust mass of 56 $M_\oplus$ for GY 91. This corresponds to a total disk mass of 0.017 $M_\odot$ for a gas-to-dust ratio of 100. Their estimate is comparable to our estimate from dust continuum radiative transfer modeling and about a factor of two larger than our disk mass estimate from N$_2$H$^+$ and C$^{18}$O. 

On the other hand, \cite{Sheehan2018} reported a disk mass of 0.17 M$_\sun$ from radiative transfer modeling, which is an order of magnitude larger than our disk mass estimate. Whereas we fixed $L_\ast$ to $1.7$ $L_\odot$ and $a_\mathrm{max}$ to 1 mm, \citet{Sheehan2018} allowed these parameters to vary freely. Their best-fit model had $L_\ast=0.11$ $L_\odot$ and $a_\mathrm{max} = 83$ mm. Lower stellar luminosities would tend to drive disk mass estimates upward due to the lower temperature, while higher values of $a_\mathrm{max}$ would drive disk mass estimates upward due to the low opacities of large dust grains.

\citet{Doppmann2005} cautioned that their estimate of $L_\ast=1.7$ $L_\odot$ is uncertain due to their extinction correction. Assuming no rotation and solar metallicity, the MIST stellar evolutionary models \citep{Paxton2011, Choi2016, Dotter2016} predict a stellar luminosity of 1.5 $L_\odot$ for a 0.6 $M_\odot$ star at an age of 0.5 Myr, which is similar to the estimated $L_\ast$ from 
\citet{Doppmann2005}. If we assume an older age of 1 Myr, the stellar luminosity drops to $0.8$ $L_\odot$. If the thermal structure is primarily set by stellar irradiation, then the disk temperature scales as $T\propto L_\ast^{0.25}$ \citep[e.g.][]{Chiang1997, DAlessio1998}. Thus, a factor of two uncertainty in the stellar luminosity should not lead to significant underestimates of the disk mass. 

Spatially resolved multi-frequency studies of HL Tau, a Class I/flat spectrum disk with millimeter continuum gaps and rings, suggest that its $a_\mathrm{max}$ values range from 1 cm in the inner disk to 300 $\mu$m in the other disk \citep[e.g.,][]{Guerra-Alvarado2024}. Therefore, the assumption of a single value of $a_\mathrm{max}$ in our work and in \cite{Sheehan2018} may be simplistic. Obtaining high-resolution, multi-band observations can enable the dust surface density, temperature, and $a_\mathrm{max}$ to be derived simultaneously as a function of disk radius \citep[e.g.,][]{Macias2021}. 

In a comparison of gas mass estimates from C$^{18}$O and N$_2$H$^+$ fluxes to disk mass estimates from kinematic modeling, \citet{Trapman2025} found that the latter tended to be larger than the former by a factor of $\sim2$. Thus, it is plausible that our disk mass for GY 91 is underestimated by a factor of a few. Given that we do not observe evidence of gravitational instability-induced spiral substructures \citep{Goldreich1965}, we use considerations of gravitational stability to estimate an upper bound on the disk mass. Protoplanetary disks are expected to become unstable at a disk-to-stellar mass ratio of M$_{\rm disk}$/$M_\ast \gtrapprox$ 0.1 \citep[e.g.,][]{Kratter2016}. Given a stellar mass of 0.6 $M_\odot$ for GY 91, we thus expect that the disk mass should not be substantially greater than $\sim0.06$ $M_\odot$. This limit is  lower than the \citet{Sheehan2018} disk mass estimate of 0.17 $M_\odot$.

\subsection{Comparison with other disks}\label{subsec:linecompare}

As in previous ALMA observations of CS in Class I and II disks \citep[e.g.,][]{LeGal2019, LeGal2021,Garufi2021, Huang2024}, our CS radial profile shows a central dip, and the moment 0 maps show azimuthally asymmetric disk emission. In a survey of Class I/flat spectrum disks, \citet{Garufi2021} detected envelope emission in CS $5-4$ in all sources. It is not clear whether the absence of obvious envelope emission in our CS $6-5$ observations of GY 91 is due to less envelope material, or because we observed a transition with a higher $E_u$. 

For GY 91, the redshifted side (southeast) is slightly brighter than the blueshifted side (northwest) of the disk in CS emission (\autoref{fig:moment0}). While C$^{18}$O and H$_2$CO $4_\mathrm{0,4}-3_\mathrm{0,3}$ show strong cloud contamination on their northwest sides that leads to the disk emission being absorbed, cloud contamination in CS is minimal. Furthermore, H$_2$CS exhibits a similar emission asymmetry as CS and shows no cloud contamination (\autoref{fig:moment0}). Azimuthal variations in surface density, temperature, or dust optical depth can contribute to CS asymmetries \citep[e.g.][]{vanderPlas2014}. The dust emission of GY 91 appears to be axisymmetric, making azimuthal surface density variations a less likely reason for the CS asymmetry. However, one cannot yet rule out possible azimuthal temperature variations due to shadowing from a misaligned inner disk, such as in HD 100546 \citep{Keyte2023}. \citet{LeGal2021} speculated that CS asymmetries could arise due to disk jets or winds, but we do not see evidence of jets or winds within our observations or previous observations of GY 91. Chemical asymmetries may also arise due to heating from of a protoplanet \citep[e.g.,][]{Cleeves2015}. Observations of other molecular lines or chemical modeling may shed light on the origins of the asymmetry in the sulfur-bearing lines.

\citet{Garufi2021} found that disk-averaged CS column densities in Class I/flat spectrum disks were typically an order of magnitude above that of Class II disks, although the number of measurements in each comparison group was only five and seven, respectively. Our disk-averaged CS column density estimate is similar to that of Class II disks \citep{LeGal2019}. In addition, we estimate a disk-averaged H$_2$CS/CS column density ratio of $\sim0.6-1.1$ (depending on the adopted excitation temperature), which is in line with values previously found for Class I and II disks \citep[e.g.][]{LeGal2019, Garufi2021, LeGal2021}. \citet{LeGal2021} interpreted such high H$_2$CS/CS ratios as evidence that organosulfur compounds contain a significant fraction of the sulfur reservoir. 

 The C$^{18}$O and H$_2$CO 4$_{0,4}$-3$_{0,3}$ moment 0 maps share similar morphologies, with a strong tail of emission beyond radii of 150 au. This may be a reflection of CO hydrogenation being one of the key pathways to forming H$_2$CO \citep[e.g.,][]{Hiraoka1994}, especially in the cold outer disk \citep[e.g.,][]{Pegues2020}. We also see a bump in the H$_2$CO radial profile at $r\sim200$ au and hints of a gap at 120 au in the channel maps. These H$_2$CO features are further out than any substructures in the high-resolution dust continuum \citep{Cieza2021} (see also \autoref{appsec:paramodel}). Gaps in H$_2$CO beyond the dust continuum have commonly been reported in Class II disks \citep{Pegues2020, Guzman2021, Facchini2021}. The H$_2$CO substructure may either reflect radial variations in the gas surface density (possibly associated with planet-disk interactions) or radial chemical variations. In the latter case, H$_2$CO substructure has been hypothesized to be due to changes in the relative importance of gas-phase and grain-surface pathways with disk radius \citep[e.g.,][]{Loomis2015} or to photodesorption outside the dust continuum \citep[e.g.,][]{Carney2017}. 

\citet{Garufi2021} noted a possible dichotomy in the behavior of the disk-averaged CS/H$_2$CO column density ratios in Class I/flat spectrum disks versus Class II disks. The former tend to have ratios below unity ($\sim0.2-0.6$), while the latter tend to have ratios of $\sim2-3$. Our values for GY 91 lie somewhere in-between, ranging from $0.5-1.1$ depending on the assumed excitation temperature. The \citet{Garufi2021} comparison was based on a small number of sources and heterogeneous observations, so a larger, uniform sample will be needed to assess the extent to which the CS/H$_2$CO column density ratio can be used as an evolutionary marker. We speculate that this trend could be linked to gas-phase C/O ratios increasing in the colder and more settled Class II disks \citep[e.g.,][]{Calahan2023}, but chemical modeling will be required to explore the factors affecting the CS/H$_2$CO ratio. 

\begin{figure}[hbt!]
    \centering
    \includegraphics[width=\linewidth]{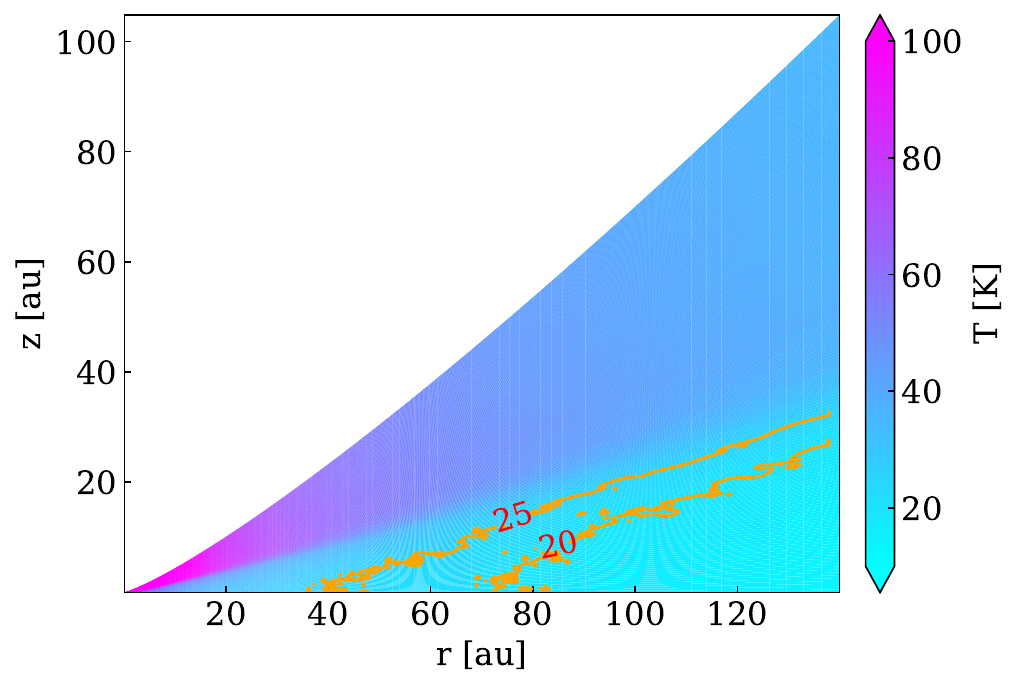}
    \caption{2D temperature structure from our best MCFOST radiative transfer model of the dust continuum. Contours correspond to 20 and 25 K in order to indicate the approximate range of locations of the CO snowline. \label{fig:tempgrid}}
\end{figure}

 N$_2$H$^+$ in GY 91 has a ring-like emission pattern, as is commonly observed in protoplanetary disks \citep[e.g.,][]{Qi2019, Trapman2025}. The peak of the N$_2$H$^+$ ring has been suggested to occur close to the CO snowline because N$_2$H$^+$ is destroyed by CO \citep[e.g.,][]{Qi2013}. To examine whether that might be the case for GY 91, we plot the 2D temperature structure from our best-fit MCFOST dust continuum model (\autoref{sec:radmodel}) along with 25 and 20 K contours in \autoref{fig:tempgrid}. Based on this temperature structure, the midplane CO snowline is expected to fall somewhere between 40 and 75 au. The peak of the N$_2$H$^+$ radial profile (\autoref{fig:radprof}) occurs at $\sim50$ au, which is compatible with the expected snowline location based on radiative transfer modeling. However, other works have shown that the relationship between the N$_2$H$^+$ distribution and the CO snowline is not necessarily straightforward \citep{vantHoff2017,Qi2019}. \citet{Qi2019} noted that N$_2$H$^+$ ring morphologies tend to fall into two categories: 1) Well-defined rings with narrow radial extents, combined with a faint, diffuse outer component and 2) broad rings. They suggested that these different morphologies were a consequence of temperature structures that are nearly vertically isothermal in the N$_2$H$^+$ emitting region in the first case and temperature structures with a steep vertical gradient in the second case. The former types of rings are expected to be more reliable tracers of the CO snowline than the latter. The broadness of GY 91's N$_2$H$^+$ ring could likewise point to a steep vertical temperature gradient, making it harder to reliably probe the midplane CO snowline location. 

We also estimated the dust continuum spectral indices ($\alpha = \frac{\ln(I_{\nu_1}/I_{\nu_2})}{\ln(\nu_1/\nu_2)}$) using our observations at 0.93 mm and 1.05 mm (\autoref{tab:imaging}) as well as the measurement of 97.5 mJy at 1.3 mm from \cite{Cieza2019}. (These correspond to frequencies of 322, 286, and 225 GHz, respectively). A 10\% systematic flux calibration uncertainty is assumed for each measurement. We find that $\alpha_\text{0.93 mm-1.3 mm} = 2.3 \pm 0.4$, $\alpha_\text{1.05 mm-1.3 mm} = 1.9 \pm 0.6$, and $\alpha_\text{0.93 mm-1.05 mm} = 3.0 \pm 1.1$. The measurement between 0.93 mm and 1.3 mm has the largest wavelength difference and therefore provides the best lever arm for estimating the spectral index. The spectral index is consistent with values measured for both young embedded disks \citep{Cacciapuoti2025} and older Class II disks \citep{Ricci2010}, with some combination of optically thick emission and grain growth playing a role in setting these values.

\subsection{What is the evolutionary status of GY 91?}
As noted in the introduction, the evolutionary status of GY 91 has been debated due to the difficulty in disentangling the effects of envelope and foreground cloud. In \autoref{sec:envelope}, we estimated the 930 $\mu$m 12-m/ACA flux ratio, finding a value of $\sim1$. \citet{Federman2023} surveyed the 870 $\mu$m 12-m/ACA flux ratios for Class 0, I, and flat spectrum disks in Orion. They find that the ratio generally increases from Class 0 to flat spectrum sources and therefore suggested that the ratio could be used as an evolutionary marker. They find that $\sim50\%$ of Class I disks have ratios $<0.6$ and $\sim70\%$ have ratios $<1$. This suggests that GY 91 is less embedded, and therefore perhaps more evolved, than the typical Class I YSO. In this case, GY 91 may not be an example of particularly early formation of disk substructures.

One caveat is that the sources in \citet{Federman2023} were more distant ($\sim400$ pc) and therefore had a maximum recoverable scale in their ACA observations of $\sim$8000 au, which is $\sim2.6\times$ greater than our MRS. Therefore, they would be more sensitive to extremely extended envelope emission. In addition, their observing wavelength is slightly shorter than ours (difference of $\sim7\%$). The ratio of envelope to disk emission tends to increase at shorter wavelengths \citep[e.g.,][]{Cacciapuoti2023}. Conducting a survey similar to that of \citet{Federman2023} in nearby star-forming regions, and possibly including the Total Power Array to increase the MRS, would provide a better benchmark for the GY 91 observations. 

While we do not observe unambiguous envelope emission in the molecular lines, our preferential targeting of transitions with $E_\mathrm{u}>25$ K may be biased against detecting colder envelope material. A challenge with observing many of the conventional envelope tracers, such as $^{12}$CO, is the severe cloud contamination \citep{Antilen2023}. However, observing SO as a tracer of accretion shocks may offer an indirect probe of envelope infall \citep[e.g.,][]{Sakai2014}.

A large fraction of Class I disks identified as having substructures, including GY 91, are located in the Ophiuchus star-forming region \citep[e.g.,][]{Sheehan2017, Sheehan2018, Segura-Cox2020, Hsieh2025, Vioque2025}. \citet{McClure2010} noted that the high extinction in Ophiuchus makes evolutionary classifications more challenging compared to other nearby star-forming regions such as Taurus. In some cases, the embedded nature of Class I disks in Ophiuchus with substructures, such as Oph IRS 63, has been well-characterized by spatially resolved molecular line observations tracing infalling streamer and envelope material \citep[e.g.,][]{Podio2024}. High quality molecular line observations of other Ophiuchus sources will be key for confirming their evolutionary status and determining how common substructures are in embedded disks. 

\subsection{Identifying good probes of disk emission for highly extincted systems}

While targeting a severely cloud-contaminated source introduces complications in our analysis, it also provides insights into good tracers of disk properties in highly extincted YSOs. Given that a large fraction of protoplanetary disks show evidence of cloud contamination, especially those in Ophiuchus \citep[e.g.,][]{Reboussin2015}, identifying tracers with minimal cloud contamination is useful for enabling studies of a broader cross-section of disks. 

Among the lines firmly detected in the GY 91 disk, CS $J=6-5$, N$_2$H$^+$ $J=3-2$, and H$_2$CS $8_{1,7}-7_{1,6}$ appear to have minimal cloud contamination. The targeted CS and H$_2$CS transitions have relatively high $E_u$ values (49.3 and 73.4 K, respectively), so they are expected to preferentially trace warm disk gas over cold foreground cloud material. In addition, N$_2$H$^+$ and CS are common dense gas tracers due to their high critical densities \citep[e.g.,][]{Shirley2015}, so they are expected to be better at tracing denser disk gas compared to more diffuse cloud material. 
 
 We observe severe cloud contamination in H$_2$CO 4$_{0,4}$-3$_{0,3}$ and C$^{18}$O, but are still able to distinguish disk emission due to the cloud contamination mostly being limited to one side of the disk. We observed no cloud contamination, but also little disk emission, from H$_2$CO 4$_{2,3}$-3$_{2,2}$. The faint H$_2$CO 4$_{2,3}$-3$_{2,2}$ line has a much higher $E_u$ (82 K) compared to the heavily cloud-contaminated H$_2$CO 4$_{0,4}$-3$_{0,3}$ line (34.9 K). Targeting an H$_2$CO line with an $E_u$ between these two values may be better suited for probing disk emission in extincted sources. $^{13}$C$^{18}$O shows no cloud contamination, but also only marginal disk emission at best. Targeting the $3-2$ transition of C$^{17}$O, which is less abundant than C$^{18}$O but more abundant than $^{13}$C$^{18}$O, may strike the right balance between being abundant enough to be readily detectable in the disk but not so abundant that cloud emission becomes significant. CO isotopologue transitions with higher upper state energy levels may also be useful to target, but have more stringent weather requirements. 

 CO isotopologues have commonly been used to estimate dynamical stellar masses \citep[e.g.][]{Simon2000, Braun2021}. Our observations show that for systems where CO isotopologues are not a good option for tracing disk kinematics due to severe cloud contamination, CS $6-5$ can be an effective dynamical stellar mass probe.

\section{Conclusions}\label{sec:conclusion}
We present new ALMA Band 7 molecular line and continuum observations to study the physical and chemical properties of the Class I YSO GY 91. Our new observations of GY 91 help to provide a roadmap for more accurate characterizations of the physical and chemical properties of disks in highly extincted environments. Our main results are as follows: 

\begin{itemize}
   \item We obtained strong detections ($>$10$\sigma$) of CS, N$_2$H$^+$, C$^{18}$O, H$_2$CO 4$_{0,4}$-3$_{0,3}$, and H$_2$CS, a weak detection ($>$5$\sigma$) of H$_2$CO 4$_{2,3}$-3$_{2,2}$, and a tentative detection of $^{13}$C$^{18}$O. 
   \item Observations of CS and N$_2$H$^+$ show kinematics consistent with Keplerian disk rotation and minimal cloud contamination, while H$_2$CO 4$_{0,4}$-3$_{0,3}$ and C$^{18}$O show significant cloud emission at velocities blue-shifted with respect to the systemic velocity. 
   \item The CS and H$_2$CS radial profiles are similar, both peaking around $\sim$25 au and flattening out past 150 au. N$_2$H$^+$ shows a more extended profile with a peak further out at $\sim$50 au. Outside the dust continuum, we also observed an intensity change that may be indicative of ringed substructure in H$_2$CO 4$_{0,4}$-3$_{0,3}$ and possibly C$^{18}$O at a radius of around 200 au. Whereas the continuum emission appears to be largely axisymmetric, we observe azimuthal asymmetries in CS and H$_2$CS emission, with brighter emission in the southeast. 
   \item By modeling CS kinematics, we inferred a dynamical stellar mass of 0.58 M$_\sun$. This mass is higher than previous estimates based on stellar evolution models (0.25 $M_\odot$), falling in line with previous findings that the dynamical masses of late M stars can be significantly larger than estimates from evolutionary models. 
   \item Disk masses were estimated using both \texttt{MCFOST} dust continuum radiative transfer modeling ($M_\mathrm{gas}$ = 0.013 $M_\odot$, given a dust mass of 44 $M_\oplus$ and gas-to-dust ratio of 100) and comparisons of N$_2$H$^+$ and C$^{18}$O fluxes to literature thermochemical models ($M_\mathrm{gas}$ = 0.008 $M_\odot$). Both values are consistent with expectations for a gravitationally stable disk ($M_\mathrm{disk} < 0.06$ $M_\odot$), given a stellar mass of 0.6 $M_\odot$. 
    \item Comparisons of our ACA and 12-m continuum observations show similar fluxes, suggesting that the dust emission is dominated by the disk rather than by envelope emission. Our molecular lines do not exhibit clear evidence of envelope emission, although the species and transitions chosen may not be optimal for tracing envelope emission, and cloud contamination may obscure envelope emission. We recommend further observations to examine the evolutionary status of GY 91 in order to clarify the timescales of the onset of disk substructures. 
\end{itemize}

\begin{acknowledgments}
We thank the anonymous referee for comments improving the clarity of the manuscript.
This paper makes use of the following ALMA data: ADS/JAO.ALMA\#2021.1.01588.S and \#2018.1.00028.S. ALMA is a partnership of ESO (representing its member states), NSF (USA) and NINS (Japan), together with NRC (Canada), MOST and ASIAA (Taiwan), and KASI (Republic of Korea), in cooperation with the Republic of Chile. The Joint ALMA Observatory is operated by ESO, AUI/NRAO and NAOJ. We thank Ryan Loomis for serving as our contact scientist, Patrick Sheehan for assisting with the observing proposal, and Alice Booth and Adolfo Carvalho for helpful discussions. J.H. and S.D.J acknowledge support from the National Science Foundation under Grant No. AST-2307916. S.D.J. also acknowledges support from the National Science Foundation Graduate Research Fellowship under Grant No. DGE-2437839. R.L.G. acknowledges funding from the French Agence Nationale de la Recherche (ANR) through the project MAPSAJE (ANR-24-CE31-2126-01). F.M. acknowledges funding from the European Research Council (ERC) under the European Union's Horizon Europe research and innovation program (grant agreement No. 101053020, project Dust2Planets). Support for C.J.L. was provided by NASA through the NASA Hubble Fellowship grant No. HST-HF2-51535.001-A awarded by the Space Telescope Science Institute, which is operated by the Association of Universities for Research in Astronomy, Inc., for NASA, under contract NAS5-26555. Support for F.L. was provided by NASA through the NASA Hubble Fellowship grant \#HST-HF2-51512.001-A awarded by the Space Telescope Science Institute, which is operated by the Association of Universities for Research in Astronomy, Incorporated, under NASA contract NAS5-26555S. This research has made use of NASA's Astrophysics Data System Bibliographic Services and the SIMBAD database, operated at CDS, Strasbourg, France.
 \end{acknowledgments}

\vspace{5mm}
\facilities{ALMA}


\software{AstroPy \citep{astropy2013, astropy2018, astropy2022};
analysisUtils \citep{Hunter2023}, 
bettermoments \citep{Teague2018bettermoments, Teague2019bettermoments},
CASA \citep{CASA2022};
DiskJockey \citep{Czekala2015};
dsharp$\textunderscore$opac \citep{Birnstiel2018}
eddy \citep{Teague2019_eddy};
emcee \citep{emcee};
FERIA \citep{Oya2022};
gofish \citep{Teague2019_gofish};
matplotlib \citep{matplotlib};
MCFOST \citep{Pinte2006,Pinte2009},
MPol \citep{mpol, Zawadzki2023};
NumPy \citep{numpy};
pymcfost (\url{https://github.com/cpinte/pymcfost});
pyro \citep{Bingham2019};
RADMC3D \citep{Dullemond2012};
VISIBLE \citep{Loomis2018};
visread \citep{Czekala2021}
          }

\appendix
\counterwithin{figure}{section}

\section{Channel Maps}\label{appsec:channelmaps}
Channel maps for all detected lines are shown in Figure~\ref{fig:cs_channelmap}.

\begin{figure*}[hbt!]
    \centering    \includegraphics[width=\linewidth]{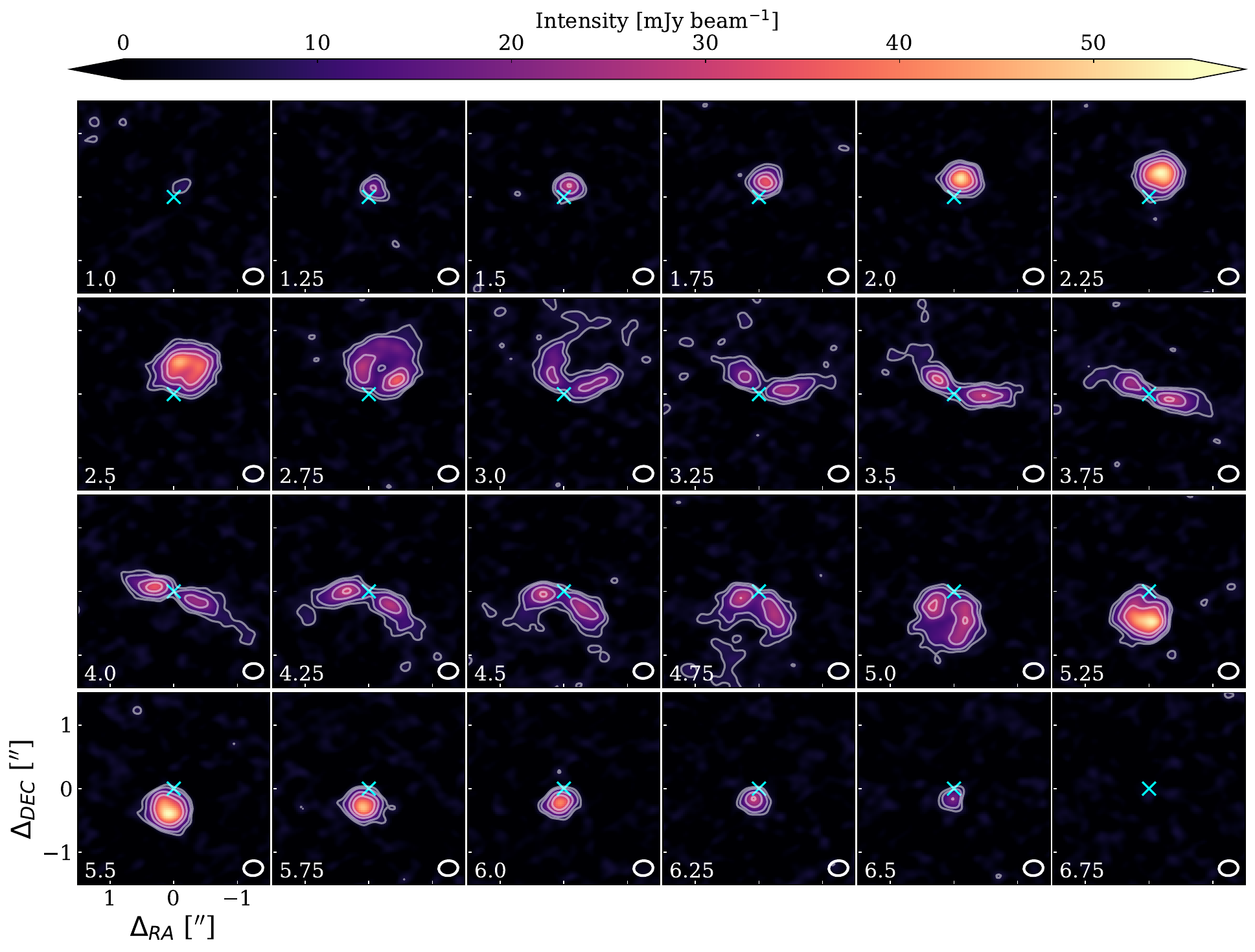}
    \caption{CS channel maps. The LSRK velocity (km s$^{-1}$) is labeled in the bottom left corner of each panel. Contours correspond to 3, 5, 10, and 15$\sigma$, where $\sigma$ is the rms listed in \autoref{tab:imaging}. The synthesized beam size is shown in the bottom right corner. The cyan cross marks the position of the continuum peak. The complete figure set (7 images) is available in the online journal. \label{fig:cs_channelmap}}
\end{figure*}

\figsetstart
\figsettitle{GY 91 channel maps}
\figsetgrpstart
\figsetgrpnum{A.1}
\figsetgrptitle{CS channel maps}
\figsetplot{CS_ChannelMap.pdf}
\figsetgrpnote{CS channel maps. The LSRK velocity (km s$^{-1}$) is labeled in the bottom left corner of each panel. Contours correspond to 3, 5, 10, and 15$\sigma$ levels, where $\sigma$ is the rms listed in \autoref{tab:imaging}. The cyan cross marks the position of the continuum peak. The synthesized beam size is shown in the bottom right corner.\label{fig:cs_channelmap}}
\figsetgrpend

\figsetend

\section{Matched Filtering Analysis} \label{appsec:matchedfilter}
\begin{figure*}[t]
    \centering
    \includegraphics[width=\linewidth]{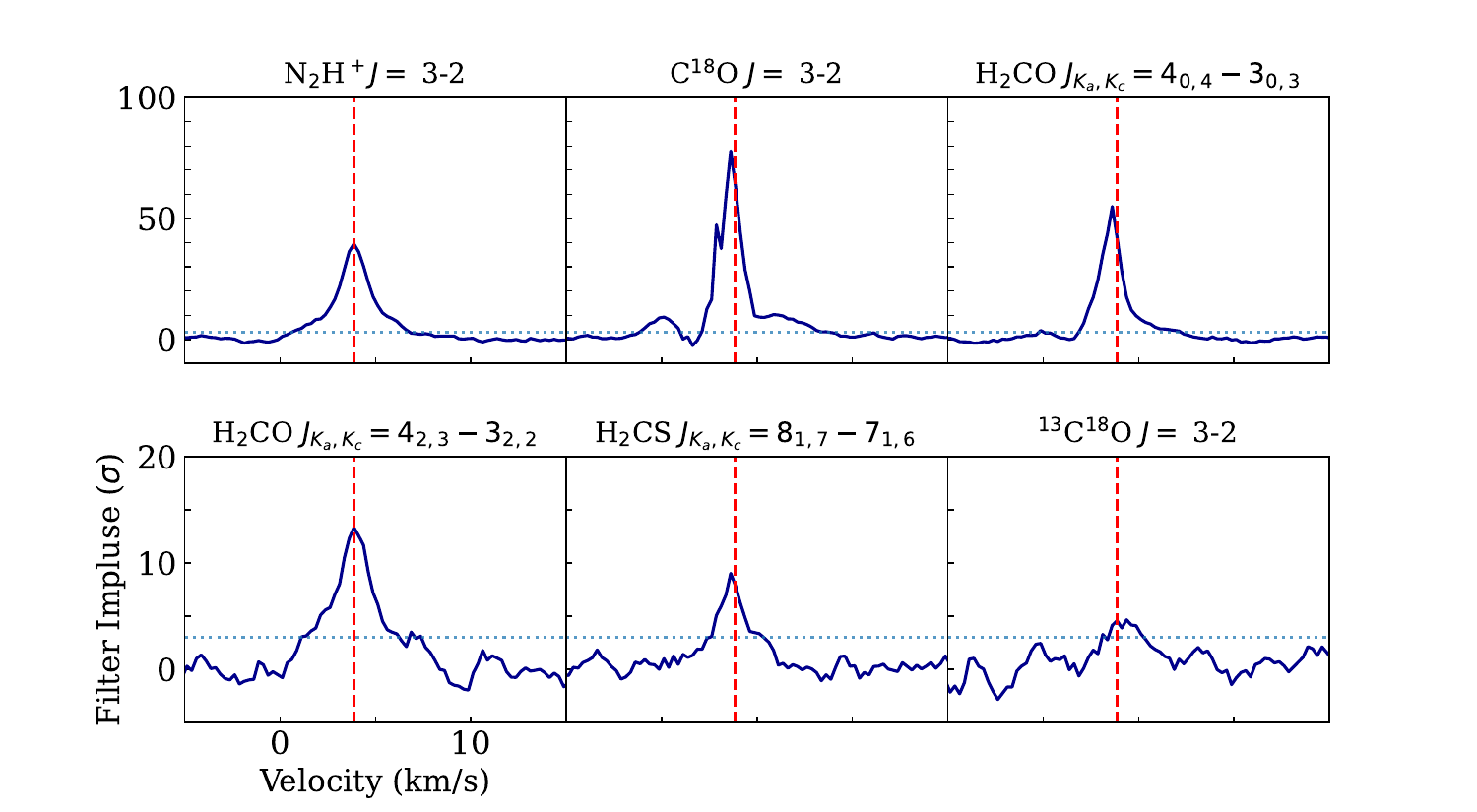}
    \caption{Matched filter response spectrum for all lines besides CS, using the CS CLEAN model as a template. The blue dotted line indicates the 3$\sigma$ impulse response level. The red dashed line corresponds to the systemic velocity (3.87 km s$^{-1}$)}. 
    \label{fig:matchedfilter}    
\end{figure*}

We applied matched filtering with the \texttt{VISIBLE} package \citep{Loomis2018} to quantify the line detection significance. \texttt{VISIBLE} creates a set of visibilities from an input template line cube and then cross-correlates them with the observed spectral line visibilities to assess how well they match. Initial imaging of CS showed high S/N emission with a Keplerian rotation signature consistent with a disk origin. We thus used the CASA-generated CLEAN model image cube as the template, since other molecules originating in the disk should have a similar kinematic pattern. The visibility weights were renormalized based on channel ranges where no line emission was expected. The matched filter responses for all targeted lines (besides CS) are shown in \autoref{fig:matchedfilter}. Nearly all lines are detected at $>5\sigma$, except for the $^{13}$C$^{18}$O line, which is tentatively detected at the $4.5\sigma$ level. For H$_2$CO $J$ = 4$_{0,4}$-3$_{0,3}$ and C$^{18}$O, the impulse response peaks at a slightly lower velocity than the systemic velocity we measure from CS in \autoref{sec:dyn_mass}. We attribute this shift in the impulse response to cloud contamination in H$_2$CO and C$^{18}$O, which is blueshifted relative to the disk systemic velocity. 

\section{Inferring the disk orientation} \label{appsec:paramodel}

\begin{figure*}[hbt!]
    \centering
    \includegraphics[width=\linewidth]{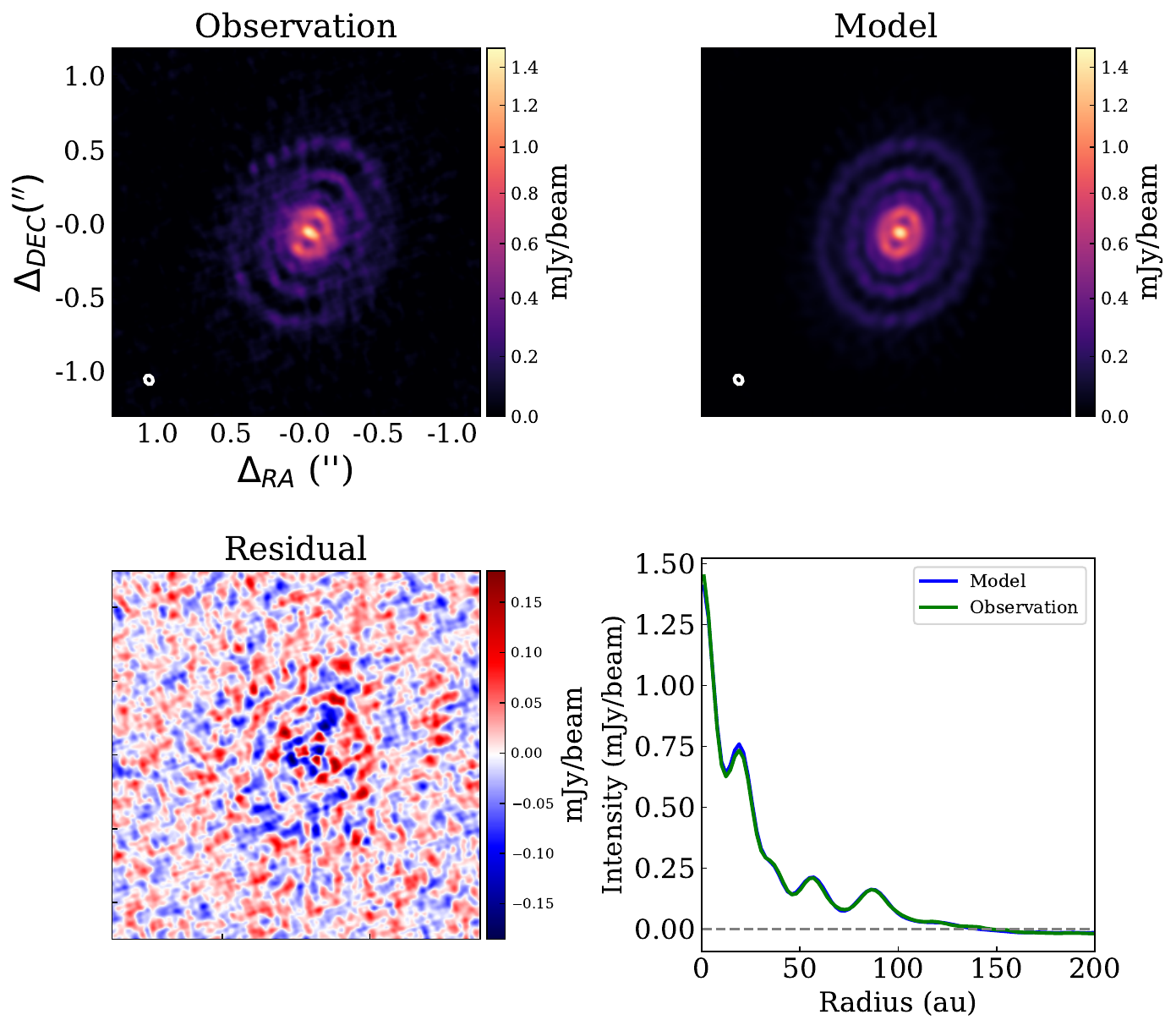}
    \caption{Top left: 1.3 millimeter dust continuum observations from \citet{Cieza2021}. The beam size is indicated in the lower left corner. Top right: CLEAN image of best-fit \texttt{pyro} model. Bottom left: Residual image made from subtracting model visibilities from data. Bottom right: Comparison of the observed and model radial profiles generated from the CLEAN images.}
    \label{fig:pyro}
\end{figure*}

\begin{table*}
    \caption{Parameter inferences for \texttt{pyro} continuum intensity model}
    \label{tab:diskparameter}
    \begin{tabular}{lcccc} \hline \hline
        Parameter& Priors & Value \\
          \\ 
          \hline 
         $\Delta x$ (arcsec) & Norm(0.0, 0.003) &$-0.0333\pm0.0005$ \\
         $\Delta y$ (arcsec) & Norm(0.0, 0.003) &$-0.0646\pm0.0004$ \\
         Position angle (deg)\tablenotemark{a} & Norm(160.0, 10.0) &$157.6\pm0.5$ \\
         Inclination (deg) & Norm(40.0, 10.0)  &$36.4\pm0.3$ \\
        log ($A_0$/Jy arcsec$^{-1}$) & Norm(0.0, 0.3) & $-0.115\substack{+0.008\\-0.009}$ \\
        log($\sigma_0$/au) & Norm(0.7, 0.1) & $0.663\pm0.007$ \\
        $r_1$ (au) & Norm(21, 10) & $20.24\pm0.07$\\
        log($A_{1}$/Jy arcsec$^{-1}$) &Norm(-1.0, 0.5) & $-0.551\substack{+0.005\\-0.004}$ \\
        log($\sigma_{1}$/au) &Norm(0.8, 0.3) & $0.644 \pm 0.007$ \\

        $r_2$ (au) & Norm(38, 10) & $36.9 \pm 0.1$\\
        log($A_{2}$/Jy arcsec$^{-1}$) &Norm(-1.0, 0.5) & $-0.97\pm0.01$\\
        log($\sigma_{2}$/au) &Norm(0.8, 0.3) & $0.6\pm0.1$ \\
        
        $r_3$ (au) & Norm(58, 10) & $56.8\pm0.1$\\
        log($A_{3}$/Jy arcsec$^{-1}$) &Norm(-1.0, 0.5) & $-1.127 \pm 0.008$\\
        log($\sigma_{3}$/au) &Norm(0.8, 0.3) & $0.78\pm 0.01$\\
        
        $r_4$ (au) & Norm(88, 10) & $85.3\pm0.2$\\
        log($A_{4}$/Jy arcsec$^{-1}$) &Norm(-1.0, 0.5) & $-1.37\pm0.01$\\
        log($\sigma_{4}$/au) &Norm(0.8, 0.3) & $0.87\pm0.02$\\
        
        $r_5$ (au) & Norm(105, 10) & $104\pm2$\\
        log($A_{5}$/Jy arcsec$^{-1}$) &Norm(0.8, 0.3) & $-1.87 \substack{+0.01\\-0.02}$ \\
        log($\sigma_{5}$/au) &Norm(0.7, 0.1) & $1.40\pm0.02$ \\
         \hline
    \end{tabular}
    \tablenotetext{a}{Defined to be east of north}
\end{table*}

\citet{Sheehan2018} inferred an inclination of $41^\circ\pm1^\circ$ from radiative transfer modeling of 870 $\mu$m continuum observations, while \citet{Cieza2021} inferred an inclination of $32\fdg5\pm0\fdg4$ by searching for the combination of P.A. and inclination that minimizes the scatter of the deprojected visibilities from high-resolution 1.3 mm continuum observations. Given the disagreement in the literature, we used \texttt{MPoL} \citep{Zawadzki2023, mpol} to perform parametric modeling of the continuum observations from \citet{Cieza2021}, which are the highest-resolution observations available. 

The raw data were retrieved from program 2018.1.00028.S (PI: L. Cieza) via the ALMA archive and calibrated with the CASA 5.4 pipeline. Details of the observations are described in \citet{Cieza2021}. We flagged channels where $^{12}$CO emission was expected and averaged the remaining channels to produce continuum-only visibilities. We applied two rounds of phase self-calibration, the first with a scan-length solution interval and the second with a 360 s solution interval. Because the absolute weights of ALMA measurement sets are known to be sometimes incorrect \citep[see, e.g.,][]{Zawadzki2023}, we used the \texttt{visread} package \citep{Czekala2021} to rescale the weights.

To infer the inclination, we modeled the disk with the following radial intensity profile: 

\begin{equation}
    I(r) = A_0 \exp \left(- \frac{r^2}{2 \sigma_0^2} \right) + \sum_{i=1}^{5} A_i \exp \left(- \frac{(r - r_i)^2}{2 \sigma_i^2} \right),
\end{equation}

\texttt{MPoL} was used to generate a model image for a given set of phase center offsets ($\Delta x$ in the east-west and $\Delta y$ in the north-south directions), inclination, and P.A. and then to generate visibilities from the model image at the same $uv$ points as the data. Thus, the model has 21 free parameters. The source distance was fixed to 140 pc. Priors were set based on the ring properties presented in \citet{Cieza2021}. The stochastic variational inference (SVI) algorithm \citep{hoffman2013} implementation in \texttt{pyro} \citep{Bingham2019} was used to estimate the posterior under the assumption that it is approximately a multivariate normal distribution. We ran SVI for 15,000 iterations and assessed convergence based on the evidence lower bound (ELBO). We then drew 2000 samples from the posterior in order to calculate the medians and 68\% confidence intervals for the model parameters, which are listed in \autoref{tab:diskparameter}. \autoref{fig:pyro} compares the CLEAN images, residuals, and radial profiles of the observations and model, showing that the observations are reproduced well. 

We then used the MPoL model results to estimate $R_\mathrm{90}$, the radius that encloses $90\%$ of the disk flux. For each set of samples from the posteriors, we first calculated the radial intensity profile $I (r)$, computed the cumulative flux for a series of radial bins up to several hundred au, where the cumulative flux leveled out, and finally calculated $R_\mathrm{90}$. Taking the median and standard deviation of the $R_\mathrm{90}$ results for all radial profiles, we estimate that $R_\mathrm{90} = 117\pm1$ au.

\bibliography{main}{}

@ARTICLE{Ohashi2023,
       author = {{Ohashi}, Nagayoshi and {Tobin}, John J. and {J{\o}rgensen}, Jes K. and {Takakuwa}, Shigehisa and {Sheehan}, Patrick and {Aikawa}, Yuri and {Li}, Zhi-Yun and {Looney}, Leslie W. and {Williams}, Jonathan P. and {Aso}, Yusuke and {Sharma}, Rajeeb and {Sai}, Jinshi (Insa Choi) and {Yamato}, Yoshihide and {Lee}, Jeong-Eun and {Tomida}, Kengo and {Yen}, Hsi-Wei and {Encalada}, Frankie J. and {Flores}, Christian and {Gavino}, Sacha and {Kido}, Miyu and {Han}, Ilseung and {Lin}, Zhe-Yu Daniel and {Narayanan}, Suchitra and {Phuong}, Nguyen Thi and {Santamar{\'\i}a-Miranda}, Alejandro and {Thieme}, Travis J. and {van't Hoff}, Merel L.~R. and {de Gregorio-Monsalvo}, Itziar and {Koch}, Patrick M. and {Kwon}, Woojin and {Lai}, Shih-Ping and {Lee}, Chang Won and {Plunkett}, Adele and {Saigo}, Kazuya and {Hirano}, Shingo and {Lam}, Ka Ho and {Mori}, Shoji},
        title = "{Early Planet Formation in Embedded Disks (eDisk). I. Overview of the Program and First Results}",
      journal = {\apj},
         year = 2023,
        month = jul,
       volume = {951},
       number = {1},
          eid = {8},
        pages = {8},
          doi = {10.3847/1538-4357/acd384},
archivePrefix = {arXiv},
       eprint = {2306.15406},
 primaryClass = {astro-ph.EP},
       adsurl = {https://ui.adsabs.harvard.edu/abs/2023ApJ...951....8O}
}

@ARTICLE{Federman2023,
       author = {{Federman}, Samuel and {Megeath}, S. Thomas and {Tobin}, John J. and {Sheehan}, Patrick D. and {Pokhrel}, Riwaj and {Habel}, Nolan and {Stutz}, Amelia M. and {Fischer}, William J. and {Hartmann}, Lee and {Stanke}, Thomas and {Narang}, Mayank and {Osorio}, Mayra and {Atnagulov}, Prabhani and {Rahatgaonkar}, Rohan},
        title = "{300: An ACA 870 {\ensuremath{\mu}}m Continuum Survey of Orion Protostars and Their Evolution}",
      journal = {\apj},
         year = 2023,
        month = feb,
       volume = {944},
       number = {1},
          eid = {49},
        pages = {49},
          doi = {10.3847/1538-4357/ac9f4b},
archivePrefix = {arXiv},
       eprint = {2210.07925},
 primaryClass = {astro-ph.SR},
       adsurl = {https://ui.adsabs.harvard.edu/abs/2023ApJ...944...49F}
}

@Misc{Hunter2023,
  author       = {{Hunter}, Todd R. and {Petry}, Dirk and {Barkats}, Denis and {Corder}, Stuartt and {Indebetouw}, Remy},
  howpublished = {Zenodo},
  month        = jan,
  title        = {{analysisUtils}},
  year         = {2023},
  adsurl       = {https://ui.adsabs.harvard.edu/abs/2023zndo...7502160H},
  doi          = {10.5281/zenodo.7502160},
  eid          = {10.5281/zenodo.7502160},
  publisher    = {Zenodo},
  version      = {2.6},
}

@ARTICLE{CASA2022,
       author = {{CASA Team} and {Bean}, Ben and {Bhatnagar}, Sanjay and {Castro}, Sandra and {Donovan Meyer}, Jennifer and {Emonts}, Bjorn and {Garcia}, Enrique and {Garwood}, Robert and {Golap}, Kumar and {Gonzalez Villalba}, Justo and {Harris}, Pamela and {Hayashi}, Yohei and {Hoskins}, Josh and {Hsieh}, Mingyu and {Jagannathan}, Preshanth and {Kawasaki}, Wataru and {Keimpema}, Aard and {Kettenis}, Mark and {Lopez}, Jorge and {Marvil}, Joshua and {Masters}, Joseph and {McNichols}, Andrew and {Mehringer}, David and {Miel}, Renaud and {Moellenbrock}, George and {Montesino}, Federico and {Nakazato}, Takeshi and {Ott}, Juergen and {Petry}, Dirk and {Pokorny}, Martin and {Raba}, Ryan and {Rau}, Urvashi and {Schiebel}, Darrell and {Schweighart}, Neal and {Sekhar}, Srikrishna and {Shimada}, Kazuhiko and {Small}, Des and {Steeb}, Jan-Willem and {Sugimoto}, Kanako and {Suoranta}, Ville and {Tsutsumi}, Takahiro and {van Bemmel}, Ilse M. and {Verkouter}, Marjolein and {Wells}, Akeem and {Xiong}, Wei and {Szomoru}, Arpad and {Griffith}, Morgan and {Glendenning}, Brian and {Kern}, Jeff},
        title = "{CASA, the Common Astronomy Software Applications for Radio Astronomy}",
      journal = {\pasp},
         year = 2022,
        month = nov,
       volume = {134},
       number = {1041},
          eid = {114501},
        pages = {114501},
          doi = {10.1088/1538-3873/ac9642},
archivePrefix = {arXiv},
       eprint = {2210.02276},
 primaryClass = {astro-ph.IM},
       adsurl = {https://ui.adsabs.harvard.edu/abs/2022PASP..134k4501C}
}

@Article{astropy2022,
  author        = {{Astropy Collaboration} and {Price-Whelan}, Adrian M. and {Lim}, Pey Lian and {Earl}, Nicholas and {Starkman}, Nathaniel and {Bradley}, Larry and {Shupe}, David L. and {Patil}, Aarya A. and {Corrales}, Lia and {Brasseur}, C.~E. and {N{\"o}the}, Maximilian and {Donath}, Axel and {Tollerud}, Erik and {Morris}, Brett M. and {Ginsburg}, Adam and {Vaher}, Eero and {Weaver}, Benjamin A. and {Tocknell}, James and {Jamieson}, William and {van Kerkwijk}, Marten H. and {Robitaille}, Thomas P. and {Merry}, Bruce and {Bachetti}, Matteo and {G{\"u}nther}, H. Moritz and {Aldcroft}, Thomas L. and {Alvarado-Montes}, Jaime A. and {Archibald}, Anne M. and {B{\'o}di}, Attila and {Bapat}, Shreyas and {Barentsen}, Geert and {Baz{\'a}n}, Juanjo and {Biswas}, Manish and {Boquien}, M{\'e}d{\'e}ric and {Burke}, D.~J. and {Cara}, Daria and {Cara}, Mihai and {Conroy}, Kyle E. and {Conseil}, Simon and {Craig}, Matthew W. and {Cross}, Robert M. and {Cruz}, Kelle L. and {D'Eugenio}, Francesco and {Dencheva}, Nadia and {Devillepoix}, Hadrien A.~R. and {Dietrich}, J{\"o}rg P. and {Eigenbrot}, Arthur Davis and {Erben}, Thomas and {Ferreira}, Leonardo and {Foreman-Mackey}, Daniel and {Fox}, Ryan and {Freij}, Nabil and {Garg}, Suyog and {Geda}, Robel and {Glattly}, Lauren and {Gondhalekar}, Yash and {Gordon}, Karl D. and {Grant}, David and {Greenfield}, Perry and {Groener}, Austen M. and {Guest}, Steve and {Gurovich}, Sebastian and {Handberg}, Rasmus and {Hart}, Akeem and {Hatfield-Dodds}, Zac and {Homeier}, Derek and {Hosseinzadeh}, Griffin and {Jenness}, Tim and {Jones}, Craig K. and {Joseph}, Prajwel and {Kalmbach}, J. Bryce and {Karamehmetoglu}, Emir and {Ka{\l}uszy{\'n}ski}, Miko{\l}aj and {Kelley}, Michael S.~P. and {Kern}, Nicholas and {Kerzendorf}, Wolfgang E. and {Koch}, Eric W. and {Kulumani}, Shankar and {Lee}, Antony and {Ly}, Chun and {Ma}, Zhiyuan and {MacBride}, Conor and {Maljaars}, Jakob M. and {Muna}, Demitri and {Murphy}, N.~A. and {Norman}, Henrik and {O'Steen}, Richard and {Oman}, Kyle A. and {Pacifici}, Camilla and {Pascual}, Sergio and {Pascual-Granado}, J. and {Patil}, Rohit R. and {Perren}, Gabriel I. and {Pickering}, Timothy E. and {Rastogi}, Tanuj and {Roulston}, Benjamin R. and {Ryan}, Daniel F. and {Rykoff}, Eli S. and {Sabater}, Jose and {Sakurikar}, Parikshit and {Salgado}, Jes{\'u}s and {Sanghi}, Aniket and {Saunders}, Nicholas and {Savchenko}, Volodymyr and {Schwardt}, Ludwig and {Seifert-Eckert}, Michael and {Shih}, Albert Y. and {Jain}, Anany Shrey and {Shukla}, Gyanendra and {Sick}, Jonathan and {Simpson}, Chris and {Singanamalla}, Sudheesh and {Singer}, Leo P. and {Singhal}, Jaladh and {Sinha}, Manodeep and {Sip{\H{o}}cz}, Brigitta M. and {Spitler}, Lee R. and {Stansby}, David and {Streicher}, Ole and {{\v{S}}umak}, Jani and {Swinbank}, John D. and {Taranu}, Dan S. and {Tewary}, Nikita and {Tremblay}, Grant R. and {de Val-Borro}, Miguel and {Van Kooten}, Samuel J. and {Vasovi{\'c}}, Zlatan and {Verma}, Shresth and {de Miranda Cardoso}, Jos{\'e} Vin{\'\i}cius and {Williams}, Peter K.~G. and {Wilson}, Tom J. and {Winkel}, Benjamin and {Wood-Vasey}, W.~M. and {Xue}, Rui and {Yoachim}, Peter and {Zhang}, Chen and {Zonca}, Andrea and {Astropy Project Contributors}},
  journal       = {\apj},
  title         = {{The Astropy Project: Sustaining and Growing a Community-oriented Open-source Project and the Latest Major Release (v5.0) of the Core Package}},
  year          = {2022},
  month         = aug,
  number        = {2},
  pages         = {167},
  volume        = {935},
  adsurl        = {https://ui.adsabs.harvard.edu/abs/2022ApJ...935..167A},
  archiveprefix = {arXiv},
  doi           = {10.3847/1538-4357/ac7c74},
  eid           = {167},
  eprint        = {2206.14220},
  primaryclass  = {astro-ph.IM},
}

@Software{Czekala2021,
  author    = {Ian Czekala and Loomis, Ryan and Andrews, Sean and Huang, Jane and Rosenfeld, Katherine},
  doi       = {10.5281/zenodo.4432501},
  month     = jan,
  publisher = {Zenodo},
  title     = {MPoL-dev/visread},
  url       = {https://doi.org/10.5281/zenodo.4432501},
  version   = {v0.0.1},
  year      = {2021},
}

@ARTICLE{Zhang2021,
       author = {{Zhang}, Ke and {Booth}, Alice S. and {Law}, Charles J. and {Bosman}, Arthur D. and {Schwarz}, Kamber R. and {Bergin}, Edwin A. and {{\"O}berg}, Karin I. and {Andrews}, Sean M. and {Guzm{\'a}n}, Viviana V. and {Walsh}, Catherine and {Qi}, Chunhua and {van't Hoff}, Merel L.~R. and {Long}, Feng and {Wilner}, David J. and {Huang}, Jane and {Czekala}, Ian and {Ilee}, John D. and {Cataldi}, Gianni and {Bergner}, Jennifer B. and {Aikawa}, Yuri and {Teague}, Richard and {Bae}, Jaehan and {Loomis}, Ryan A. and {Calahan}, Jenny K. and {Alarc{\'o}n}, Felipe and {M{\'e}nard}, Fran{\c{c}}ois and {Le Gal}, Romane and {Sierra}, Anibal and {Yamato}, Yoshihide and {Nomura}, Hideko and {Tsukagoshi}, Takashi and {P{\'e}rez}, Laura M. and {Trapman}, Leon and {Liu}, Yao and {Furuya}, Kenji},
        title = "{Molecules with ALMA at Planet-forming Scales (MAPS). V. CO Gas Distributions}",
      journal = {\apjs},
         year = 2021,
        month = nov,
       volume = {257},
       number = {1},
          eid = {5},
        pages = {5},
          doi = {10.3847/1538-4365/ac1580},
archivePrefix = {arXiv},
       eprint = {2109.06233},
 primaryClass = {astro-ph.EP},
       adsurl = {https://ui.adsabs.harvard.edu/abs/2021ApJS..257....5Z}
}

@ARTICLE{Cieza2021,
       author = {{Cieza}, Lucas A. and {Gonz{\'a}lez-Ruilova}, Camilo and {Hales}, Antonio S. and {Pinilla}, Paola and {Ru{\'\i}z-Rodr{\'\i}guez}, Dary and {Zurlo}, Alice and {Casassus}, Sim{\'o}n and {P{\'e}rez}, Sebasti{\'a}n and {C{\'a}novas}, Hector and {Arce-Tord}, Carla and {Flock}, Mario and {Kurtovic}, Nicolas and {Marino}, Sebastian and {Nogueira}, Pedro H. and {Perez}, Laura and {Price}, Daniel J. and {Principe}, David A. and {Williams}, Jonathan P.},
        title = "{The Ophiuchus DIsc Survey Employing ALMA (ODISEA) - III. The evolution of substructures in massive discs at 3-5 au resolution}",
      journal = {\mnras},
         year = 2021,
        month = feb,
       volume = {501},
       number = {2},
        pages = {2934-2953},
          doi = {10.1093/mnras/staa3787},
archivePrefix = {arXiv},
       eprint = {2012.00189},
 primaryClass = {astro-ph.EP},
       adsurl = {https://ui.adsabs.harvard.edu/abs/2021MNRAS.501.2934C}
}

@ARTICLE{Segura-Cox2020,
       author = {{Segura-Cox}, Dominique M. and {Schmiedeke}, Anika and {Pineda}, Jaime E. and {Stephens}, Ian W. and {Fern{\'a}ndez-L{\'o}pez}, Manuel and {Looney}, Leslie W. and {Caselli}, Paola and {Li}, Zhi-Yun and {Mundy}, Lee G. and {Kwon}, Woojin and {Harris}, Robert J.},
        title = "{Four annular structures in a protostellar disk less than 500,000 years old}",
      journal = {\nat},
         year = 2020,
        month = oct,
       volume = {586},
       number = {7828},
        pages = {228-231},
          doi = {10.1038/s41586-020-2779-6},
archivePrefix = {arXiv},
       eprint = {2010.03657},
 primaryClass = {astro-ph.EP},
       adsurl = {https://ui.adsabs.harvard.edu/abs/2020Natur.586..228S}
}

@ARTICLE{Teague2019_gofish,
       author = {{Teague}, Richard},
        title = "{GoFish: Fishing for Line Observations in Protoplanetary Disks}",
      journal = {The Journal of Open Source Software},
         year = 2019,
        month = sep,
       volume = {4},
       number = {41},
          eid = {1632},
        pages = {1632},
          doi = {10.21105/joss.01632},
       adsurl = {https://ui.adsabs.harvard.edu/abs/2019JOSS....4.1632T}
}

@ARTICLE{Teague2019_eddy,
       author = {{Teague}, Richard},
        title = "{eddy: Extracting Protoplanetary Disk Dynamics with Python}",
      journal = {The Journal of Open Source Software},
         year = 2019,
        month = feb,
       volume = {4},
       number = {34},
        pages = {1220},
          doi = {10.21105/joss.01220},
       adsurl = {https://ui.adsabs.harvard.edu/abs/2019JOSS....4.1220T}
}

@Article{astropy2018,
     author = {{Astropy Collaboration} and {Price-Whelan}, A.~M. and {Sip{\H{o}}cz}, B.~M. and {G{\"u}nther}, H.~M. and {Lim}, P.~L. and {Crawford}, S.~M. and {Conseil}, S. and {Shupe}, D.~L. and {Craig}, M.~W. and {Dencheva}, N. and {Ginsburg}, A. and {VanderPlas}, J.~T. and {Bradley}, L.~D. and {P{\'e}rez-Su{\'a}rez}, D. and {de Val-Borro}, M. and {Aldcroft}, T.~L. and {Cruz}, K.~L. and {Robitaille}, T.~P. and {Tollerud}, E.~J. and {Ardelean}, C. and {Babej}, T. and {Bach}, Y.~P. and {Bachetti}, M. and {Bakanov}, A.~V. and {Bamford}, S.~P. and {Barentsen}, G. and {Barmby}, P. and {Baumbach}, A. and {Berry}, K.~L. and {Biscani}, F. and {Boquien}, M. and {Bostroem}, K.~A. and {Bouma}, L.~G. and {Brammer}, G.~B. and {Bray}, E.~M. and {Breytenbach}, H. and {Buddelmeijer}, H. and {Burke}, D.~J. and {Calderone}, G. and {Cano Rodr{\'\i}guez}, J.~L. and {Cara}, M. and {Cardoso}, J.~V.~M. and {Cheedella}, S. and {Copin}, Y. and {Corrales}, L. and {Crichton}, D. and {D'Avella}, D. and {Deil}, C. and {Depagne}, {\'E}. and {Dietrich}, J.~P. and {Donath}, A. and {Droettboom}, M. and {Earl}, N. and {Erben}, T. and {Fabbro}, S. and {Ferreira}, L.~A. and {Finethy}, T. and {Fox}, R.~T. and {Garrison}, L.~H. and {Gibbons}, S.~L.~J. and {Goldstein}, D.~A. and {Gommers}, R. and {Greco}, J.~P. and {Greenfield}, P. and {Groener}, A.~M. and {Grollier}, F. and {Hagen}, A. and {Hirst}, P. and {Homeier}, D. and {Horton}, A.~J. and {Hosseinzadeh}, G. and {Hu}, L. and {Hunkeler}, J.~S. and {Ivezi{\'c}}, {\v{Z}}. and {Jain}, A. and {Jenness}, T. and {Kanarek}, G. and {Kendrew}, S. and {Kern}, N.~S. and {Kerzendorf}, W.~E. and {Khvalko}, A. and {King}, J. and {Kirkby}, D. and {Kulkarni}, A.~M. and {Kumar}, A. and {Lee}, A. and {Lenz}, D. and {Littlefair}, S.~P. and {Ma}, Z. and {Macleod}, D.~M. and {Mastropietro}, M. and {McCully}, C. and {Montagnac}, S. and {Morris}, B.~M. and {Mueller}, M. and {Mumford}, S.~J. and {Muna}, D. and {Murphy}, N.~A. and {Nelson}, S. and {Nguyen}, G.~H. and {Ninan}, J.~P. and {N{\"o}the}, M. and {Ogaz}, S. and {Oh}, S. and {Parejko}, J.~K. and {Parley}, N. and {Pascual}, S. and {Patil}, R. and {Patil}, A.~A. and {Plunkett}, A.~L. and {Prochaska}, J.~X. and {Rastogi}, T. and {Reddy Janga}, V. and {Sabater}, J. and {Sakurikar}, P. and {Seifert}, M. and {Sherbert}, L.~E. and {Sherwood-Taylor}, H. and {Shih}, A.~Y. and {Sick}, J. and {Silbiger}, M.~T. and {Singanamalla}, S. and {Singer}, L.~P. and {Sladen}, P.~H. and {Sooley}, K.~A. and {Sornarajah}, S. and {Streicher}, O. and {Teuben}, P. and {Thomas}, S.~W. and {Tremblay}, G.~R. and {Turner}, J.~E.~H. and {Terr{\'o}n}, V. and {van Kerkwijk}, M.~H. and {de la Vega}, A. and {Watkins}, L.~L. and {Weaver}, B.~A. and {Whitmore}, J.~B. and {Woillez}, J. and {Zabalza}, V. and {Astropy Contributors}},
  journal       = {\aj},
  title         = {{The Astropy Project: Building an Open-science Project and Status of the v2.0 Core Package}},
  year          = {2018},
  month         = sep,
  number        = {3},
  pages         = {123},
  volume        = {156},
  adsurl        = {https://ui.adsabs.harvard.edu/abs/2018AJ....156..123A},
  archiveprefix = {arXiv},
  doi           = {10.3847/1538-3881/aabc4f},
  eid           = {123},
  eprint        = {1801.02634},
  primaryclass  = {astro-ph.IM},
}

@ARTICLE{Andrews2018,
       author = {{Andrews}, Sean M. and {Huang}, Jane and {P{\'e}rez}, Laura M. and {Isella}, Andrea and {Dullemond}, Cornelis P. and {Kurtovic}, Nicol{\'a}s T. and {Guzm{\'a}n}, Viviana V. and {Carpenter}, John M. and {Wilner}, David J. and {Zhang}, Shangjia and {Zhu}, Zhaohuan and {Birnstiel}, Tilman and {Bai}, Xue-Ning and {Benisty}, Myriam and {Hughes}, A. Meredith and {{\"O}berg}, Karin I. and {Ricci}, Luca},
        title = "{The Disk Substructures at High Angular Resolution Project (DSHARP). I. Motivation, Sample, Calibration, and Overview}",
      journal = {\apjl},
         year = 2018,
        month = dec,
       volume = {869},
       number = {2},
          eid = {L41},
        pages = {L41},
          doi = {10.3847/2041-8213/aaf741},
archivePrefix = {arXiv},
       eprint = {1812.04040},
 primaryClass = {astro-ph.SR},
       adsurl = {https://ui.adsabs.harvard.edu/abs/2018ApJ...869L..41A}
}

@ARTICLE{Huang2018,
       author = {{Huang}, Jane and {Andrews}, Sean M. and {Dullemond}, Cornelis P. and {Isella}, Andrea and {P{\'e}rez}, Laura M. and {Guzm{\'a}n}, Viviana V. and {{\"O}berg}, Karin I. and {Zhu}, Zhaohuan and {Zhang}, Shangjia and {Bai}, Xue-Ning and {Benisty}, Myriam and {Birnstiel}, Tilman and {Carpenter}, John M. and {Hughes}, A. Meredith and {Ricci}, Luca and {Weaver}, Erik and {Wilner}, David J.},
        title = "{The Disk Substructures at High Angular Resolution Project (DSHARP). II. Characteristics of Annular Substructures}",
      journal = {\apjl},
         year = 2018,
        month = dec,
       volume = {869},
       number = {2},
          eid = {L42},
        pages = {L42},
          doi = {10.3847/2041-8213/aaf740},
archivePrefix = {arXiv},
       eprint = {1812.04041},
 primaryClass = {astro-ph.EP},
       adsurl = {https://ui.adsabs.harvard.edu/abs/2018ApJ...869L..42H}
}

@ARTICLE{Birnstiel2018,
       author = {{Birnstiel}, Tilman and {Dullemond}, Cornelis P. and {Zhu}, Zhaohuan and {Andrews}, Sean M. and {Bai}, Xue-Ning and {Wilner}, David J. and {Carpenter}, John M. and {Huang}, Jane and {Isella}, Andrea and {Benisty}, Myriam and {P{\'e}rez}, Laura M. and {Zhang}, Shangjia},
        title = "{The Disk Substructures at High Angular Resolution Project (DSHARP). V. Interpreting ALMA Maps of Protoplanetary Disks in Terms of a Dust Model}",
      journal = {\apjl},
         year = 2018,
        month = dec,
       volume = {869},
       number = {2},
          eid = {L45},
        pages = {L45},
          doi = {10.3847/2041-8213/aaf743},
archivePrefix = {arXiv},
       eprint = {1812.04043},
 primaryClass = {astro-ph.SR},
       adsurl = {https://ui.adsabs.harvard.edu/abs/2018ApJ...869L..45B}
}

@ARTICLE{Long2018,
       author = {{Long}, Feng and {Pinilla}, Paola and {Herczeg}, Gregory J. and {Harsono}, Daniel and {Dipierro}, Giovanni and {Pascucci}, Ilaria and {Hendler}, Nathan and {Tazzari}, Marco and {Ragusa}, Enrico and {Salyk}, Colette and {Edwards}, Suzan and {Lodato}, Giuseppe and {van de Plas}, Gerrit and {Johnstone}, Doug and {Liu}, Yao and {Boehler}, Yann and {Cabrit}, Sylvie and {Manara}, Carlo F. and {Menard}, Francois and {Mulders}, Gijs D. and {Nisini}, Brunella and {Fischer}, William J. and {Rigliaco}, Elisabetta and {Banzatti}, Andrea and {Avenhaus}, Henning and {Gully-Santiago}, Michael},
        title = "{Gaps and Rings in an ALMA Survey of Disks in the Taurus Star-forming Region}",
      journal = {\apj},
         year = 2018,
        month = dec,
       volume = {869},
       number = {1},
          eid = {17},
        pages = {17},
          doi = {10.3847/1538-4357/aae8e1},
archivePrefix = {arXiv},
       eprint = {1810.06044},
 primaryClass = {astro-ph.SR},
       adsurl = {https://ui.adsabs.harvard.edu/abs/2018ApJ...869...17L}
}

@software{Teague2018bettermoments,
       author = {{Teague}, Richard and {Foreman-Mackey}, Daniel},
        title = "{bettermoments: A robust method to measure line centroids}",
         year = 2018,
        month = sep,
          eid = {10.5281/zenodo.1419754},
          doi = {10.5281/zenodo.1419754},
      version = {v1.0},
    publisher = {Zenodo},
       adsurl = {https://ui.adsabs.harvard.edu/abs/2018zndo...1419754T}
}

@ARTICLE{Suriano2018,
       author = {{Suriano}, Scott S. and {Li}, Zhi-Yun and {Krasnopolsky}, Ruben and {Shang}, Hsien},
        title = "{The formation of rings and gaps in magnetically coupled disc-wind systems: ambipolar diffusion and reconnection}",
      journal = {\mnras},
         year = 2018,
        month = jun,
       volume = {477},
       number = {1},
        pages = {1239-1257},
          doi = {10.1093/mnras/sty717},
archivePrefix = {arXiv},
       eprint = {1712.06217},
 primaryClass = {astro-ph.SR},
       adsurl = {https://ui.adsabs.harvard.edu/abs/2018MNRAS.477.1239S}
}

@ARTICLE{Loomis2018,
       author = {{Loomis}, Ryan A. and {{\"O}berg}, Karin I. and {Andrews}, Sean M. and {Walsh}, Catherine and {Czekala}, Ian and {Huang}, Jane and {Rosenfeld}, Katherine A.},
        title = "{Detecting Weak Spectral Lines in Interferometric Data through Matched Filtering}",
      journal = {\aj},
         year = 2018,
        month = apr,
       volume = {155},
       number = {4},
          eid = {182},
        pages = {182},
          doi = {10.3847/1538-3881/aab604},
archivePrefix = {arXiv},
       eprint = {1803.04987},
 primaryClass = {astro-ph.IM},
       adsurl = {https://ui.adsabs.harvard.edu/abs/2018AJ....155..182L}
}

@ARTICLE{Sheehan2018,
       author = {{Sheehan}, Patrick D. and {Eisner}, Josh A.},
        title = "{Multiple Gaps in the Disk of the Class I Protostar GY 91}",
      journal = {\apj},
         year = 2018,
        month = apr,
       volume = {857},
       number = {1},
          eid = {18},
        pages = {18},
          doi = {10.3847/1538-4357/aaae65},
archivePrefix = {arXiv},
       eprint = {1803.02847},
 primaryClass = {astro-ph.SR},
       adsurl = {https://ui.adsabs.harvard.edu/abs/2018ApJ...857...18S}
}

@ARTICLE{Zhang2015,
       author = {{Zhang}, Ke and {Blake}, Geoffrey A. and {Bergin}, Edwin A.},
        title = "{Evidence of Fast Pebble Growth Near Condensation Fronts in the HL Tau Protoplanetary Disk}",
      journal = {\apjl},
         year = 2015,
        month = jun,
       volume = {806},
       number = {1},
          eid = {L7},
        pages = {L7},
          doi = {10.1088/2041-8205/806/1/L7},
archivePrefix = {arXiv},
       eprint = {1505.00882},
 primaryClass = {astro-ph.EP},
       adsurl = {https://ui.adsabs.harvard.edu/abs/2015ApJ...806L...7Z}
}

@ARTICLE{Takahashi2014,
       author = {{Takahashi}, Sanemichi Z. and {Inutsuka}, Shu-ichiro},
        title = "{Two-component Secular Gravitational Instability in a Protoplanetary Disk: A Possible Mechanism for Creating Ring-like Structures}",
      journal = {\apj},
         year = 2014,
        month = oct,
       volume = {794},
       number = {1},
          eid = {55},
        pages = {55},
          doi = {10.1088/0004-637X/794/1/55},
archivePrefix = {arXiv},
       eprint = {1312.6870},
 primaryClass = {astro-ph.EP},
       adsurl = {https://ui.adsabs.harvard.edu/abs/2014ApJ...794...55T}
}

@ARTICLE{Rosenfeld2012,
       author = {{Rosenfeld}, Katherine A. and {Andrews}, Sean M. and {Wilner}, David J. and {Stempels}, H.~C.},
        title = "{A Disk-based Dynamical Mass Estimate for the Young Binary V4046 Sgr}",
      journal = {\apj},
         year = 2012,
        month = nov,
       volume = {759},
       number = {2},
          eid = {119},
        pages = {119},
          doi = {10.1088/0004-637X/759/2/119},
archivePrefix = {arXiv},
       eprint = {1209.4407},
 primaryClass = {astro-ph.SR},
       adsurl = {https://ui.adsabs.harvard.edu/abs/2012ApJ...759..119R}
}

@ARTICLE{astropy2013,
       author = {{Astropy Collaboration} and {Robitaille}, Thomas P. and {Tollerud}, Erik J. and {Greenfield}, Perry and {Droettboom}, Michael and {Bray}, Erik and {Aldcroft}, Tom and {Davis}, Matt and {Ginsburg}, Adam and {Price-Whelan}, Adrian M. and {Kerzendorf}, Wolfgang E. and {Conley}, Alexander and {Crighton}, Neil and {Barbary}, Kyle and {Muna}, Demitri and {Ferguson}, Henry and {Grollier}, Fr{\'e}d{\'e}ric and {Parikh}, Madhura M. and {Nair}, Prasanth H. and {Unther}, Hans M. and {Deil}, Christoph and {Woillez}, Julien and {Conseil}, Simon and {Kramer}, Roban and {Turner}, James E.~H. and {Singer}, Leo and {Fox}, Ryan and {Weaver}, Benjamin A. and {Zabalza}, Victor and {Edwards}, Zachary I. and {Azalee Bostroem}, K. and {Burke}, D.~J. and {Casey}, Andrew R. and {Crawford}, Steven M. and {Dencheva}, Nadia and {Ely}, Justin and {Jenness}, Tim and {Labrie}, Kathleen and {Lim}, Pey Lian and {Pierfederici}, Francesco and {Pontzen}, Andrew and {Ptak}, Andy and {Refsdal}, Brian and {Servillat}, Mathieu and {Streicher}, Ole},
        title = "{Astropy: A community Python package for astronomy}",
      journal = {\aap},
         year = 2013,
        month = oct,
       volume = {558},
          eid = {A33},
        pages = {A33},
          doi = {10.1051/0004-6361/201322068},
archivePrefix = {arXiv},
       eprint = {1307.6212},
 primaryClass = {astro-ph.IM},
       adsurl = {https://ui.adsabs.harvard.edu/abs/2013A&A...558A..33A}
}

@ARTICLE{Paxton2011,
       author = {{Paxton}, Bill and {Bildsten}, Lars and {Dotter}, Aaron and {Herwig}, Falk and {Lesaffre}, Pierre and {Timmes}, Frank},
        title = "{Modules for Experiments in Stellar Astrophysics (MESA)}",
      journal = {\apjs},
         year = 2011,
        month = jan,
       volume = {192},
       number = {1},
          eid = {3},
        pages = {3},
          doi = {10.1088/0067-0049/192/1/3},
archivePrefix = {arXiv},
       eprint = {1009.1622},
 primaryClass = {astro-ph.SR},
       adsurl = {https://ui.adsabs.harvard.edu/abs/2011ApJS..192....3P}
}

@ARTICLE{Choi2016,
       author = {{Choi}, Jieun and {Dotter}, Aaron and {Conroy}, Charlie and {Cantiello}, Matteo and {Paxton}, Bill and {Johnson}, Benjamin D.},
        title = "{Mesa Isochrones and Stellar Tracks (MIST). I. Solar-scaled Models}",
      journal = {\apj},
         year = 2016,
        month = jun,
       volume = {823},
       number = {2},
          eid = {102},
        pages = {102},
          doi = {10.3847/0004-637X/823/2/102},
archivePrefix = {arXiv},
       eprint = {1604.08592},
 primaryClass = {astro-ph.SR},
       adsurl = {https://ui.adsabs.harvard.edu/abs/2016ApJ...823..102C}
}

@ARTICLE{Dotter2016,
       author = {{Dotter}, Aaron},
        title = "{MESA Isochrones and Stellar Tracks (MIST) 0: Methods for the Construction of Stellar Isochrones}",
      journal = {\apjs},
         year = 2016,
        month = jan,
       volume = {222},
       number = {1},
          eid = {8},
        pages = {8},
          doi = {10.3847/0067-0049/222/1/8},
archivePrefix = {arXiv},
       eprint = {1601.05144},
 primaryClass = {astro-ph.SR},
       adsurl = {https://ui.adsabs.harvard.edu/abs/2016ApJS..222....8D}
}

@ARTICLE{Cieza2019,
       author = {{Cieza}, Lucas A. and {Ru{\'\i}z-Rodr{\'\i}guez}, Dary and {Hales}, Antonio and {Casassus}, Simon and {P{\'e}rez}, Sebastian and {Gonzalez-Ruilova}, Camilo and {C{\'a}novas}, Hector and {Williams}, Jonathan P. and {Zurlo}, Alice and {Ansdell}, Megan and {Avenhaus}, Henning and {Bayo}, Amelia and {Bertrang}, Gesa H. -M. and {Christiaens}, Valentin and {Dent}, William and {Ferrero}, Gabriel and {Gamen}, Roberto and {Olofsson}, Johan and {Orcajo}, Santiago and {Pe{\~n}a Ram{\'\i}rez}, Karla and {Principe}, David and {Schreiber}, Matthias R. and {van der Plas}, Gerrit},
        title = "{The Ophiuchus DIsc Survey Employing ALMA (ODISEA) - I: project description and continuum images at 28 au resolution}",
      journal = {\mnras},
         year = 2019,
        month = jan,
       volume = {482},
       number = {1},
        pages = {698-714},
          doi = {10.1093/mnras/sty2653},
archivePrefix = {arXiv},
       eprint = {1809.08844},
 primaryClass = {astro-ph.EP},
       adsurl = {https://ui.adsabs.harvard.edu/abs/2019MNRAS.482..698C}
}

@ARTICLE{matplotlib,
       author = {{Hunter}, John D.},
        title = "{Matplotlib: A 2D Graphics Environment}",
      journal = {Computing in Science and Engineering},
         year = 2007,
        month = may,
       volume = {9},
       number = {3},
        pages = {90-95},
          doi = {10.1109/MCSE.2007.55},
       adsurl = {https://ui.adsabs.harvard.edu/abs/2007CSE.....9...90H}
}

@ARTICLE{numpy,
       author = {{Harris}, Charles R. and {Millman}, K. Jarrod and {van der Walt}, St{\'e}fan J. and {Gommers}, Ralf and {Virtanen}, Pauli and {Cournapeau}, David and {Wieser}, Eric and {Taylor}, Julian and {Berg}, Sebastian and {Smith}, Nathaniel J. and {Kern}, Robert and {Picus}, Matti and {Hoyer}, Stephan and {van Kerkwijk}, Marten H. and {Brett}, Matthew and {Haldane}, Allan and {del R{\'\i}o}, Jaime Fern{\'a}ndez and {Wiebe}, Mark and {Peterson}, Pearu and {G{\'e}rard-Marchant}, Pierre and {Sheppard}, Kevin and {Reddy}, Tyler and {Weckesser}, Warren and {Abbasi}, Hameer and {Gohlke}, Christoph and {Oliphant}, Travis E.},
        title = "{Array programming with NumPy}",
      journal = {\nat},
         year = 2020,
        month = sep,
       volume = {585},
       number = {7825},
        pages = {357-362},
          doi = {10.1038/s41586-020-2649-2},
archivePrefix = {arXiv},
       eprint = {2006.10256},
 primaryClass = {cs.MS},
       adsurl = {https://ui.adsabs.harvard.edu/abs/2020Natur.585..357H}
}

@software{mpol,
author       = {Ian Czekala and
                Jeff Jennings and   
                Brianna Zawadzki and
                Ryan Loomis and
                Kadri Nizam and 
                Megan Delamer and 
                Kaylee de Soto and
                Robert Frazier and
                Hannah Grzybowski and
                Mary Ogborn and                    
                Tyler Quinn},
title        = {MPoL-dev/MPoL: v0.2.0 Release},
month        = nov,
year         = 2023,
publisher    = {Zenodo},
version      = {v0.2.0},
doi          = {10.5281/zenodo.3594081},
url          = {https://doi.org/10.5281/zenodo.3594081}
}

@ARTICLE{Zawadzki2023,
    author = {{Zawadzki}, Brianna and {Czekala}, Ian and {Loomis}, Ryan A. and {Quinn}, Tyler and {Grzybowski}, Hannah and {Frazier}, Robert C. and {Jennings}, Jeff and {Nizam}, Kadri M. and {Jian}, Yina},
        title = "{Regularized Maximum Likelihood Image Synthesis and Validation for ALMA Continuum Observations of Protoplanetary Disks}",
    journal = {\pasp},
        year = 2023,
        month = jun,
    volume = {135},
    number = {1048},
        eid = {064503},
        pages = {064503},
        doi = {10.1088/1538-3873/acdf84},
archivePrefix = {arXiv},
    eprint = {2209.11813},
primaryClass = {astro-ph.EP},
    adsurl = {https://ui.adsabs.harvard.edu/abs/2023PASP..135f4503Z}
}

@ARTICLE{Braun2021,
       author = {{Braun}, Teresa A.~M. and {Yen}, Hsi-Wei and {Koch}, Patrick M. and {Manara}, Carlo F. and {Miotello}, Anna and {Testi}, Leonardo},
        title = "{Dynamical Stellar Masses of Pre-main-sequence Stars in Lupus and Taurus Obtained with ALMA Surveys in Comparison with Stellar Evolutionary Models}",
      journal = {\apj},
         year = 2021,
        month = feb,
       volume = {908},
       number = {1},
          eid = {46},
        pages = {46},
          doi = {10.3847/1538-4357/abd24f},
archivePrefix = {arXiv},
       eprint = {2012.07441},
 primaryClass = {astro-ph.SR},
       adsurl = {https://ui.adsabs.harvard.edu/abs/2021ApJ...908...46B}
}

@ARTICLE{Simon2000,
       author = {{Simon}, M. and {Dutrey}, A. and {Guilloteau}, S.},
        title = "{Dynamical Masses of T Tauri Stars and Calibration of Pre-Main-Sequence Evolution}",
      journal = {\apj},
         year = 2000,
        month = dec,
       volume = {545},
       number = {2},
        pages = {1034-1043},
          doi = {10.1086/317838},
archivePrefix = {arXiv},
       eprint = {astro-ph/0008370},
 primaryClass = {astro-ph},
       adsurl = {https://ui.adsabs.harvard.edu/abs/2000ApJ...545.1034S}
}

@software{Teague2020keplerianmask,
       author = {{Teague}, Rich},
        title = "{richteague/keplerian\_mask: Initial Release}",
         year = 2020,
        month = dec,
          eid = {10.5281/zenodo.4321137},
          doi = {10.5281/zenodo.4321137},
      version = {1.0},
    publisher = {Zenodo},
       adsurl = {https://ui.adsabs.harvard.edu/abs/2020zndo...4321137T}
}

@ARTICLE{Draine2003,
       author = {{Draine}, B.~T.},
        title = "{Interstellar Dust Grains}",
      journal = {\araa},
         year = 2003,
        month = jan,
       volume = {41},
        pages = {241-289},
          doi = {10.1146/annurev.astro.41.011802.094840},
archivePrefix = {arXiv},
       eprint = {astro-ph/0304489},
 primaryClass = {astro-ph},
       adsurl = {https://ui.adsabs.harvard.edu/abs/2003ARA&A..41..241D}
}

@ARTICLE{Warren2008,
       author = {{Warren}, Stephen G. and {Brandt}, Richard E.},
        title = "{Optical constants of ice from the ultraviolet to the microwave: A revised compilation}",
      journal = {Journal of Geophysical Research (Atmospheres)},
         year = 2008,
        month = jul,
       volume = {113},
       number = {D14},
          eid = {D14220},
        pages = {D14220},
          doi = {10.1029/2007JD009744},
       adsurl = {https://ui.adsabs.harvard.edu/abs/2008JGRD..11314220W}
}

@ARTICLE{Henning1996,
       author = {{Henning}, T. and {Stognienko}, R.},
        title = "{Dust opacities for protoplanetary accretion disks: influence of dust aggregates.}",
      journal = {\aap},
         year = 1996,
        month = jul,
       volume = {311},
        pages = {291-303},
       adsurl = {https://ui.adsabs.harvard.edu/abs/1996A&A...311..291H}
}

@ARTICLE{Goldreich1980,
       author = {{Goldreich}, P. and {Tremaine}, S.},
        title = "{Disk-satellite interactions.}",
      journal = {\apj},
         year = 1980,
        month = oct,
       volume = {241},
        pages = {425-441},
          doi = {10.1086/158356},
       adsurl = {https://ui.adsabs.harvard.edu/abs/1980ApJ...241..425G}
}

@ARTICLE{Zhang2018,
       author = {{Zhang}, Shangjia and {Zhu}, Zhaohuan and {Huang}, Jane and {Guzm{\'a}n}, Viviana V. and {Andrews}, Sean M. and {Birnstiel}, Tilman and {Dullemond}, Cornelis P. and {Carpenter}, John M. and {Isella}, Andrea and {P{\'e}rez}, Laura M. and {Benisty}, Myriam and {Wilner}, David J. and {Baruteau}, Cl{\'e}ment and {Bai}, Xue-Ning and {Ricci}, Luca},
        title = "{The Disk Substructures at High Angular Resolution Project (DSHARP). VII. The Planet-Disk Interactions Interpretation}",
      journal = {\apjl},
         year = 2018,
        month = dec,
       volume = {869},
       number = {2},
          eid = {L47},
        pages = {L47},
          doi = {10.3847/2041-8213/aaf744},
archivePrefix = {arXiv},
       eprint = {1812.04045},
 primaryClass = {astro-ph.EP},
       adsurl = {https://ui.adsabs.harvard.edu/abs/2018ApJ...869L..47Z}
}

@INPROCEEDINGS{Baruteau2014,
       author = {{Baruteau}, C. and {Crida}, A. and {Paardekooper}, S. -J. and {Masset}, F. and {Guilet}, J. and {Bitsch}, B. and {Nelson}, R. and {Kley}, W. and {Papaloizou}, J.},
        title = "{Planet-Disk Interactions and Early Evolution of Planetary Systems}",
    booktitle = {Protostars and Planets VI},
         year = 2014,
       editor = {{Beuther}, Henrik and {Klessen}, Ralf S. and {Dullemond}, Cornelis P. and {Henning}, Thomas},
        month = jan,
        pages = {667-689},
          doi = {10.2458/azu_uapress_9780816531240-ch029},
archivePrefix = {arXiv},
       eprint = {1312.4293},
 primaryClass = {astro-ph.EP},
       adsurl = {https://ui.adsabs.harvard.edu/abs/2014prpl.conf..667B}
}

@INPROCEEDINGS{Paardekooper2023,
       author = {{Paardekooper}, S. and {Dong}, R. and {Duffell}, P. and {Fung}, J. and {Masset}, F.~S. and {Ogilvie}, G. and {Tanaka}, H.},
        title = "{Planet-Disk Interactions and Orbital Evolution}",
    booktitle = {Protostars and Planets VII},
         year = 2023,
       editor = {{Inutsuka}, S. and {Aikawa}, Y. and {Muto}, T. and {Tomida}, K. and {Tamura}, M.},
       series = {Astronomical Society of the Pacific Conference Series},
       volume = {534},
        month = jul,
        pages = {685},
          doi = {10.48550/arXiv.2203.09595},
archivePrefix = {arXiv},
       eprint = {2203.09595},
 primaryClass = {astro-ph.EP},
       adsurl = {https://ui.adsabs.harvard.edu/abs/2023ASPC..534..685P}
}

@ARTICLE{Canovas2019,
       author = {{C{\'a}novas}, H. and {Cantero}, C. and {Cieza}, L. and {Bombrun}, A. and {Lammers}, U. and {Mer{\'\i}n}, B. and {Mora}, A. and {Ribas}, {\'A}. and {Ru{\'\i}z-Rodr{\'\i}guez}, D.},
        title = "{Census of {\ensuremath{\rho}} Ophiuchi candidate members from Gaia Data Release 2}",
      journal = {\aap},
         year = 2019,
        month = jun,
       volume = {626},
          eid = {A80},
        pages = {A80},
          doi = {10.1051/0004-6361/201935321},
archivePrefix = {arXiv},
       eprint = {1902.07600},
 primaryClass = {astro-ph.EP},
       adsurl = {https://ui.adsabs.harvard.edu/abs/2019A&A...626A..80C}
}

@ARTICLE{McClure2010,
       author = {{McClure}, M.~K. and {Furlan}, E. and {Manoj}, P. and {Luhman}, K.~L. and {Watson}, D.~M. and {Forrest}, W.~J. and {Espaillat}, C. and {Calvet}, N. and {D'Alessio}, P. and {Sargent}, B. and {Tobin}, J.~J. and {Chiang}, Hsin-Fang},
        title = "{The Evolutionary State of the Pre-main Sequence Population in Ophiuchus: A Large Infrared Spectrograph Survey}",
      journal = {\apjs},
         year = 2010,
        month = may,
       volume = {188},
       number = {1},
        pages = {75-122},
          doi = {10.1088/0067-0049/188/1/75},
       adsurl = {https://ui.adsabs.harvard.edu/abs/2010ApJS..188...75M}
}

@ARTICLE{Enoch2009,
       author = {{Enoch}, Melissa L. and {Evans}, Neal J., II and {Sargent}, Anneila I. and {Glenn}, Jason},
        title = "{Properties of the Youngest Protostars in Perseus, Serpens, and Ophiuchus}",
      journal = {\apj},
         year = 2009,
        month = feb,
       volume = {692},
       number = {2},
        pages = {973-997},
          doi = {10.1088/0004-637X/692/2/973},
archivePrefix = {arXiv},
       eprint = {0809.4012},
 primaryClass = {astro-ph},
       adsurl = {https://ui.adsabs.harvard.edu/abs/2009ApJ...692..973E}
}

@ARTICLE{Pinte2006,
       author = {{Pinte}, C. and {M{\'e}nard}, F. and {Duch{\^e}ne}, G. and {Bastien}, P.},
        title = "{Monte Carlo radiative transfer in protoplanetary disks}",
      journal = {\aap},
         year = 2006,
        month = dec,
       volume = {459},
       number = {3},
        pages = {797-804},
          doi = {10.1051/0004-6361:20053275},
archivePrefix = {arXiv},
       eprint = {astro-ph/0606550},
 primaryClass = {astro-ph},
       adsurl = {https://ui.adsabs.harvard.edu/abs/2006A&A...459..797P}
}

@ARTICLE{Pinte2009,
       author = {{Pinte}, C. and {Harries}, T.~J. and {Min}, M. and {Watson}, A.~M. and {Dullemond}, C.~P. and {Woitke}, P. and {M{\'e}nard}, F. and {Dur{\'a}n-Rojas}, M.~C.},
        title = "{Benchmark problems for continuum radiative transfer. High optical depths, anisotropic scattering, and polarisation}",
      journal = {\aap},
         year = 2009,
        month = may,
       volume = {498},
       number = {3},
        pages = {967-980},
          doi = {10.1051/0004-6361/200811555},
archivePrefix = {arXiv},
       eprint = {0903.1231},
 primaryClass = {astro-ph.SR},
       adsurl = {https://ui.adsabs.harvard.edu/abs/2009A&A...498..967P}
}

@ARTICLE{Sheehan2017,
       author = {{Sheehan}, Patrick D. and {Eisner}, Josh A.},
        title = "{WL 17: A Young Embedded Transition Disk}",
      journal = {\apjl},
         year = 2017,
        month = may,
       volume = {840},
       number = {2},
          eid = {L12},
        pages = {L12},
          doi = {10.3847/2041-8213/aa6df8},
archivePrefix = {arXiv},
       eprint = {1704.07916},
 primaryClass = {astro-ph.SR},
       adsurl = {https://ui.adsabs.harvard.edu/abs/2017ApJ...840L..12S}
}

@ARTICLE{Endres2016,
       author = {{Endres}, Christian P. and {Schlemmer}, Stephan and {Schilke}, Peter and {Stutzki}, J{\"u}rgen and {M{\"u}ller}, Holger S.~P.},
        title = "{The Cologne Database for Molecular Spectroscopy, CDMS, in the Virtual Atomic and Molecular Data Centre, VAMDC}",
      journal = {Journal of Molecular Spectroscopy},
         year = 2016,
        month = sep,
       volume = {327},
        pages = {95-104},
          doi = {10.1016/j.jms.2016.03.005},
archivePrefix = {arXiv},
       eprint = {1603.03264},
 primaryClass = {astro-ph.IM},
       adsurl = {https://ui.adsabs.harvard.edu/abs/2016JMoSp.327...95E}
}

@ARTICLE{Kepley2020,
       author = {{Kepley}, Amanda A. and {Tsutsumi}, Takahiro and {Brogan}, Crystal L. and {Indebetouw}, Remy and {Yoon}, Ilsang and {Mason}, Brian and {Donovan Meyer}, Jennifer},
        title = "{Auto-multithresh: A General Purpose Automasking Algorithm}",
      journal = {\pasp},
         year = 2020,
        month = feb,
       volume = {132},
       number = {1008},
          eid = {024505},
        pages = {024505},
          doi = {10.1088/1538-3873/ab5e14},
archivePrefix = {arXiv},
       eprint = {1912.04970},
 primaryClass = {astro-ph.IM},
       adsurl = {https://ui.adsabs.harvard.edu/abs/2020PASP..132b4505K}
}

@ARTICLE{Doppmann2005,
       author = {{Doppmann}, G.~W. and {Greene}, T.~P. and {Covey}, K.~R. and {Lada}, C.~J.},
        title = "{The Physical Natures of Class I and Flat-Spectrum Protostellar Photospheres: A Near-Infrared Spectroscopic Study}",
      journal = {\aj},
         year = 2005,
        month = sep,
       volume = {130},
       number = {3},
        pages = {1145-1170},
          doi = {10.1086/431954},
archivePrefix = {arXiv},
       eprint = {astro-ph/0505295},
 primaryClass = {astro-ph},
       adsurl = {https://ui.adsabs.harvard.edu/abs/2005AJ....130.1145D}
}

@ARTICLE{Law2021_profile,
       author = {{Law}, Charles J. and {Loomis}, Ryan A. and {Teague}, Richard and {{\"O}berg}, Karin I. and {Czekala}, Ian and {Andrews}, Sean M. and {Huang}, Jane and {Aikawa}, Yuri and {Alarc{\'o}n}, Felipe and {Bae}, Jaehan and {Bergin}, Edwin A. and {Bergner}, Jennifer B. and {Boehler}, Yann and {Booth}, Alice S. and {Bosman}, Arthur D. and {Calahan}, Jenny K. and {Cataldi}, Gianni and {Cleeves}, L. Ilsedore and {Furuya}, Kenji and {Guzm{\'a}n}, Viviana V. and {Ilee}, John D. and {Le Gal}, Romane and {Liu}, Yao and {Long}, Feng and {M{\'e}nard}, Fran{\c{c}}ois and {Nomura}, Hideko and {Qi}, Chunhua and {Schwarz}, Kamber R. and {Sierra}, Anibal and {Tsukagoshi}, Takashi and {Yamato}, Yoshihide and {van't Hoff}, Merel L.~R. and {Walsh}, Catherine and {Wilner}, David J. and {Zhang}, Ke},
        title = "{Molecules with ALMA at Planet-forming Scales (MAPS). III. Characteristics of Radial Chemical Substructures}",
      journal = {\apjs},
         year = 2021,
        month = nov,
       volume = {257},
       number = {1},
          eid = {3},
        pages = {3},
          doi = {10.3847/1538-4365/ac1434},
archivePrefix = {arXiv},
       eprint = {2109.06210},
 primaryClass = {astro-ph.EP},
       adsurl = {https://ui.adsabs.harvard.edu/abs/2021ApJS..257....3L}
}

@ARTICLE{Huang2024,
       author = {{Huang}, Jane and {Bergin}, Edwin A. and {Le Gal}, Romane and {Andrews}, Sean M. and {Bae}, Jaehan and {Keyte}, Luke and {Sturm}, J.~A.},
        title = "{Constraints on the Gas-phase C/O Ratio of DR Tau's Outer Disk from CS, SO, and C$_{2}$H Observations}",
      journal = {\apj},
         year = 2024,
        month = oct,
       volume = {973},
       number = {2},
          eid = {135},
        pages = {135},
          doi = {10.3847/1538-4357/ad6447},
archivePrefix = {arXiv},
       eprint = {2407.01679},
 primaryClass = {astro-ph.EP},
       adsurl = {https://ui.adsabs.harvard.edu/abs/2024ApJ...973..135H}
}

@ARTICLE{Oberg2021,
       author = {{{\"O}berg}, Karin I. and {Guzm{\'a}n}, Viviana V. and {Walsh}, Catherine and {Aikawa}, Yuri and {Bergin}, Edwin A. and {Law}, Charles J. and {Loomis}, Ryan A. and {Alarc{\'o}n}, Felipe and {Andrews}, Sean M. and {Bae}, Jaehan and {Bergner}, Jennifer B. and {Boehler}, Yann and {Booth}, Alice S. and {Bosman}, Arthur D. and {Calahan}, Jenny K. and {Cataldi}, Gianni and {Cleeves}, L. Ilsedore and {Czekala}, Ian and {Furuya}, Kenji and {Huang}, Jane and {Ilee}, John D. and {Kurtovic}, Nicolas T. and {Le Gal}, Romane and {Liu}, Yao and {Long}, Feng and {M{\'e}nard}, Fran{\c{c}}ois and {Nomura}, Hideko and {P{\'e}rez}, Laura M. and {Qi}, Chunhua and {Schwarz}, Kamber R. and {Sierra}, Anibal and {Teague}, Richard and {Tsukagoshi}, Takashi and {Yamato}, Yoshihide and {van't Hoff}, Merel L.~R. and {Waggoner}, Abygail R. and {Wilner}, David J. and {Zhang}, Ke},
        title = "{Molecules with ALMA at Planet-forming Scales (MAPS). I. Program Overview and Highlights}",
      journal = {\apjs},
         year = 2021,
        month = nov,
       volume = {257},
       number = {1},
          eid = {1},
        pages = {1},
          doi = {10.3847/1538-4365/ac1432},
archivePrefix = {arXiv},
       eprint = {2109.06268},
 primaryClass = {astro-ph.EP},
       adsurl = {https://ui.adsabs.harvard.edu/abs/2021ApJS..257....1O}
}

@ARTICLE{Kratter2016,
       author = {{Kratter}, Kaitlin and {Lodato}, Giuseppe},
        title = "{Gravitational Instabilities in Circumstellar Disks}",
      journal = {\araa},
         year = 2016,
        month = sep,
       volume = {54},
        pages = {271-311},
          doi = {10.1146/annurev-astro-081915-023307},
archivePrefix = {arXiv},
       eprint = {1603.01280},
 primaryClass = {astro-ph.SR},
       adsurl = {https://ui.adsabs.harvard.edu/abs/2016ARA&A..54..271K}
}

@ARTICLE{Podio2024,
       author = {{Podio}, L. and {Ceccarelli}, C. and {Codella}, C. and {Sabatini}, G. and {Segura-Cox}, D. and {Balucani}, N. and {Rimola}, A. and {Ugliengo}, P. and {Chandler}, C.~J. and {Sakai}, N. and {Svoboda}, B. and {Pineda}, J. and {De Simone}, M. and {Bianchi}, E. and {Caselli}, P. and {Isella}, A. and {Aikawa}, Y. and {Bouvier}, M. and {Caux}, E. and {Chahine}, L. and {Charnley}, S.~B. and {Cuello}, N. and {Dulieu}, F. and {Evans}, L. and {Fedele}, D. and {Feng}, S. and {Fontani}, F. and {Hama}, T. and {Hanawa}, T. and {Herbst}, E. and {Hirota}, T. and {Jim{\'e}nez-Serra}, I. and {Johnstone}, D. and {Lefloch}, B. and {Le Gal}, R. and {Loinard}, L. and {Liu}, H. Baobab and {L{\'o}pez-Sepulcre}, A. and {Maud}, L.~T. and {Maureira}, M.~J. and {Menard}, F. and {Miotello}, A. and {Moellenbrock}, G. and {Nomura}, H. and {Oba}, Y. and {Ohashi}, S. and {Okoda}, Y. and {Oya}, Y. and {Sakai}, T. and {Shirley}, Y. and {Testi}, L. and {Vastel}, C. and {Viti}, S. and {Watanabe}, N. and {Watanabe}, Y. and {Zhang}, Y. and {Zhang}, Z.~E. and {Yamamoto}, S.},
        title = "{FAUST. XVII. Super deuteration in the planet-forming system IRS 63 where the streamer strikes the disk}",
      journal = {\aap},
         year = 2024,
        month = aug,
       volume = {688},
          eid = {L22},
        pages = {L22},
          doi = {10.1051/0004-6361/202450742},
archivePrefix = {arXiv},
       eprint = {2407.04813},
 primaryClass = {astro-ph.EP},
       adsurl = {https://ui.adsabs.harvard.edu/abs/2024A&A...688L..22P}
}

@ARTICLE{Cleeves2015,
       author = {{Cleeves}, L. Ilsedore and {Bergin}, Edwin A. and {Harries}, Tim J.},
        title = "{Indirect Detection of Forming Protoplanets via Chemical Asymmetries in Disks}",
      journal = {\apj},
         year = 2015,
        month = jul,
       volume = {807},
       number = {1},
          eid = {2},
        pages = {2},
          doi = {10.1088/0004-637X/807/1/2},
archivePrefix = {arXiv},
       eprint = {1505.07470},
 primaryClass = {astro-ph.SR},
       adsurl = {https://ui.adsabs.harvard.edu/abs/2015ApJ...807....2C}
}

@ARTICLE{LeGal2019,
       author = {{Le Gal}, Romane and {{\"O}berg}, Karin I. and {Loomis}, Ryan A. and {Pegues}, Jamila and {Bergner}, Jennifer B.},
        title = "{Sulfur Chemistry in Protoplanetary Disks: CS and H$_{2}$CS}",
      journal = {\apj},
         year = 2019,
        month = may,
       volume = {876},
       number = {1},
          eid = {72},
        pages = {72},
          doi = {10.3847/1538-4357/ab1416},
archivePrefix = {arXiv},
       eprint = {1903.11105},
 primaryClass = {astro-ph.GA},
       adsurl = {https://ui.adsabs.harvard.edu/abs/2019ApJ...876...72L}
}

@ARTICLE{LeGal2021,
       author = {{Le Gal}, Romane and {{\"O}berg}, Karin I. and {Teague}, Richard and {Loomis}, Ryan A. and {Law}, Charles J. and {Walsh}, Catherine and {Bergin}, Edwin A. and {M{\'e}nard}, Fran{\c{c}}ois and {Wilner}, David J. and {Andrews}, Sean M. and {Aikawa}, Yuri and {Booth}, Alice S. and {Cataldi}, Gianni and {Bergner}, Jennifer B. and {Bosman}, Arthur D. and {Cleeves}, L. Ilse and {Czekala}, Ian and {Furuya}, Kenji and {Guzm{\'a}n}, Viviana V. and {Huang}, Jane and {Ilee}, John D. and {Nomura}, Hideko and {Qi}, Chunhua and {Schwarz}, Kamber R. and {Tsukagoshi}, Takashi and {Yamato}, Yoshihide and {Zhang}, Ke},
        title = "{Molecules with ALMA at Planet-forming Scales (MAPS). XII. Inferring the C/O and S/H Ratios in Protoplanetary Disks with Sulfur Molecules}",
      journal = {\apjs},
         year = 2021,
        month = nov,
       volume = {257},
       number = {1},
          eid = {12},
        pages = {12},
          doi = {10.3847/1538-4365/ac2583},
archivePrefix = {arXiv},
       eprint = {2109.06286},
 primaryClass = {astro-ph.GA},
       adsurl = {https://ui.adsabs.harvard.edu/abs/2021ApJS..257...12L}
}

@ARTICLE{vanderPlas2014,
       author = {{van der Plas}, G. and {Casassus}, S. and {M{\'e}nard}, F. and {Perez}, S. and {Thi}, W.~F. and {Pinte}, C. and {Christiaens}, V.},
        title = "{Spatially Resolved HCN J = 4-3 and CS J = 7-6 Emission from the Disk around HD 142527}",
      journal = {\apjl},
         year = 2014,
        month = sep,
       volume = {792},
       number = {2},
          eid = {L25},
        pages = {L25},
          doi = {10.1088/2041-8205/792/2/L25},
archivePrefix = {arXiv},
       eprint = {1407.1735},
 primaryClass = {astro-ph.SR},
       adsurl = {https://ui.adsabs.harvard.edu/abs/2014ApJ...792L..25V}
}

@ARTICLE{Pegues2020,
       author = {{Pegues}, Jamila and {{\"O}berg}, Karin I. and {Bergner}, Jennifer B. and {Loomis}, Ryan A. and {Qi}, Chunhua and {Le Gal}, Romane and {Cleeves}, L. Ilsedore and {Guzm{\'a}n}, Viviana V. and {Huang}, Jane and {J{\o}rgensen}, Jes K. and {Andrews}, Sean M. and {Blake}, Geoffrey A. and {Carpenter}, John M. and {Schwarz}, Kamber R. and {Williams}, Jonathan P. and {Wilner}, David J.},
        title = "{An ALMA Survey of H$_{2}$CO in Protoplanetary Disks}",
      journal = {\apj},
         year = 2020,
        month = feb,
       volume = {890},
       number = {2},
          eid = {142},
        pages = {142},
          doi = {10.3847/1538-4357/ab64d9},
archivePrefix = {arXiv},
       eprint = {2002.12525},
 primaryClass = {astro-ph.SR},
       adsurl = {https://ui.adsabs.harvard.edu/abs/2020ApJ...890..142P}
}

@ARTICLE{Pegues2021,
       author = {{Pegues}, Jamila and {Czekala}, Ian and {Andrews}, Sean M. and {{\"O}berg}, Karin I. and {Herczeg}, Gregory J. and {Bergner}, Jennifer B. and {Ilsedore Cleeves}, L. and {Guzm{\'a}n}, Viviana V. and {Huang}, Jane and {Long}, Feng and {Teague}, Richard and {Wilner}, David J.},
        title = "{Dynamical Masses and Stellar Evolutionary Model Predictions of M Stars}",
      journal = {\apj},
         year = 2021,
        month = feb,
       volume = {908},
       number = {1},
          eid = {42},
        pages = {42},
          doi = {10.3847/1538-4357/abd4eb},
archivePrefix = {arXiv},
       eprint = {2101.05838},
 primaryClass = {astro-ph.SR},
       adsurl = {https://ui.adsabs.harvard.edu/abs/2021ApJ...908...42P}
}

@ARTICLE{Facchini2021,
       author = {{Facchini}, Stefano and {Teague}, Richard and {Bae}, Jaehan and {Benisty}, Myriam and {Keppler}, Miriam and {Isella}, Andrea},
        title = "{The Chemical Inventory of the Planet-hosting Disk PDS 70}",
      journal = {\aj},
         year = 2021,
        month = sep,
       volume = {162},
       number = {3},
          eid = {99},
        pages = {99},
          doi = {10.3847/1538-3881/abf0a4},
archivePrefix = {arXiv},
       eprint = {2101.08369},
 primaryClass = {astro-ph.EP},
       adsurl = {https://ui.adsabs.harvard.edu/abs/2021AJ....162...99F}
}

@ARTICLE{Guzman2021,
       author = {{Guzm{\'a}n}, Viviana V. and {Bergner}, Jennifer B. and {Law}, Charles J. and {{\"O}berg}, Karin I. and {Walsh}, Catherine and {Cataldi}, Gianni and {Aikawa}, Yuri and {Bergin}, Edwin A. and {Czekala}, Ian and {Huang}, Jane and {Andrews}, Sean M. and {Loomis}, Ryan A. and {Zhang}, Ke and {Le Gal}, Romane and {Alarc{\'o}n}, Felipe and {Ilee}, John D. and {Teague}, Richard and {Cleeves}, L. Ilsedore and {Wilner}, David J. and {Long}, Feng and {Schwarz}, Kamber R. and {Bosman}, Arthur D. and {P{\'e}rez}, Laura M. and {M{\'e}nard}, Fran{\c{c}}ois and {Liu}, Yao},
        title = "{Molecules with ALMA at Planet-forming Scales (MAPS). VI. Distribution of the Small Organics HCN, C$_{2}$H, and H$_{2}$CO}",
      journal = {\apjs},
         year = 2021,
        month = nov,
       volume = {257},
       number = {1},
          eid = {6},
        pages = {6},
          doi = {10.3847/1538-4365/ac1440},
archivePrefix = {arXiv},
       eprint = {2109.06391},
 primaryClass = {astro-ph.EP},
       adsurl = {https://ui.adsabs.harvard.edu/abs/2021ApJS..257....6G}
}

@ARTICLE{Gaia2016,
       author = {{Gaia Collaboration} and {Prusti}, T. and {de Bruijne}, J.~H.~J. and {Brown}, A.~G.~A. and {Vallenari}, A. and {Babusiaux}, C. and {Bailer-Jones}, C.~A.~L. and {Bastian}, U. and {Biermann}, M. and {Evans}, D.~W. and {Eyer}, L. and {Jansen}, F. and {Jordi}, C. and {Klioner}, S.~A. and {Lammers}, U. and {Lindegren}, L. and {Luri}, X. and {Mignard}, F. and {Milligan}, D.~J. and {Panem}, C. and {Poinsignon}, V. and {Pourbaix}, D. and {Randich}, S. and {Sarri}, G. and {Sartoretti}, P. and {Siddiqui}, H.~I. and {Soubiran}, C. and {Valette}, V. and {van Leeuwen}, F. and {Walton}, N.~A. and {Aerts}, C. and {Arenou}, F. and {Cropper}, M. and {Drimmel}, R. and {H{\o}g}, E. and {Katz}, D. and {Lattanzi}, M.~G. and {O'Mullane}, W. and {Grebel}, E.~K. and {Holland}, A.~D. and {Huc}, C. and {Passot}, X. and {Bramante}, L. and {Cacciari}, C. and {Casta{\~n}eda}, J. and {Chaoul}, L. and {Cheek}, N. and {De Angeli}, F. and {Fabricius}, C. and {Guerra}, R. and {Hern{\'a}ndez}, J. and {Jean-Antoine-Piccolo}, A. and {Masana}, E. and {Messineo}, R. and {Mowlavi}, N. and {Nienartowicz}, K. and {Ord{\'o}{\~n}ez-Blanco}, D. and {Panuzzo}, P. and {Portell}, J. and {Richards}, P.~J. and {Riello}, M. and {Seabroke}, G.~M. and {Tanga}, P. and {Th{\'e}venin}, F. and {Torra}, J. and {Els}, S.~G. and {Gracia-Abril}, G. and {Comoretto}, G. and {Garcia-Reinaldos}, M. and {Lock}, T. and {Mercier}, E. and {Altmann}, M. and {Andrae}, R. and {Astraatmadja}, T.~L. and {Bellas-Velidis}, I. and {Benson}, K. and {Berthier}, J. and {Blomme}, R. and {Busso}, G. and {Carry}, B. and {Cellino}, A. and {Clementini}, G. and {Cowell}, S. and {Creevey}, O. and {Cuypers}, J. and {Davidson}, M. and {De Ridder}, J. and {de Torres}, A. and {Delchambre}, L. and {Dell'Oro}, A. and {Ducourant}, C. and {Fr{\'e}mat}, Y. and {Garc{\'\i}a-Torres}, M. and {Gosset}, E. and {Halbwachs}, J. -L. and {Hambly}, N.~C. and {Harrison}, D.~L. and {Hauser}, M. and {Hestroffer}, D. and {Hodgkin}, S.~T. and {Huckle}, H.~E. and {Hutton}, A. and {Jasniewicz}, G. and {Jordan}, S. and {Kontizas}, M. and {Korn}, A.~J. and {Lanzafame}, A.~C. and {Manteiga}, M. and {Moitinho}, A. and {Muinonen}, K. and {Osinde}, J. and {Pancino}, E. and {Pauwels}, T. and {Petit}, J. -M. and {Recio-Blanco}, A. and {Robin}, A.~C. and {Sarro}, L.~M. and {Siopis}, C. and {Smith}, M. and {Smith}, K.~W. and {Sozzetti}, A. and {Thuillot}, W. and {van Reeven}, W. and {Viala}, Y. and {Abbas}, U. and {Abreu Aramburu}, A. and {Accart}, S. and {Aguado}, J.~J. and {Allan}, P.~M. and {Allasia}, W. and {Altavilla}, G. and {{\'A}lvarez}, M.~A. and {Alves}, J. and {Anderson}, R.~I. and {Andrei}, A.~H. and {Anglada Varela}, E. and {Antiche}, E. and {Antoja}, T. and {Ant{\'o}n}, S. and {Arcay}, B. and {Atzei}, A. and {Ayache}, L. and {Bach}, N. and {Baker}, S.~G. and {Balaguer-N{\'u}{\~n}ez}, L. and {Barache}, C. and {Barata}, C. and {Barbier}, A. and {Barblan}, F. and {Baroni}, M. and {Barrado y Navascu{\'e}s}, D. and {Barros}, M. and {Barstow}, M.~A. and {Becciani}, U. and {Bellazzini}, M. and {Bellei}, G. and {Bello Garc{\'\i}a}, A. and {Belokurov}, V. and {Bendjoya}, P. and {Berihuete}, A. and {Bianchi}, L. and {Bienaym{\'e}}, O. and {Billebaud}, F. and {Blagorodnova}, N. and {Blanco-Cuaresma}, S. and {Boch}, T. and {Bombrun}, A. and {Borrachero}, R. and {Bouquillon}, S. and {Bourda}, G. and {Bouy}, H. and {Bragaglia}, A. and {Breddels}, M.~A. and {Brouillet}, N. and {Br{\"u}semeister}, T. and {Bucciarelli}, B. and {Budnik}, F. and {Burgess}, P. and {Burgon}, R. and {Burlacu}, A. and {Busonero}, D. and {Buzzi}, R. and {Caffau}, E. and {Cambras}, J. and {Campbell}, H. and {Cancelliere}, R. and {Cantat-Gaudin}, T. and {Carlucci}, T. and {Carrasco}, J.~M. and {Castellani}, M. and {Charlot}, P. and {Charnas}, J. and {Charvet}, P. and {Chassat}, F. and {Chiavassa}, A. and {Clotet}, M. and {Cocozza}, G. and {Collins}, R.~S. and {Collins}, P. and {Costigan}, G. and {Crifo}, F. and {Cross}, N.~J.~G. and {Crosta}, M. and {Crowley}, C. and {Dafonte}, C. and {Damerdji}, Y. and {Dapergolas}, A. and {David}, P. and {David}, M. and {De Cat}, P. and {de Felice}, F. and {de Laverny}, P. and {De Luise}, F. and {De March}, R. and {de Martino}, D. and {de Souza}, R. and {Debosscher}, J. and {del Pozo}, E. and {Delbo}, M. and {Delgado}, A. and {Delgado}, H.~E. and {di Marco}, F. and {Di Matteo}, P. and {Diakite}, S. and {Distefano}, E. and {Dolding}, C. and {Dos Anjos}, S. and {Drazinos}, P. and {Dur{\'a}n}, J. and {Dzigan}, Y. and {Ecale}, E. and {Edvardsson}, B. and {Enke}, H. and {Erdmann}, M. and {Escolar}, D. and {Espina}, M. and {Evans}, N.~W. and {Eynard Bontemps}, G. and {Fabre}, C. and {Fabrizio}, M. and {Faigler}, S. and {Falc{\~a}o}, A.~J. and {Farr{\`a}s Casas}, M. and {Faye}, F. and {Federici}, L. and {Fedorets}, G. and {Fern{\'a}ndez-Hern{\'a}ndez}, J. and {Fernique}, P. and {Fienga}, A. and {Figueras}, F. and {Filippi}, F. and {Findeisen}, K. and {Fonti}, A. and {Fouesneau}, M. and {Fraile}, E. and {Fraser}, M. and {Fuchs}, J. and {Furnell}, R. and {Gai}, M. and {Galleti}, S. and {Galluccio}, L. and {Garabato}, D. and {Garc{\'\i}a-Sedano}, F. and {Gar{\'e}}, P. and {Garofalo}, A. and {Garralda}, N. and {Gavras}, P. and {Gerssen}, J. and {Geyer}, R. and {Gilmore}, G. and {Girona}, S. and {Giuffrida}, G. and {Gomes}, M. and {Gonz{\'a}lez-Marcos}, A. and {Gonz{\'a}lez-N{\'u}{\~n}ez}, J. and {Gonz{\'a}lez-Vidal}, J.~J. and {Granvik}, M. and {Guerrier}, A. and {Guillout}, P. and {Guiraud}, J. and {G{\'u}rpide}, A. and {Guti{\'e}rrez-S{\'a}nchez}, R. and {Guy}, L.~P. and {Haigron}, R. and {Hatzidimitriou}, D. and {Haywood}, M. and {Heiter}, U. and {Helmi}, A. and {Hobbs}, D. and {Hofmann}, W. and {Holl}, B. and {Holland}, G. and {Hunt}, J.~A.~S. and {Hypki}, A. and {Icardi}, V. and {Irwin}, M. and {Jevardat de Fombelle}, G. and {Jofr{\'e}}, P. and {Jonker}, P.~G. and {Jorissen}, A. and {Julbe}, F. and {Karampelas}, A. and {Kochoska}, A. and {Kohley}, R. and {Kolenberg}, K. and {Kontizas}, E. and {Koposov}, S.~E. and {Kordopatis}, G. and {Koubsky}, P. and {Kowalczyk}, A. and {Krone-Martins}, A. and {Kudryashova}, M. and {Kull}, I. and {Bachchan}, R.~K. and {Lacoste-Seris}, F. and {Lanza}, A.~F. and {Lavigne}, J. -B. and {Le Poncin-Lafitte}, C. and {Lebreton}, Y. and {Lebzelter}, T. and {Leccia}, S. and {Leclerc}, N. and {Lecoeur-Taibi}, I. and {Lemaitre}, V. and {Lenhardt}, H. and {Leroux}, F. and {Liao}, S. and {Licata}, E. and {Lindstr{\o}m}, H.~E.~P. and {Lister}, T.~A. and {Livanou}, E. and {Lobel}, A. and {L{\"o}ffler}, W. and {L{\'o}pez}, M. and {Lopez-Lozano}, A. and {Lorenz}, D. and {Loureiro}, T. and {MacDonald}, I. and {Magalh{\~a}es Fernandes}, T. and {Managau}, S. and {Mann}, R.~G. and {Mantelet}, G. and {Marchal}, O. and {Marchant}, J.~M. and {Marconi}, M. and {Marie}, J. and {Marinoni}, S. and {Marrese}, P.~M. and {Marschalk{\'o}}, G. and {Marshall}, D.~J. and {Mart{\'\i}n-Fleitas}, J.~M. and {Martino}, M. and {Mary}, N. and {Matijevi{\v{c}}}, G. and {Mazeh}, T. and {McMillan}, P.~J. and {Messina}, S. and {Mestre}, A. and {Michalik}, D. and {Millar}, N.~R. and {Miranda}, B.~M.~H. and {Molina}, D. and {Molinaro}, R. and {Molinaro}, M. and {Moln{\'a}r}, L. and {Moniez}, M. and {Montegriffo}, P. and {Monteiro}, D. and {Mor}, R. and {Mora}, A. and {Morbidelli}, R. and {Morel}, T. and {Morgenthaler}, S. and {Morley}, T. and {Morris}, D. and {Mulone}, A.~F. and {Muraveva}, T. and {Musella}, I. and {Narbonne}, J. and {Nelemans}, G. and {Nicastro}, L. and {Noval}, L. and {Ord{\'e}novic}, C. and {Ordieres-Mer{\'e}}, J. and {Osborne}, P. and {Pagani}, C. and {Pagano}, I. and {Pailler}, F. and {Palacin}, H. and {Palaversa}, L. and {Parsons}, P. and {Paulsen}, T. and {Pecoraro}, M. and {Pedrosa}, R. and {Pentik{\"a}inen}, H. and {Pereira}, J. and {Pichon}, B. and {Piersimoni}, A.~M. and {Pineau}, F. -X. and {Plachy}, E. and {Plum}, G. and {Poujoulet}, E. and {Pr{\v{s}}a}, A. and {Pulone}, L. and {Ragaini}, S. and {Rago}, S. and {Rambaux}, N. and {Ramos-Lerate}, M. and {Ranalli}, P. and {Rauw}, G. and {Read}, A. and {Regibo}, S. and {Renk}, F. and {Reyl{\'e}}, C. and {Ribeiro}, R.~A. and {Rimoldini}, L. and {Ripepi}, V. and {Riva}, A. and {Rixon}, G. and {Roelens}, M. and {Romero-G{\'o}mez}, M. and {Rowell}, N. and {Royer}, F. and {Rudolph}, A. and {Ruiz-Dern}, L. and {Sadowski}, G. and {Sagrist{\`a} Sell{\'e}s}, T. and {Sahlmann}, J. and {Salgado}, J. and {Salguero}, E. and {Sarasso}, M. and {Savietto}, H. and {Schnorhk}, A. and {Schultheis}, M. and {Sciacca}, E. and {Segol}, M. and {Segovia}, J.~C. and {Segransan}, D. and {Serpell}, E. and {Shih}, I. -C. and {Smareglia}, R. and {Smart}, R.~L. and {Smith}, C. and {Solano}, E. and {Solitro}, F. and {Sordo}, R. and {Soria Nieto}, S. and {Souchay}, J. and {Spagna}, A. and {Spoto}, F. and {Stampa}, U. and {Steele}, I.~A. and {Steidelm{\"u}ller}, H. and {Stephenson}, C.~A. and {Stoev}, H. and {Suess}, F.~F. and {S{\"u}veges}, M. and {Surdej}, J. and {Szabados}, L. and {Szegedi-Elek}, E. and {Tapiador}, D. and {Taris}, F. and {Tauran}, G. and {Taylor}, M.~B. and {Teixeira}, R. and {Terrett}, D. and {Tingley}, B. and {Trager}, S.~C. and {Turon}, C. and {Ulla}, A. and {Utrilla}, E. and {Valentini}, G. and {van Elteren}, A. and {Van Hemelryck}, E. and {van Leeuwen}, M. and {Varadi}, M. and {Vecchiato}, A. and {Veljanoski}, J. and {Via}, T. and {Vicente}, D. and {Vogt}, S. and {Voss}, H. and {Votruba}, V. and {Voutsinas}, S. and {Walmsley}, G. and {Weiler}, M. and {Weingrill}, K. and {Werner}, D. and {Wevers}, T. and {Whitehead}, G. and {Wyrzykowski}, {\L}. and {Yoldas}, A. and {{\v{Z}}erjal}, M. and {Zucker}, S. and {Zurbach}, C. and {Zwitter}, T. and {Alecu}, A. and {Allen}, M. and {Allende Prieto}, C. and {Amorim}, A. and {Anglada-Escud{\'e}}, G. and {Arsenijevic}, V. and {Azaz}, S. and {Balm}, P. and {Beck}, M. and {Bernstein}, H. -H. and {Bigot}, L. and {Bijaoui}, A. and {Blasco}, C. and {Bonfigli}, M. and {Bono}, G. and {Boudreault}, S. and {Bressan}, A. and {Brown}, S. and {Brunet}, P. -M. and {Bunclark}, P. and {Buonanno}, R. and {Butkevich}, A.~G. and {Carret}, C. and {Carrion}, C. and {Chemin}, L. and {Ch{\'e}reau}, F. and {Corcione}, L. and {Darmigny}, E. and {de Boer}, K.~S. and {de Teodoro}, P. and {de Zeeuw}, P.~T. and {Delle Luche}, C. and {Domingues}, C.~D. and {Dubath}, P. and {Fodor}, F. and {Fr{\'e}zouls}, B. and {Fries}, A. and {Fustes}, D. and {Fyfe}, D. and {Gallardo}, E. and {Gallegos}, J. and {Gardiol}, D. and {Gebran}, M. and {Gomboc}, A. and {G{\'o}mez}, A. and {Grux}, E. and {Gueguen}, A. and {Heyrovsky}, A. and {Hoar}, J. and {Iannicola}, G. and {Isasi Parache}, Y. and {Janotto}, A. -M. and {Joliet}, E. and {Jonckheere}, A. and {Keil}, R. and {Kim}, D. -W. and {Klagyivik}, P. and {Klar}, J. and {Knude}, J. and {Kochukhov}, O. and {Kolka}, I. and {Kos}, J. and {Kutka}, A. and {Lainey}, V. and {LeBouquin}, D. and {Liu}, C. and {Loreggia}, D. and {Makarov}, V.~V. and {Marseille}, M.~G. and {Martayan}, C. and {Martinez-Rubi}, O. and {Massart}, B. and {Meynadier}, F. and {Mignot}, S. and {Munari}, U. and {Nguyen}, A. -T. and {Nordlander}, T. and {Ocvirk}, P. and {O'Flaherty}, K.~S. and {Olias Sanz}, A. and {Ortiz}, P. and {Osorio}, J. and {Oszkiewicz}, D. and {Ouzounis}, A. and {Palmer}, M. and {Park}, P. and {Pasquato}, E. and {Peltzer}, C. and {Peralta}, J. and {P{\'e}turaud}, F. and {Pieniluoma}, T. and {Pigozzi}, E. and {Poels}, J. and {Prat}, G. and {Prod'homme}, T. and {Raison}, F. and {Rebordao}, J.~M. and {Risquez}, D. and {Rocca-Volmerange}, B. and {Rosen}, S. and {Ruiz-Fuertes}, M.~I. and {Russo}, F. and {Sembay}, S. and {Serraller Vizcaino}, I. and {Short}, A. and {Siebert}, A. and {Silva}, H. and {Sinachopoulos}, D. and {Slezak}, E. and {Soffel}, M. and {Sosnowska}, D. and {Strai{\v{z}}ys}, V. and {ter Linden}, M. and {Terrell}, D. and {Theil}, S. and {Tiede}, C. and {Troisi}, L. and {Tsalmantza}, P. and {Tur}, D. and {Vaccari}, M. and {Vachier}, F. and {Valles}, P. and {Van Hamme}, W. and {Veltz}, L. and {Virtanen}, J. and {Wallut}, J. -M. and {Wichmann}, R. and {Wilkinson}, M.~I. and {Ziaeepour}, H. and {Zschocke}, S.},
        title = "{The Gaia mission}",
      journal = {\aap},
         year = 2016,
        month = nov,
       volume = {595},
          eid = {A1},
        pages = {A1},
          doi = {10.1051/0004-6361/201629272},
archivePrefix = {arXiv},
       eprint = {1609.04153},
 primaryClass = {astro-ph.IM},
       adsurl = {https://ui.adsabs.harvard.edu/abs/2016A&A...595A...1G}
}

@ARTICLE{Gaia2021,
       author = {{Gaia Collaboration} and {Brown}, A.~G.~A. and {Vallenari}, A. and {Prusti}, T. and {de Bruijne}, J.~H.~J. and {Babusiaux}, C. and {Biermann}, M. and {Creevey}, O.~L. and {Evans}, D.~W. and {Eyer}, L. and {Hutton}, A. and {Jansen}, F. and {Jordi}, C. and {Klioner}, S.~A. and {Lammers}, U. and {Lindegren}, L. and {Luri}, X. and {Mignard}, F. and {Panem}, C. and {Pourbaix}, D. and {Randich}, S. and {Sartoretti}, P. and {Soubiran}, C. and {Walton}, N.~A. and {Arenou}, F. and {Bailer-Jones}, C.~A.~L. and {Bastian}, U. and {Cropper}, M. and {Drimmel}, R. and {Katz}, D. and {Lattanzi}, M.~G. and {van Leeuwen}, F. and {Bakker}, J. and {Cacciari}, C. and {Casta{\~n}eda}, J. and {De Angeli}, F. and {Ducourant}, C. and {Fabricius}, C. and {Fouesneau}, M. and {Fr{\'e}mat}, Y. and {Guerra}, R. and {Guerrier}, A. and {Guiraud}, J. and {Jean-Antoine Piccolo}, A. and {Masana}, E. and {Messineo}, R. and {Mowlavi}, N. and {Nicolas}, C. and {Nienartowicz}, K. and {Pailler}, F. and {Panuzzo}, P. and {Riclet}, F. and {Roux}, W. and {Seabroke}, G.~M. and {Sordo}, R. and {Tanga}, P. and {Th{\'e}venin}, F. and {Gracia-Abril}, G. and {Portell}, J. and {Teyssier}, D. and {Altmann}, M. and {Andrae}, R. and {Bellas-Velidis}, I. and {Benson}, K. and {Berthier}, J. and {Blomme}, R. and {Brugaletta}, E. and {Burgess}, P.~W. and {Busso}, G. and {Carry}, B. and {Cellino}, A. and {Cheek}, N. and {Clementini}, G. and {Damerdji}, Y. and {Davidson}, M. and {Delchambre}, L. and {Dell'Oro}, A. and {Fern{\'a}ndez-Hern{\'a}ndez}, J. and {Galluccio}, L. and {Garc{\'\i}a-Lario}, P. and {Garcia-Reinaldos}, M. and {Gonz{\'a}lez-N{\'u}{\~n}ez}, J. and {Gosset}, E. and {Haigron}, R. and {Halbwachs}, J. -L. and {Hambly}, N.~C. and {Harrison}, D.~L. and {Hatzidimitriou}, D. and {Heiter}, U. and {Hern{\'a}ndez}, J. and {Hestroffer}, D. and {Hodgkin}, S.~T. and {Holl}, B. and {Jan{\ss}en}, K. and {Jevardat de Fombelle}, G. and {Jordan}, S. and {Krone-Martins}, A. and {Lanzafame}, A.~C. and {L{\"o}ffler}, W. and {Lorca}, A. and {Manteiga}, M. and {Marchal}, O. and {Marrese}, P.~M. and {Moitinho}, A. and {Mora}, A. and {Muinonen}, K. and {Osborne}, P. and {Pancino}, E. and {Pauwels}, T. and {Petit}, J. -M. and {Recio-Blanco}, A. and {Richards}, P.~J. and {Riello}, M. and {Rimoldini}, L. and {Robin}, A.~C. and {Roegiers}, T. and {Rybizki}, J. and {Sarro}, L.~M. and {Siopis}, C. and {Smith}, M. and {Sozzetti}, A. and {Ulla}, A. and {Utrilla}, E. and {van Leeuwen}, M. and {van Reeven}, W. and {Abbas}, U. and {Abreu Aramburu}, A. and {Accart}, S. and {Aerts}, C. and {Aguado}, J.~J. and {Ajaj}, M. and {Altavilla}, G. and {{\'A}lvarez}, M.~A. and {{\'A}lvarez Cid-Fuentes}, J. and {Alves}, J. and {Anderson}, R.~I. and {Anglada Varela}, E. and {Antoja}, T. and {Audard}, M. and {Baines}, D. and {Baker}, S.~G. and {Balaguer-N{\'u}{\~n}ez}, L. and {Balbinot}, E. and {Balog}, Z. and {Barache}, C. and {Barbato}, D. and {Barros}, M. and {Barstow}, M.~A. and {Bartolom{\'e}}, S. and {Bassilana}, J. -L. and {Bauchet}, N. and {Baudesson-Stella}, A. and {Becciani}, U. and {Bellazzini}, M. and {Bernet}, M. and {Bertone}, S. and {Bianchi}, L. and {Blanco-Cuaresma}, S. and {Boch}, T. and {Bombrun}, A. and {Bossini}, D. and {Bouquillon}, S. and {Bragaglia}, A. and {Bramante}, L. and {Breedt}, E. and {Bressan}, A. and {Brouillet}, N. and {Bucciarelli}, B. and {Burlacu}, A. and {Busonero}, D. and {Butkevich}, A.~G. and {Buzzi}, R. and {Caffau}, E. and {Cancelliere}, R. and {C{\'a}novas}, H. and {Cantat-Gaudin}, T. and {Carballo}, R. and {Carlucci}, T. and {Carnerero}, M.~I. and {Carrasco}, J.~M. and {Casamiquela}, L. and {Castellani}, M. and {Castro-Ginard}, A. and {Castro Sampol}, P. and {Chaoul}, L. and {Charlot}, P. and {Chemin}, L. and {Chiavassa}, A. and {Cioni}, M. -R.~L. and {Comoretto}, G. and {Cooper}, W.~J. and {Cornez}, T. and {Cowell}, S. and {Crifo}, F. and {Crosta}, M. and {Crowley}, C. and {Dafonte}, C. and {Dapergolas}, A. and {David}, M. and {David}, P. and {de Laverny}, P. and {De Luise}, F. and {De March}, R. and {De Ridder}, J. and {de Souza}, R. and {de Teodoro}, P. and {de Torres}, A. and {del Peloso}, E.~F. and {del Pozo}, E. and {Delbo}, M. and {Delgado}, A. and {Delgado}, H.~E. and {Delisle}, J. -B. and {Di Matteo}, P. and {Diakite}, S. and {Diener}, C. and {Distefano}, E. and {Dolding}, C. and {Eappachen}, D. and {Edvardsson}, B. and {Enke}, H. and {Esquej}, P. and {Fabre}, C. and {Fabrizio}, M. and {Faigler}, S. and {Fedorets}, G. and {Fernique}, P. and {Fienga}, A. and {Figueras}, F. and {Fouron}, C. and {Fragkoudi}, F. and {Fraile}, E. and {Franke}, F. and {Gai}, M. and {Garabato}, D. and {Garcia-Gutierrez}, A. and {Garc{\'\i}a-Torres}, M. and {Garofalo}, A. and {Gavras}, P. and {Gerlach}, E. and {Geyer}, R. and {Giacobbe}, P. and {Gilmore}, G. and {Girona}, S. and {Giuffrida}, G. and {Gomel}, R. and {Gomez}, A. and {Gonzalez-Santamaria}, I. and {Gonz{\'a}lez-Vidal}, J.~J. and {Granvik}, M. and {Guti{\'e}rrez-S{\'a}nchez}, R. and {Guy}, L.~P. and {Hauser}, M. and {Haywood}, M. and {Helmi}, A. and {Hidalgo}, S.~L. and {Hilger}, T. and {H{\l}adczuk}, N. and {Hobbs}, D. and {Holland}, G. and {Huckle}, H.~E. and {Jasniewicz}, G. and {Jonker}, P.~G. and {Juaristi Campillo}, J. and {Julbe}, F. and {Karbevska}, L. and {Kervella}, P. and {Khanna}, S. and {Kochoska}, A. and {Kontizas}, M. and {Kordopatis}, G. and {Korn}, A.~J. and {Kostrzewa-Rutkowska}, Z. and {Kruszy{\'n}ska}, K. and {Lambert}, S. and {Lanza}, A.~F. and {Lasne}, Y. and {Le Campion}, J. -F. and {Le Fustec}, Y. and {Lebreton}, Y. and {Lebzelter}, T. and {Leccia}, S. and {Leclerc}, N. and {Lecoeur-Taibi}, I. and {Liao}, S. and {Licata}, E. and {Lindstr{\o}m}, E.~P. and {Lister}, T.~A. and {Livanou}, E. and {Lobel}, A. and {Madrero Pardo}, P. and {Managau}, S. and {Mann}, R.~G. and {Marchant}, J.~M. and {Marconi}, M. and {Marcos Santos}, M.~M.~S. and {Marinoni}, S. and {Marocco}, F. and {Marshall}, D.~J. and {Martin Polo}, L. and {Mart{\'\i}n-Fleitas}, J.~M. and {Masip}, A. and {Massari}, D. and {Mastrobuono-Battisti}, A. and {Mazeh}, T. and {McMillan}, P.~J. and {Messina}, S. and {Michalik}, D. and {Millar}, N.~R. and {Mints}, A. and {Molina}, D. and {Molinaro}, R. and {Moln{\'a}r}, L. and {Montegriffo}, P. and {Mor}, R. and {Morbidelli}, R. and {Morel}, T. and {Morris}, D. and {Mulone}, A.~F. and {Munoz}, D. and {Muraveva}, T. and {Murphy}, C.~P. and {Musella}, I. and {Noval}, L. and {Ord{\'e}novic}, C. and {Orr{\`u}}, G. and {Osinde}, J. and {Pagani}, C. and {Pagano}, I. and {Palaversa}, L. and {Palicio}, P.~A. and {Panahi}, A. and {Pawlak}, M. and {Pe{\~n}alosa Esteller}, X. and {Penttil{\"a}}, A. and {Piersimoni}, A.~M. and {Pineau}, F. -X. and {Plachy}, E. and {Plum}, G. and {Poggio}, E. and {Poretti}, E. and {Poujoulet}, E. and {Pr{\v{s}}a}, A. and {Pulone}, L. and {Racero}, E. and {Ragaini}, S. and {Rainer}, M. and {Raiteri}, C.~M. and {Rambaux}, N. and {Ramos}, P. and {Ramos-Lerate}, M. and {Re Fiorentin}, P. and {Regibo}, S. and {Reyl{\'e}}, C. and {Ripepi}, V. and {Riva}, A. and {Rixon}, G. and {Robichon}, N. and {Robin}, C. and {Roelens}, M. and {Rohrbasser}, L. and {Romero-G{\'o}mez}, M. and {Rowell}, N. and {Royer}, F. and {Rybicki}, K.~A. and {Sadowski}, G. and {Sagrist{\`a} Sell{\'e}s}, A. and {Sahlmann}, J. and {Salgado}, J. and {Salguero}, E. and {Samaras}, N. and {Sanchez Gimenez}, V. and {Sanna}, N. and {Santove{\~n}a}, R. and {Sarasso}, M. and {Schultheis}, M. and {Sciacca}, E. and {Segol}, M. and {Segovia}, J.~C. and {S{\'e}gransan}, D. and {Semeux}, D. and {Shahaf}, S. and {Siddiqui}, H.~I. and {Siebert}, A. and {Siltala}, L. and {Slezak}, E. and {Smart}, R.~L. and {Solano}, E. and {Solitro}, F. and {Souami}, D. and {Souchay}, J. and {Spagna}, A. and {Spoto}, F. and {Steele}, I.~A. and {Steidelm{\"u}ller}, H. and {Stephenson}, C.~A. and {S{\"u}veges}, M. and {Szabados}, L. and {Szegedi-Elek}, E. and {Taris}, F. and {Tauran}, G. and {Taylor}, M.~B. and {Teixeira}, R. and {Thuillot}, W. and {Tonello}, N. and {Torra}, F. and {Torra}, J. and {Turon}, C. and {Unger}, N. and {Vaillant}, M. and {van Dillen}, E. and {Vanel}, O. and {Vecchiato}, A. and {Viala}, Y. and {Vicente}, D. and {Voutsinas}, S. and {Weiler}, M. and {Wevers}, T. and {Wyrzykowski}, {\L}. and {Yoldas}, A. and {Yvard}, P. and {Zhao}, H. and {Zorec}, J. and {Zucker}, S. and {Zurbach}, C. and {Zwitter}, T.},
        title = "{Gaia Early Data Release 3. Summary of the contents and survey properties}",
      journal = {\aap},
         year = 2021,
        month = may,
       volume = {649},
          eid = {A1},
        pages = {A1},
          doi = {10.1051/0004-6361/202039657},
archivePrefix = {arXiv},
       eprint = {2012.01533},
 primaryClass = {astro-ph.GA},
       adsurl = {https://ui.adsabs.harvard.edu/abs/2021A&A...649A...1G}
}

@ARTICLE{Mathis1977,
       author = {{Mathis}, J.~S. and {Rumpl}, W. and {Nordsieck}, K.~H.},
        title = "{The size distribution of interstellar grains.}",
      journal = {\apj},
         year = 1977,
        month = oct,
       volume = {217},
        pages = {425-433},
          doi = {10.1086/155591},
       adsurl = {https://ui.adsabs.harvard.edu/abs/1977ApJ...217..425M}
}

@ARTICLE{Goldreich1965,
       author = {{Goldreich}, P. and {Lynden-Bell}, D.},
        title = "{II. Spiral arms as sheared gravitational instabilities}",
      journal = {\mnras},
         year = 1965,
        month = jan,
       volume = {130},
        pages = {125},
          doi = {10.1093/mnras/130.2.125},
       adsurl = {https://ui.adsabs.harvard.edu/abs/1965MNRAS.130..125G}
}

@Article{hoffman2013,
  author     = {Hoffman, Matthew D. and Blei, David M. and Wang, Chong and Paisley, John},
  journal    = {J. Mach. Learn. Res.},
  title      = {Stochastic variational inference},
  year       = {2013},
  issn       = {1532-4435},
  month      = {may},
  number     = {1},
  pages      = {1303–1347},
  volume     = {14},
  issue_date = {January 2013},
  numpages   = {45},
  publisher  = {JMLR.org},
}

@Article{Bingham2019,
  author  = {Eli Bingham and Jonathan P. Chen and Martin Jankowiak and Fritz Obermeyer and Neeraj Pradhan and Theofanis Karaletsos and Rohit Singh and Paul Szerlip and Paul Horsfall and Noah D. Goodman},
  journal = {Journal of Machine Learning Research},
  title   = {Pyro: Deep Universal Probabilistic Programming},
  year    = {2019},
  number  = {28},
  pages   = {1--6},
  volume  = {20},
  url     = {http://jmlr.org/papers/v20/18-403.html},
}

@ARTICLE{emcee,
       author = {{Foreman-Mackey}, Daniel and {Hogg}, David W. and {Lang}, Dustin and
        {Goodman}, Jonathan},
        title = "{emcee: The MCMC Hammer}",
      journal = {Publications of the Astronomical Society of the Pacific},
         year = 2013,
        month = Mar,
       volume = {125},
        pages = {306},
          doi = {10.1086/670067},
archivePrefix = {arXiv},
       eprint = {1202.3665},
       adsurl = {https://ui.adsabs.harvard.edu/#abs/2013PASP..125..306F}
}

@ARTICLE{Oya2022,
       author = {{Oya}, Yoko and {Kibukawa}, Hirofumi and {Miyake}, Shota and {Yamamoto}, Satoshi},
        title = "{FERIA: Flat Envelope Model with Rotation and Infall under Angular Momentum Conservation}",
      journal = {\pasp},
         year = 2022,
        month = sep,
       volume = {134},
       number = {1039},
          eid = {094301},
        pages = {094301},
          doi = {10.1088/1538-3873/ac8839},
archivePrefix = {arXiv},
       eprint = {2208.04581},
 primaryClass = {astro-ph.IM},
       adsurl = {https://ui.adsabs.harvard.edu/abs/2022PASP..134i4301O}
}

@INPROCEEDINGS{Pineda2023,
       author = {{Pineda}, J.~E. and {Arzoumanian}, D. and {Andre}, P. and {Friesen}, R.~K. and {Zavagno}, A. and {Clarke}, S.~D. and {Inoue}, T. and {Chen}, C. and {Lee}, Y. and {Soler}, J.~D. and {Kuffmeier}, M.},
        title = "{From Bubbles and Filaments to Cores and Disks: Gas Gathering and Growth of Structure Leading to the Formation of Stellar Systems}",
    booktitle = {Protostars and Planets VII},
         year = 2023,
       editor = {{Inutsuka}, S. and {Aikawa}, Y. and {Muto}, T. and {Tomida}, K. and {Tamura}, M.},
       series = {Astronomical Society of the Pacific Conference Series},
       volume = {534},
        month = jul,
        pages = {233},
          doi = {10.48550/arXiv.2205.03935},
archivePrefix = {arXiv},
       eprint = {2205.03935},
 primaryClass = {astro-ph.GA},
       adsurl = {https://ui.adsabs.harvard.edu/abs/2023ASPC..534..233P}
}

@ARTICLE{Valdivia-Mena2022,
       author = {{Valdivia-Mena}, M.~T. and {Pineda}, J.~E. and {Segura-Cox}, D.~M. and {Caselli}, P. and {Neri}, R. and {L{\'o}pez-Sepulcre}, A. and {Cunningham}, N. and {Bouscasse}, L. and {Semenov}, D. and {Henning}, Th. and {Pi{\'e}tu}, V. and {Chapillon}, E. and {Dutrey}, A. and {Fuente}, A. and {Guilloteau}, S. and {Hsieh}, T.~H. and {Jim{\'e}nez-Serra}, I. and {Marino}, S. and {Maureira}, M.~J. and {Smirnov-Pinchukov}, G.~V. and {Tafalla}, M. and {Zhao}, B.},
        title = "{PRODIGE - envelope to disk with NOEMA. I. A 3000 au streamer feeding a Class I protostar}",
      journal = {\aap},
         year = 2022,
        month = nov,
       volume = {667},
          eid = {A12},
        pages = {A12},
          doi = {10.1051/0004-6361/202243310},
archivePrefix = {arXiv},
       eprint = {2208.01023},
 primaryClass = {astro-ph.GA},
       adsurl = {https://ui.adsabs.harvard.edu/abs/2022A&A...667A..12V}
}

@ARTICLE{Ginski2021,
       author = {{Ginski}, Christian and {Facchini}, Stefano and {Huang}, Jane and {Benisty}, Myriam and {Vaendel}, Dennis and {Stapper}, Lucas and {Dominik}, Carsten and {Bae}, Jaehan and {M{\'e}nard}, Fran{\c{c}}ois and {Muro-Arena}, Gabriela and {Hogerheijde}, Michiel R. and {McClure}, Melissa and {van Holstein}, Rob G. and {Birnstiel}, Tilman and {Boehler}, Yann and {Bohn}, Alexander and {Flock}, Mario and {Mamajek}, Eric E. and {Manara}, Carlo F. and {Pinilla}, Paola and {Pinte}, Christophe and {Ribas}, {\'A}lvaro},
        title = "{Disk Evolution Study Through Imaging of Nearby Young Stars (DESTINYS): Late Infall Causing Disk Misalignment and Dynamic Structures in SU Aur}",
      journal = {\apjl},
         year = 2021,
        month = feb,
       volume = {908},
       number = {2},
          eid = {L25},
        pages = {L25},
          doi = {10.3847/2041-8213/abdf57},
archivePrefix = {arXiv},
       eprint = {2102.08781},
 primaryClass = {astro-ph.EP},
       adsurl = {https://ui.adsabs.harvard.edu/abs/2021ApJ...908L..25G}
}

@ARTICLE{Taniguchi2024,
       author = {{Taniguchi}, Kotomi and {Pineda}, Jaime E. and {Caselli}, Paola and {Shimoikura}, Tomomi and {Friesen}, Rachel K. and {Segura-Cox}, Dominique M. and {Schmiedeke}, Anika},
        title = "{The Reservoir of the Per-emb-2 Streamer}",
      journal = {\apj},
         year = 2024,
        month = apr,
       volume = {965},
       number = {2},
          eid = {162},
        pages = {162},
          doi = {10.3847/1538-4357/ad2fa1},
archivePrefix = {arXiv},
       eprint = {2402.19099},
 primaryClass = {astro-ph.GA},
       adsurl = {https://ui.adsabs.harvard.edu/abs/2024ApJ...965..162T}
}

@ARTICLE{Anderson2019,
       author = {{Anderson}, Dana E. and {Blake}, Geoffrey A. and {Bergin}, Edwin A. and {Zhang}, Ke and {Carpenter}, John M. and {Schwarz}, Kamber R. and {Huang}, Jane and {{\"O}berg}, Karin I.},
        title = "{Probing the Gas Content of Late-stage Protoplanetary Disks with N$_{2}$H$^{+}$}",
      journal = {\apj},
         year = 2019,
        month = aug,
       volume = {881},
       number = {2},
          eid = {127},
        pages = {127},
          doi = {10.3847/1538-4357/ab2cb5},
       adsurl = {https://ui.adsabs.harvard.edu/abs/2019ApJ...881..127A}
}

@ARTICLE{Trapman2018,
       author = {{Trapman}, Leon and {Zhang}, Ke and {van't Hoff}, Merel L.~R. and {Hogerheijde}, Michiel R. and {Bergin}, Edwin A.},
        title = "{A Novel Way of Measuring the Gas Disk Mass of Protoplanetary Disks Using N$_{2}$H$^{+}$ and C$^{18}$O}",
      journal = {\apjl},
         year = 2022,
        month = feb,
       volume = {926},
       number = {1},
          eid = {L2},
        pages = {L2},
          doi = {10.3847/2041-8213/ac4f47},
archivePrefix = {arXiv},
       eprint = {2201.09900},
 primaryClass = {astro-ph.EP},
       adsurl = {https://ui.adsabs.harvard.edu/abs/2022ApJ...926L...2T}
}

@ARTICLE{Trapman2025,
       author = {{Trapman}, Leon and {Longarini}, Cristiano and {Rosotti}, Giovanni P. and {Andrews}, Sean M. and {Bae}, Jaehan and {Barraza-Alfaro}, Marcelo and {Benisty}, Myriam and {Cataldi}, Gianni and {Curone}, Pietro and {Czekala}, Ian and {Facchini}, Stefano and {Fasano}, Daniele and {Flock}, Mario and {Fukagawa}, Misato and {Galloway-Sprietsma}, Maria and {Garg}, Himanshi and {Hall}, Cassandra and {Huang}, Jane and {Ilee}, John D. and {Izquierdo}, Andres F. and {Kanagawa}, Kazuhiro and {Lesur}, Geoffroy and {Lodato}, Giuseppe and {Loomis}, Ryan A. and {Orihara}, Ryuta and {Paneque-Carreno}, Teresa and {Pinte}, Christophe and {Price}, Daniel and {Stadler}, Jochen and {Teague}, Richard and {van Terwisga}, Sierk and {Testi}, Leonardo and {Yen}, Hsi-Wei and {Wafflard-Fernandez}, Gaylor and {Wilner}, David J. and {Winter}, Andrew J. and {W{\"o}lfer}, Lisa and {Yoshida}, Tomohiro C. and {Zawadzki}, Brianna and {Zhang}, Ke},
        title = "{exoALMA. XIII. Gas Masses from N$_{2}$H$^{+}$ and C$^{18}$O: A Comparison of Measurement Techniques for Protoplanetary Gas Disk Masses}",
      journal = {\apjl},
         year = 2025,
        month = may,
       volume = {984},
       number = {1},
          eid = {L18},
        pages = {L18},
          doi = {10.3847/2041-8213/adc430},
       adsurl = {https://ui.adsabs.harvard.edu/abs/2025ApJ...984L..18T}
}

@Article{Bruderer2012,
  author        = {{Bruderer}, S. and {van Dishoeck}, E.~F. and {Doty}, S.~D. and {Herczeg}, G.~J.},
  title         = {{The warm gas atmosphere of the HD 100546 disk seen by Herschel. Evidence of a gas-rich, carbon-poor atmosphere?}},
  journal       = {\aap},
  year          = {2012},
  volume        = {541},
  pages         = {A91},
  month         = may,
  adsurl        = {http://adsabs.harvard.edu/abs/2012A%26A...541A..91B},
  archiveprefix = {arXiv},
  doi           = {10.1051/0004-6361/201118218},
  eid           = {A91},
  eprint        = {1201.4860},
  primaryclass  = {astro-ph.SR},
}

@Article{Bruderer2013,
  author        = {{Bruderer}, S.},
  title         = {{Survival of molecular gas in cavities of transition disks. I. CO}},
  journal       = {\aap},
  year          = {2013},
  volume        = {559},
  pages         = {A46},
  month         = nov,
  adsurl        = {http://adsabs.harvard.edu/abs/2013A%26A...559A..46B},
  archiveprefix = {arXiv},
  doi           = {10.1051/0004-6361/201321171},
  eid           = {A46},
  eprint        = {1308.2966},
  primaryclass  = {astro-ph.SR},
}

@ARTICLE{Hsieh2024,
       author = {{Hsieh}, Cheng-Han and {Arce}, H{\'e}ctor G. and {Maureira}, Mar{\'\i}a Jos{\'e} and {Pineda}, Jaime E. and {Segura-Cox}, Dominique and {Mardones}, Diego and {Dunham}, Michael M. and {Arun}, Aiswarya},
        title = "{The ALMA Legacy Survey of Class 0/I Disks in Corona australis, Aquila, chaMaeleon, oPhiuchus north, Ophiuchus, Serpens (CAMPOS). I. Evolution of Protostellar Disk Radii}",
      journal = {\apj},
         year = 2024,
        month = oct,
       volume = {973},
       number = {2},
          eid = {138},
        pages = {138},
          doi = {10.3847/1538-4357/ad6152},
archivePrefix = {arXiv},
       eprint = {2404.02809},
 primaryClass = {astro-ph.SR},
       adsurl = {https://ui.adsabs.harvard.edu/abs/2024ApJ...973..138H}
}

@ARTICLE{Hsieh2025,
       author = {{Hsieh}, Cheng-Han and {Arce}, H{\'e}ctor G. and {Jos{\'e} Maureira}, Mar{\'\i}a and {Pineda}, Jaime E. and {Segura-Cox}, Dominique and {Mardones}, Diego and {Dunham}, Michael M. and {Li}, Hui and {Offner}, Stella S.~R.},
        title = "{CAMPOS II. The onset of protostellar disk substructures and planet formation}",
      journal = {arXiv e-prints},
         year = 2025,
        month = apr,
          eid = {arXiv:2504.11577},
        pages = {arXiv:2504.11577},
          doi = {10.48550/arXiv.2504.11577},
archivePrefix = {arXiv},
       eprint = {2504.11577},
 primaryClass = {astro-ph.EP},
       adsurl = {https://ui.adsabs.harvard.edu/abs/2025arXiv250411577H}
}

@ARTICLE{Morbidelli2020,
       author = {{Morbidelli}, A.},
        title = "{Planet formation by pebble accretion in ringed disks}",
      journal = {\aap},
         year = 2020,
        month = jun,
       volume = {638},
          eid = {A1},
        pages = {A1},
          doi = {10.1051/0004-6361/202037983},
archivePrefix = {arXiv},
       eprint = {2004.04942},
 primaryClass = {astro-ph.EP},
       adsurl = {https://ui.adsabs.harvard.edu/abs/2020A&A...638A...1M}
}

@ARTICLE{vanKempen2007,
       author = {{van Kempen}, T.~A. and {van Dishoeck}, E.~F. and {Brinch}, C. and
        {Hogerheijde}, M.~R.},
        title = "{Searching for gas-rich disks around T Tauri stars in Lupus}",
      journal = {\aap},
         year = 2007,
        month = Jan,
       volume = {461},
        pages = {983-990},
          doi = {10.1051/0004-6361:20065174},
archivePrefix = {arXiv},
       eprint = {astro-ph/0606302},
 primaryClass = {astro-ph},
       adsurl = {https://ui.adsabs.harvard.edu/#abs/2007A&A...461..983V}
}

@ARTICLE{Antilen2023,
       author = {{Antilen}, Juanita and {Casassus}, Simon and {Cieza}, Lucas A. and {Gonz{\'a}lez-Ruilova}, Camilo},
        title = "{Gas distribution in ODISEA sources from ALMA long-baseline observations in $^{12}$CO(2-1)}",
      journal = {\mnras},
         year = 2023,
        month = jun,
       volume = {522},
       number = {2},
        pages = {2611-2627},
          doi = {10.1093/mnras/stad975},
archivePrefix = {arXiv},
       eprint = {2304.15002},
 primaryClass = {astro-ph.EP},
       adsurl = {https://ui.adsabs.harvard.edu/abs/2023MNRAS.522.2611A}
}

@ARTICLE{Keppler2019,
       author = {{Keppler}, M. and {Teague}, R. and {Bae}, J. and {Benisty}, M. and {Henning}, T. and {van Boekel}, R. and {Chapillon}, E. and {Pinilla}, P. and {Williams}, J.~P. and {Bertrang}, G.~H. -M. and {Facchini}, S. and {Flock}, M. and {Ginski}, Ch. and {Juhasz}, A. and {Klahr}, H. and {Liu}, Y. and {M{\"u}ller}, A. and {P{\'e}rez}, L.~M. and {Pohl}, A. and {Rosotti}, G. and {Samland}, M. and {Semenov}, D.},
        title = "{Highly structured disk around the planet host PDS 70 revealed by high-angular resolution observations with ALMA}",
      journal = {\aap},
         year = 2019,
        month = may,
       volume = {625},
          eid = {A118},
        pages = {A118},
          doi = {10.1051/0004-6361/201935034},
archivePrefix = {arXiv},
       eprint = {1902.07639},
 primaryClass = {astro-ph.EP},
       adsurl = {https://ui.adsabs.harvard.edu/abs/2019A&A...625A.118K}
}

@ARTICLE{Loomis2025,
       author = {{Loomis}, Ryan A. and {Facchini}, Stefano and {Benisty}, Myriam and {Curone}, Pietro and {Ilee}, John D. and {Cataldi}, Gianni and {Yen}, Hsi-Wei and {Teague}, Richard and {Pinte}, Christophe and {Huang}, Jane and {Garg}, Himanshi and {Orihara}, Ryuta and {Czekala}, Ian and {Zawadzki}, Brianna and {Andrews}, Sean M. and {Wilner}, David J. and {Bae}, Jaehan and {Barraza-Alfaro}, Marcelo and {Fasano}, Daniele and {Flock}, Mario and {Fukagawa}, Misato and {Galloway-Sprietsma}, Maria and {Izquierdo}, Andr{\'e}s F. and {Kanagawa}, Kazuhiro and {Lesur}, Geoffroy and {Longarini}, Cristiano and {Menard}, Francois and {Price}, Daniel J. and {Rosotti}, Giovanni and {Stadler}, Jochen and {Wafflard-Fernandez}, Gaylor and {W{\"o}lfer}, Lisa and {Yoshida}, Tomohiro C.},
        title = "{exoALMA. II. Data Calibration and Imaging Pipeline}",
      journal = {\apjl},
         year = 2025,
        month = may,
       volume = {984},
       number = {1},
          eid = {L7},
        pages = {L7},
          doi = {10.3847/2041-8213/adc43a},
archivePrefix = {arXiv},
       eprint = {2504.19870},
 primaryClass = {astro-ph.EP},
       adsurl = {https://ui.adsabs.harvard.edu/abs/2025ApJ...984L...7L}
}

@Article{Macias2021,
  author        = {{Mac{\'\i}as}, E. and {Guerra-Alvarado}, O. and {Carrasco-Gonz{\'a}lez}, C. and {Ribas}, {\'A}. and {Espaillat}, C.~C. and {Huang}, J. and {Andrews}, S.~M.},
  journal       = {\aap},
  title         = {{Characterizing the dust content of disk substructures in TW Hydrae}},
  year          = {2021},
  month         = apr,
  pages         = {A33},
  volume        = {648},
  adsurl        = {https://ui.adsabs.harvard.edu/abs/2021A&A...648A..33M},
  archiveprefix = {arXiv},
  doi           = {10.1051/0004-6361/202039812},
  eid           = {A33},
  eprint        = {2102.04648},
  primaryclass  = {astro-ph.EP},
}

@ARTICLE{Garufi2021,
       author = {{Garufi}, A. and {Podio}, L. and {Codella}, C. and {Fedele}, D. and {Bianchi}, E. and {Favre}, C. and {Bacciotti}, F. and {Ceccarelli}, C. and {Mercimek}, S. and {Rygl}, K. and {Teague}, R. and {Testi}, L.},
        title = "{ALMA chemical survey of disk-outflow sources in Taurus (ALMA-DOT). V. Sample, overview, and demography of disk molecular emission}",
      journal = {\aap},
         year = 2021,
        month = jan,
       volume = {645},
          eid = {A145},
        pages = {A145},
          doi = {10.1051/0004-6361/202039483},
archivePrefix = {arXiv},
       eprint = {2012.07667},
 primaryClass = {astro-ph.GA},
       adsurl = {https://ui.adsabs.harvard.edu/abs/2021A&A...645A.145G}
}

@Article{Keyte2023,
  author        = {{Keyte}, Luke and {Kama}, Mihkel and {Booth}, Alice S. and {Bergin}, Edwin A. and {Cleeves}, L. Ilsedore and {van Dishoeck}, Ewine F. and {Drozdovskaya}, Maria N. and {Furuya}, Kenji and {Rawlings}, Jonathan and {Shorttle}, Oliver and {Walsh}, Catherine},
  journal       = {Nature Astronomy},
  title         = {{Azimuthal C/O variations in a planet-forming disk}},
  year          = {2023},
  month         = jun,
  pages         = {684-693},
  volume        = {7},
  adsurl        = {https://ui.adsabs.harvard.edu/abs/2023NatAs...7..684K},
  archiveprefix = {arXiv},
  doi           = {10.1038/s41550-023-01951-9},
  eprint        = {2303.08927},
  primaryclass  = {astro-ph.EP},
}

@ARTICLE{Reboussin2015,
       author = {{Reboussin}, L. and {Guilloteau}, S. and {Simon}, M. and {Grosso}, N. and {Wakelam}, V. and {Di Folco}, E. and {Dutrey}, A. and {Pi{\'e}tu}, V.},
        title = "{Sensitive survey for $^{13}$CO, CN, H$_{2}$CO, and SO in the disks of T Tauri and Herbig Ae stars. II. Stars in {\ensuremath{\rho}} Ophiuchi and upper Scorpius}",
      journal = {\aap},
         year = 2015,
        month = jun,
       volume = {578},
          eid = {A31},
        pages = {A31},
          doi = {10.1051/0004-6361/201525705},
archivePrefix = {arXiv},
       eprint = {1504.04542},
 primaryClass = {astro-ph.SR},
       adsurl = {https://ui.adsabs.harvard.edu/abs/2015A&A...578A..31R}
}

@ARTICLE{Shirley2015,
       author = {{Shirley}, Yancy L.},
        title = "{The Critical Density and the Effective Excitation Density of Commonly Observed Molecular Dense Gas Tracers}",
      journal = {\pasp},
         year = 2015,
        month = mar,
       volume = {127},
       number = {949},
        pages = {299},
          doi = {10.1086/680342},
archivePrefix = {arXiv},
       eprint = {1501.01629},
 primaryClass = {astro-ph.IM},
       adsurl = {https://ui.adsabs.harvard.edu/abs/2015PASP..127..299S}
}

@ARTICLE{Hiraoka1994,
       author = {{Hiraoka}, Kenzo and {Ohashi}, Nagayuki and {Kihara}, Yosihide and {Yamamoto}, Kazuyosi and {Sato}, Tetsuya and {Yamashita}, Akihiro},
        title = "{Formation of formaldehyde and methanol from the reactions of H atoms with solid CO at 10-20 K}",
      journal = {Chemical Physics Letters},
         year = 1994,
        month = nov,
       volume = {229},
       number = {4},
        pages = {408-414},
          doi = {10.1016/0009-2614(94)01066-8},
       adsurl = {https://ui.adsabs.harvard.edu/abs/1994CPL...229..408H}
}
\bibliographystyle{aasjournal}

\end{document}